\documentclass[fleqn,usenatbib]{mnras}

\usepackage{newtxtext}
\usepackage[varvw]{newtxmath}

\usepackage[T1]{fontenc}
\usepackage{paralist}
\usepackage{multirow}
\usepackage{siunitx}
\usepackage{makecell}

\usepackage{graphicx}	
\usepackage{amsmath}	

\usepackage[dvipsnames]{xcolor}

\usepackage[normalem]{ulem}

\newcommand{\kms}{km\,s$^{-1}$}
\newcommand{\Htwo}{H$_{2}$}
\newcommand{\hi}{H\,{\sc i}}
\newcommand{\co}{CO}
\newcommand{\hix}{H\,{\sc ix}}

\title[The H IX galaxy survey IV]{The H IX galaxy survey IV: What does the molecular gas reveal in \hi\ eXtreme galaxies?}

\author[V. A. Capa et al.]{
V. A. Capa,$^{1}$
H. D\'enes,$^{1, 2}$\thanks{E-mail: helgadenes@gmail.com}
K. A. Lutz, $^{3}$
T. H. Brown $^{4, 5}$
\\
$^{1}$School of Physical Sciences and Nanotechnology, Yachay Tech University, Hacienda San Jos\'e S\/N, 100119, Urcuqu\'i, Ecuador \\
$^{2}$ College of Sciences and Engineering, Universidad San Francisco de Quito, Quito, 170901, Ecuador\\
$^{3}$ DLR Galileo Competence Center, Oberpfaffenhofen, 82234 Wessling, Germany\\
$^{4}$ National Research Council of Canada, 
Herzberg Astronomy and Astrophysics Research Centre, 
5071 W. Saanich Rd. Victoria, BC, V9E 2E7, Canada \\
$^{5}$ Department of Physics \& Astronomy, University of Victoria, Finnerty Road, Victoria, BC, V8P 1A1, Canada
}

\date{Accepted XXX. Received YYY; in original form ZZZ}

\pubyear{2024}

\begin{document}
\label{firstpage}
\pagerange{\pageref{firstpage}--\pageref{lastpage}}
\maketitle

\begin{abstract}

Galaxies with extreme amounts of neutral atomic hydrogen (\hi), known as \hi-eXtreme (\hix) galaxies, paradoxically exhibit normal or low star formation rates. To investigate the bottleneck in their star formation, we present ALMA CO(1-0) observations for five \hix\ survey galaxies: ESO208-G026, ESO378-G003, ESO381-G005, ESO417-G018, and ESO055-G013. While these systems possess massive, extended \hi\ discs, their molecular gas discs are compact with normal \Htwo\ masses relative to their stellar mass. Dynamical modelling with \textsc{3DBarolo} reveals that the \hi\ and molecular gas are kinematically consistent. Both gas phases show high velocity dispersions, indicating enhanced turbulence compared to regular late-type galaxies. We calculate the specific angular momentum for the stellar, atomic, and molecular components alongside the global stability parameter ($q$). Most of the system angular momentum is stored in the \hi\ discs, rendering them globally stable against collapse. This stability suppresses the atomic-to-molecular gas conversion, as evidenced by extremely long \hi\ depletion times contrasted with typical \Htwo\ depletion times. The high angular momentum and \hi\ excess likely originate from past gas accretion or recent tidal interactions; this is especially apparent in ESO378-G003 and ESO381-G005, which exhibit extended \hi\ structures and positive tidal indices.
 
\end{abstract}




\section{Introduction}
\label{sec:Introduction}

The gas reservoir of a galaxy is one of the most defining factors in its evolution, since it is the fuel for star formation. Stars are primarily formed in giant molecular clouds (GMCs, e.g. \citealt{Lada2010}) and the star formation rate of galaxies is strongly correlated with their molecular gas surface density, also known as the molecular-hydrogen (\Htwo) Schmidt-Kennicutt law (e.g. \citealt{Bigiel2008, Leroy2009}). In the low surface density outer regions of galactic discs, where the interstellar medium becomes dominated by atomic gas, an \hi-based or total gas-based Schmidt-Kennicutt relation is also used \citep{Leroy2008}. To form and maintain the molecular gas and the star formation in a galaxy a reservoir of neutral atomic hydrogen (\hi) is essential. 

Typical late-type, star-forming galaxies tend to have large \hi\ reservoirs distributed in a disc. The \hi\ content of these galaxies correlates with their stellar properties, such as the optical luminosity, diameter and specific star formation rate (e.g. \citealt{Toribio2011, Catinella2012, Denes2014}). However, due to various physical processes, such as gas stripping and gas accretion, there is a substantial scatter in these scaling relations and several galaxies are significantly removed from the relations. While the \hi-deficient population has been extensively studied and can be relatively well explained by gas stripping or strangulation in high density environments (e.g.\citealt{Haynes1984, Hibbard1996, Chung2007, Cortese2011, Hess2013}), we are only starting to understand \hi-rich galaxies.

Gas accretion onto galaxies can happen from cosmic gas filaments or from other galaxies. An example of the latter is the Milky Way - Magellanic Clouds system, where the Milky Way is stripping gas from the Magellanic Clouds through the Magellanic Stream (e.g. \citealt{Besla2007, DiTeodoro2019}). The stripped gas will eventually be accreted onto the Milky Way and grow the \hi\ reservoir of our Galaxy. Other examples are the gas rich galaxies in the xGASS survey \citep{Gereb2016}. The prototype of these extremely gas-rich galaxies is an early-type galaxy with a large \hi\ reservoir, which also indicates that the gas was acquired through interactions or a merger. Gas rich galaxies often have extended ultraviolet (XUV) discs (e.g. \citealt{Bigiel_2010, Holwerda_2012}). The origin of the XUV disc is also attributed to gas accretion, gas rich mergers or tidal interactions (e.g. \citealt{Holwerda_2012, Das_2021}).

In addition to gas accretion, another option for a large \hi\ reservoir could be inefficient gas conversion, which means that the \hi\ gas is for some reason, not cooling and converting efficiently into molecular gas. Less molecular gas would then result in a lower star formation rate compared to the size of the \hi\ reservoir. A possible explanation for the inefficient gas conversion could be high angular momentum (e.g. \citealt{Obreschkow2016, Stevens2018, Lutz2018, Murugeshan2019, Romeo2020}). 

To understand which of these two processes is responsible for the unusually high \hi\ content in some galaxies, the \hix\ survey \citep{Lutz2017,Lutz2018} mapped the \hi\ content of 12 nearby \hi-eXtreme galaxies with the Australia Telescope Compact Array (ATCA). These galaxies were selected to contain at least 2.5\,times more \hi\ than expected based on the $R$-band \hi\ mass scaling relation \citep{Denes2014}. Based on their stellar properties, the \hix\ survey galaxies look like average star-forming spirals, but they reside in halos with higher than average angular momentum (spin) \citep{Lutz2018}. Thus, they are able to build up and maintain large \hi\ discs, with \hi\ masses up to $\sim 10^{10.8}\,\mathrm{M}_\odot$ and \hi\ disc sizes 2.3 to 5 times the 25\,mag\,arcsec$^{-2}$ isophotal radius \citep{Lutz2018}. Due to the large amount of \hi\ located outside the stellar disc, \hix\ galaxies have a lower global, \hi-based star formation efficiency than average galaxies in the control sample of the survey. 

The \Htwo\ to \hi\ ratio increases both with stellar mass and with morphological type, when moving from disc to bulge dominated systems \citep{Catinella2018}. In addition, studies of spatially resolved nearby galaxies also found that the \Htwo\ to \hi\ fraction correlates with the stellar surface brightness and anti-correlates with the star formation efficiency (e.g. \citealt{Bigiel2016}). This suggests that the \Htwo\ fraction is a limiting factor in the star formation efficiency. Traditionally, the most \hi-rich galaxies known were Giant Low Surface Brightness (GLSB) galaxies (e.g. \citealt{Matthews2001, Das_2013}). However, the \hix\ galaxies, despite their massive \hi\ reservoirs, are not GLSB galaxies; they exhibit normal optical surface brightnesses and typical specific star formation rates for their stellar masses. The \hix\ sample contains one confirmed XUV galaxy, NGC 289, and one potential XUV galaxy, ESO075-G006. However, most of the galaxies in the \hix\ sample, including the galaxies presented in this paper, do not possess an XUV disc. \citet{Hallenbeck2016} studied three nearby \hi-rich galaxies and found that their sample has typical \Htwo\ discs and star formation rates, which suggests that these galaxies are inefficient in converting their \hi\ into \Htwo\ rather than being inefficient at forming stars. Another example of an extreme, \hi-rich system is Rubin's Galaxy (UGC 2885), which is an isolated giant spiral that also exhibits an enormous \hi\ disc but a relatively low star formation efficiency \citep{Carvalho2024}.

To find the bottleneck of the \hi\ to star formation process for the \hix\ galaxies, we mapped the molecular gas content of 5 \hix\ galaxies with the ALMA telescope by observing the CO(1-0) rotational transition. In this paper, we present the ALMA observations, which enable us to measure the amount and kinematic properties of the \co\ in the disc. We convert this into the total molecular gas mass and compare the result to the star formation rate and the \hi\ content of the galaxies. Considering the small sample size of our data, we also include the three \hi\ rich galaxies presented in \citet{Hallenbeck2016}, UGC 6168, UGC 7899, NGC 5230, and UGC 2885 \citep{Carvalho2024} in our scaling relations analysis. We employ 3D dynamical modelling with \textsc{3DBarolo}, to measure the kinematics of both the CO and \hi\ gas discs, including their velocity dispersion, the specific angular momentum of the gas components and assess the global disc stability against gravitational collapse using a stability parameter analogous to the Toomre criterion. Additionally, we compare the molecular and atomic gas content, along with their depletion times, to scaling relations from other galaxies. By combining these kinematic, dynamical, and scaling relation analyses with an investigation of potential tidal interactions, we can identify the mechanisms related to the \hi\ excess.

This paper is organised as follows: in Section \ref{sec:data}, we present the ALMA CO data and auxiliary observations for the sample; in Section \ref{sec:Results}, we present results on the molecular gas properties, scaling relations, dynamical modelling, velocity dispersion, angular momentum, and disc stability; in Section \ref{sec:Discussion}, we discuss the implications of our findings, including the role of tidal interactions and the connection between gas kinematics and star formation efficiency; and in Section \ref{sec:Conclusions}, we summarise our main conclusions. Throughout this paper we assume a flat $\Lambda$CDM cosmology with the following cosmological parameters $H_0 = 70\,\mathrm{km\,s^{-1}\,Mpc^{-1}}$, $\Omega_\mathrm{m}$ = 0.3. All velocities are given in the optical convention ($cz$).

\section{Sample selection and data}
\label{sec:data}
\subsection{The \hix\ galaxy survey} 

The sample of \hi\ eXtreme (\hix) galaxies \citep{Lutz2017} have been selected to contain at least 2.5 times more \hi\ than expected from scaling relations by \citet{Denes2014}. In addition, we also applied the following selection criteria: an absolute 2MASX $K$-band magnitude smaller than $-22.0$\,mag to select primarily massive spirals, and a location south of $-30\,\mathrm{deg}$ in declination for good observation with the Australia Telescope Compact Array (ATCA). For more details of the sample selection see \citet{Lutz2017}.

From the \hix\ survey we selected 6 \hi-rich galaxies to be observed in the \co(1-0) line transition. We selected these galaxies to be representative of the whole \hix\ survey in terms of \hi\ mass, size and morphology. We estimated the expected molecular gas content based on the \Htwo\ gas mass fraction to specific star formation rate scaling relation by \citet{Saintonge2017} and aimed to obtain a \Htwo\ column density of $1\, \mathrm{M}_\odot\,\mathrm{pc}^{-2}$ to probe the more diffuse interstellar medium in the outskirts of the stellar disc. Out of the 6 proposed galaxies observed with ALMA, we detected 5 galaxies in the \co(1-0) rotational transition. 

In the following section we describe the obtained \co\ data.

\subsection{ALMA-ACA data}

We have obtained ALMA \co(1-0) (referred to as \co\ hereafter) line data for 5 \hix\ galaxies: ESO208-G026, ESO378-G003, ESO381-G005, ESO417-G018, ESO055-G013, under the ALMA project code: 2019.2.00169.S. All five galaxies have been observed with the 7m array and the three more extended galaxies (ESO378-G003, ESO381-G005 and ESO417-G018) have been observed with the total power array as well. ALMA observations were taken between the 25th of December 2019 and the 9th of March 2020 in band 3. We centred the spectral line observations assuming the same systemic velocity for the \co\ as for the \hi\ line from the previous \hi\ observations \citep{Lutz2017}. We used a frequency resolution of $\sim$ 1.1290 MHz, which corresponds to a velocity resolution of $\sim$ 2.6 \kms. Table~\ref{tab:ALMA-data} shows the details of the ALMA observations.

In this work we refer to the interferometer only data as ``pbcorr'' and the combined interferometer and total power data as ``combined'' data.
 
The default velocity resolution \co\ data cubes delivered by the ALMA automatic pipeline have relatively low signal-to-noise ratios. To obtain robust \co\ maps we binned the data cubes by 5-channels, which results in a velocity resolution of $\sim$ 13 \kms. These data cubes are referred to as ``binned'' cubes.

\begin{table*}
    \centering
    \begin{tabular}{cccccccc}
    \hline
        ID & Array & $T_{\mathrm{int}}$ & $\nu_{\mathrm{c}}$ & $\theta_{\mathrm{maj}}\times\theta_{\mathrm{min}}$ & \makecell{$\Delta v$\\ pbcorr / binned} & \makecell{RMS noise\\ pbcorr / binned} & \makecell{$3\sigma_\mathrm{SB}$\\ pbcorr / binned}\\
        & & (h) & (GHz) & (arcsec) & (\kms) & ($\mathrm{mJy\,beam^{-1}}$) & $(\mathrm{M}_\odot\,\mathrm{pc^{-2}})$ \\
        (1) & (2) & (3) & (4) & (5) & (6) & (7) & (8)\\
        \hline
        ESO055-G013 & 7-m & 0.487 & 112.406 & 14.1 $\times$ 11.0 & 2.71 / 13.56 & 18.45 / 10.29 & 3.14 / 4.00\\
        ESO208-G026 & 7-m & 0.361 & 114.125 & 14.1 $\times$ 8.8 & 2.63 / 13.17 & 24.12 / 14.44 & 6.18 / 8.25\\
        ESO378-G003 & 7-m & 0.354 & 114.113 & 14.1 $\times$ 8.6 & 2.63 / 13.17 & 16.16 / 9.55 & 3.91 / 5.09\\
        ESO381-G005 & 7-m & 0.239 & 113.069 & 16.6 $\times$ 8.4 & 2.68 / 13.41 & 21.23 / 11.28 & 4.31 / 5.15\\
        ESO417-G018 & 7-m & 0.285 & 113.436 & 14.7 $\times$ 7.8 & 2.66 / 13.33 & 20.25 / 12.49 & 5.16 / 7.08\\
        \hline
        ESO378-G003 & TP + 7-m & 3.100 & 114.113 & 56.6 $\times$ 56.6 & 2.63 / 13.17 & 16.26 / 9.51 & 3.92 / 4.93\\
        ESO381-G005 & TP + 7-m & 2.067 & 113.069 & 56.8 $\times$ 56.8 & 2.68 / 13.41 & 21.53 / 12.47 & 4.59 / 6.05\\
        ESO417-G018 & TP + 7-m & 2.300 & 113.436 & 56.1 $\times$ 56.1 & 2.66 / 13.33 & 20.82 / 12.33 & 5.71 / 7.25\\
        \hline
    \end{tabular}
    \caption{ALMA-ACA \co\ observations. (1): The galaxy ID. (2): Array of ALMA used for observation. (3): Integration time in hours. (4): Central frequency of the observation. (5): Angular resolution (Beam major axis $\times$ beam minor axis). (6): The velocity resolution of the pbcorr/binned data cubes. (7): The RMS noise of the pbcorr/binned data cubes, measured by \textsc{SoFiA}. (8): The $3\sigma$ surface brightness (SB) sensitivity, expressed as molecular gas surface density of the pbcorr/binned data cubes.}
    \label{tab:ALMA-data}
\end{table*}

\begin{figure*}
     \centering
         \includegraphics[width=8cm]{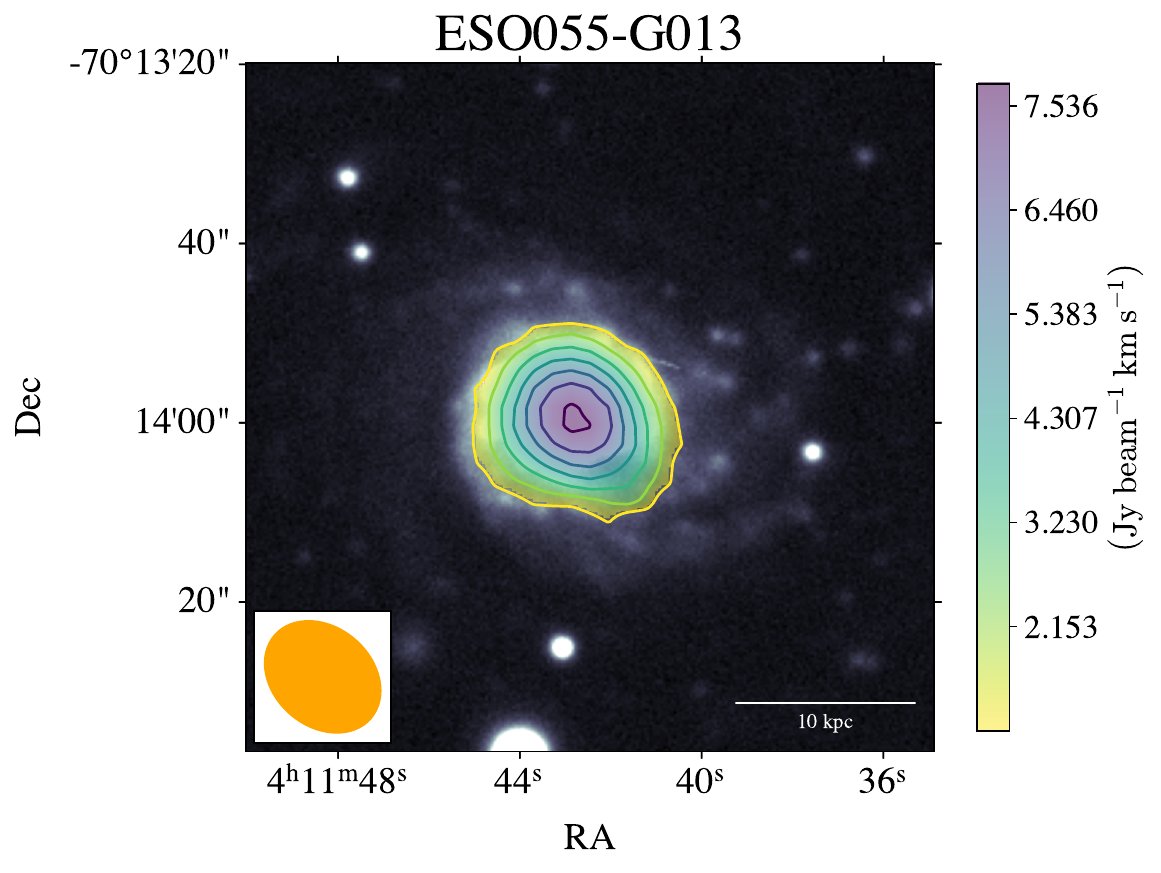}
         \includegraphics[width=8cm]{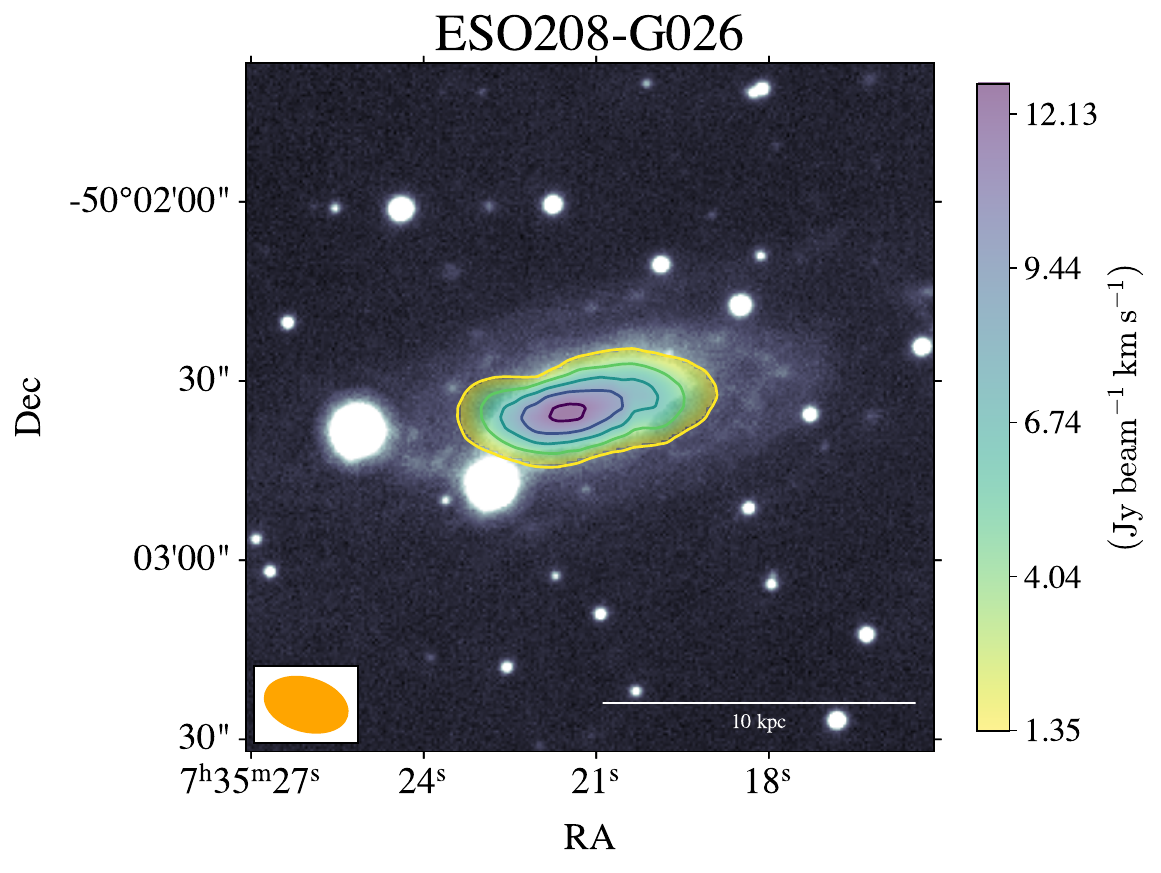}
         \includegraphics[width=8cm]{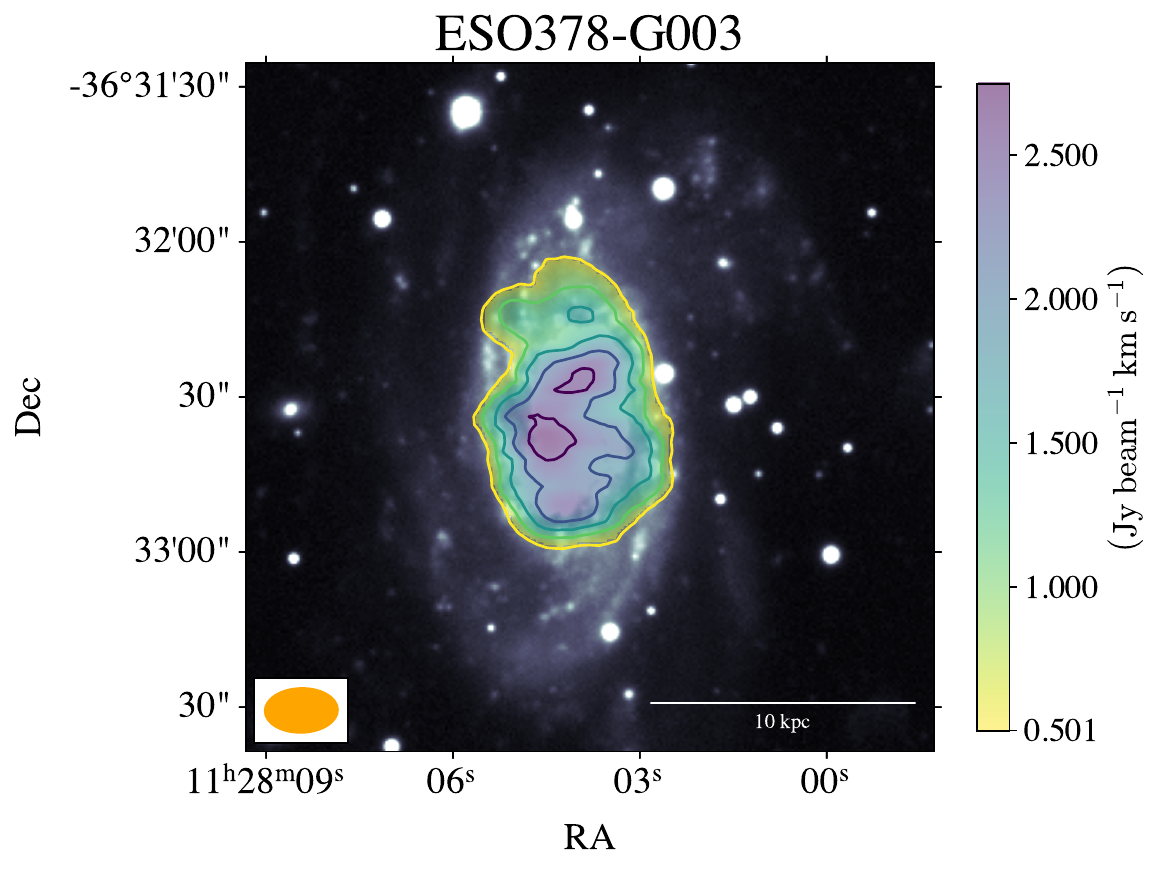}
         \includegraphics[width=8cm]{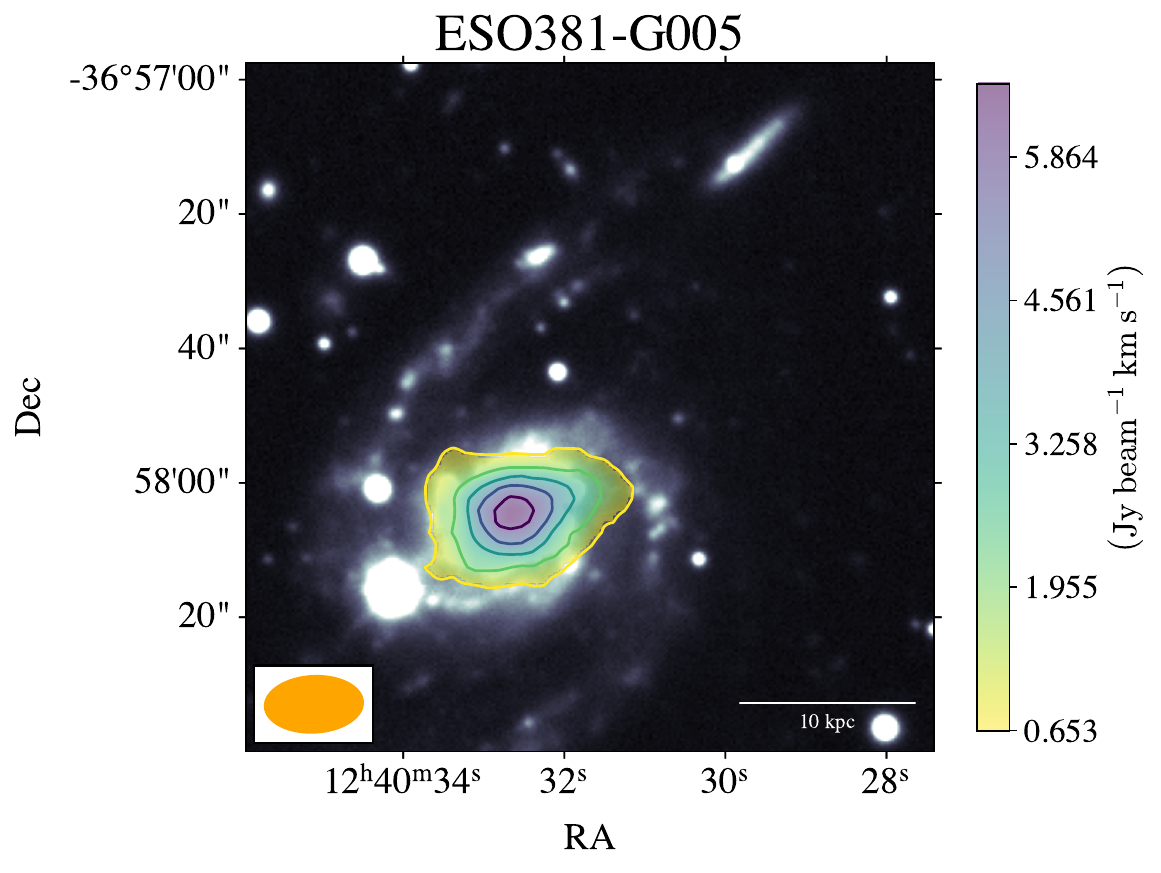}
         \includegraphics[width=8cm]{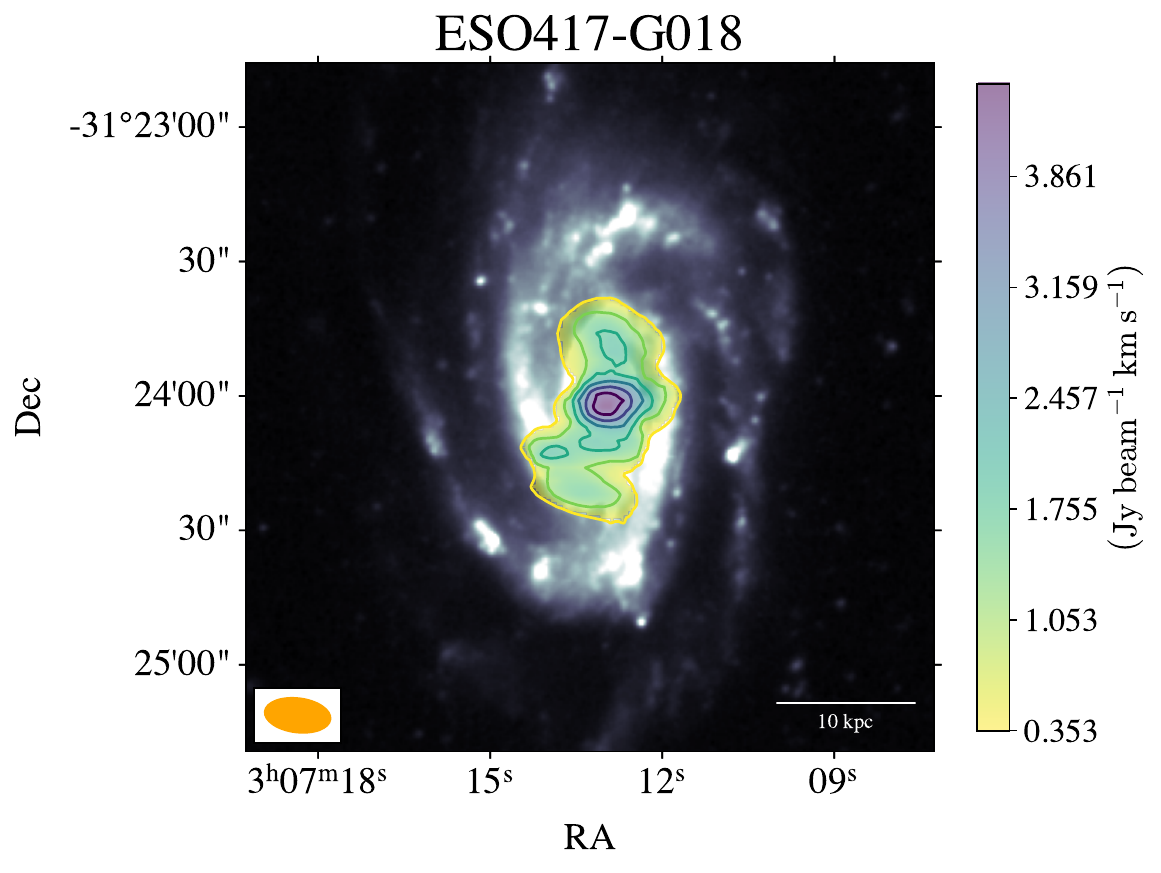}
        \caption{Legacy Sky Survey (DR10) $g$-band image overlaid with ALMA \co\ contour levels (binned, dilated data set). The synthesised beam of the \co\ data is shown in the lower left corner of each image. Contour levels are colour coded according to the integrated \co\ intensity in units of the rms. All galaxies show a slight asymmetry or distortion in their optical image.}
        \label{fig:overlays}
\end{figure*}

\begin{table*}
    \centering
    \begin{tabular}{cccccccc}
    \hline
        ID & RA & Dec. & $D_\text{L}$ & $z$ & $v_\text{sys}$ &  $\mathrm{INC}$ &  $\mathrm{PA}$ \\
         & $(\mathrm{deg})$ & $(\mathrm{deg})$ & $(\mathrm{Mpc})$ & & $(\mathrm{km\,s^{-1}})$ & $(\mathrm{deg})$ & $(\mathrm{deg})$\\
        (1) & (2) & (3) & (4) & (5) & (6) & (7) & (8)\\
        \hline
        ESO055-G013 & $62.9290$  & $-70.2331$ & $104.98$ & $0.024636$ & $7338.82$ & $28.6$ & $67.4$\\
        ESO208-G026 & $113.8280$ & $-50.0430$ & $39.72$  & $0.009951$ & $2917.14$ & $72.0$ & $285.9$\\
        ESO378-G003 & $172.0167$ & $-36.5427$ & $40.56$  & $0.010080$ & $2994.13$ & $61.5$ & $358.4$\\
        ESO381-G005 & $190.1363$ & $-36.9681$ & $80.04$  & $0.018993$ & $5659.46$ & $49.9$ & $303.9$\\
        ESO417-G018 & $46.8050$  & $-31.4007$ & $67.35$  & $0.015848$ & $4705.73$ & $62.4$ & $357.4$\\
        \hline
    \end{tabular}
    \caption{General properties of the \co-observed sample. (1): Galaxy ID. (2): Right ascension. (3): Declination. (4): Luminosity distance. (5): Redshift. (6): Systemic velocity (optical convention). (7): Average fitted inclination from \textsc{3DBarolo}, and (8): average fitted position angle also from \textsc{3DBarolo}.}
    \label{tab:generalprop}
\end{table*}

\subsection{CO moment maps and integrated spectrum}
\label{sec:molecular_properties}

\begin{figure*}
    \centering
    \includegraphics[width=0.42\textwidth]{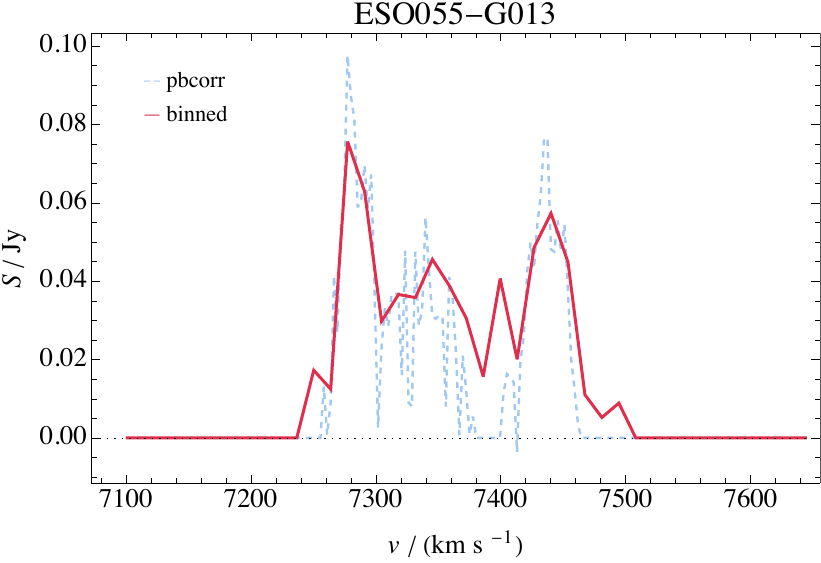}
    \includegraphics[width=0.42\textwidth]{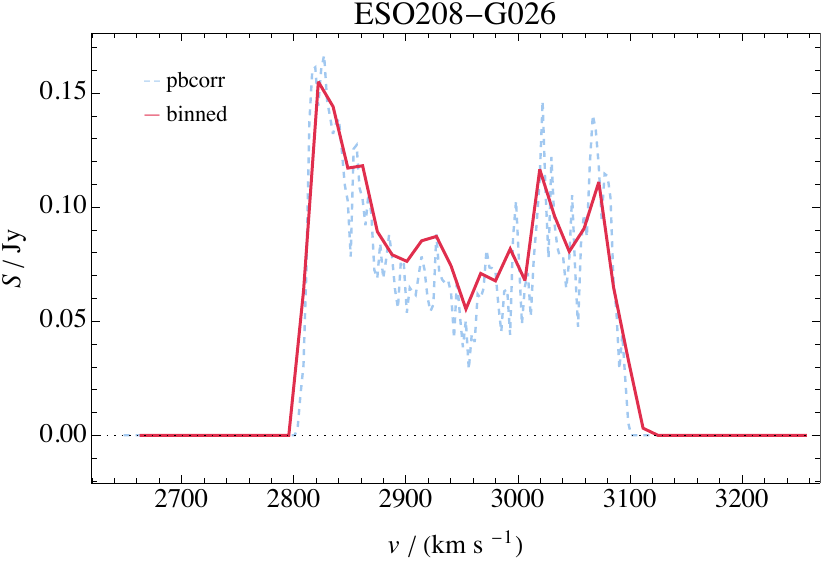}
    \includegraphics[width=0.42\textwidth]{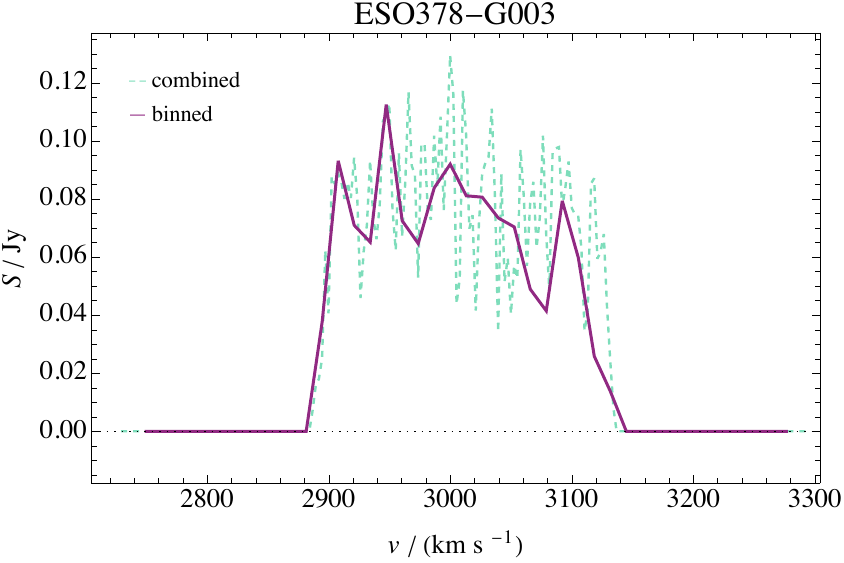}
    \includegraphics[width=0.42\textwidth]{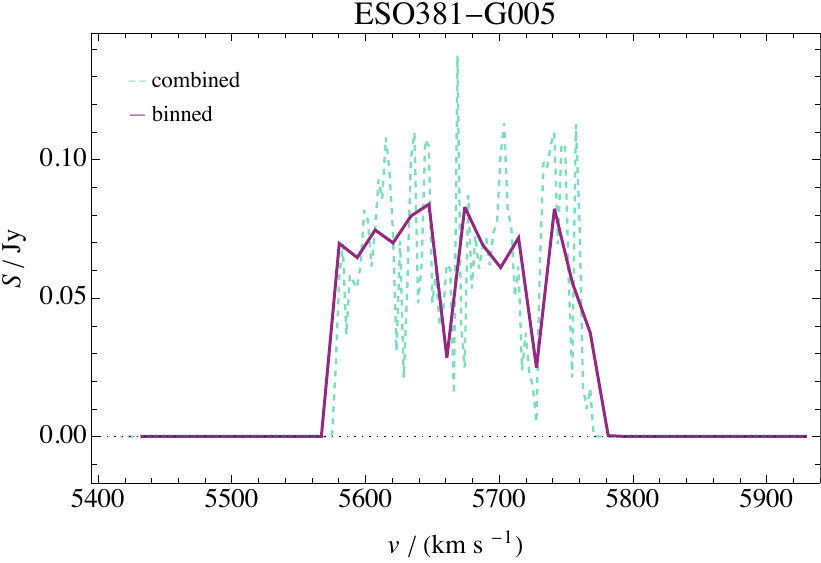}
    \includegraphics[width=0.42\textwidth]{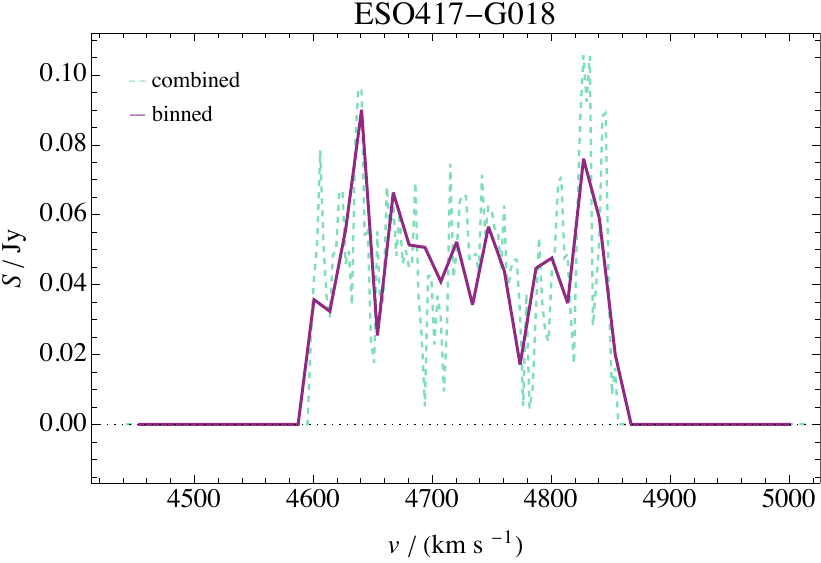}
    \caption{Integrated spectrum of each galaxy using the 3D \textsc{SoFiA} mask. Dashed line spectra are from the pipeline data cubes, and solid lines are from the binned cubes. We use pbcorr (7m array only) data for ESO055-G013 and ESO208-G026, and combined (7m array and total power) data of ESO378-G003, ESO381-G005 and ESO417-G018.}
    \label{fig:spectra}
\end{figure*}

\begin{figure*}
     \centering
         \includegraphics[width=8cm]{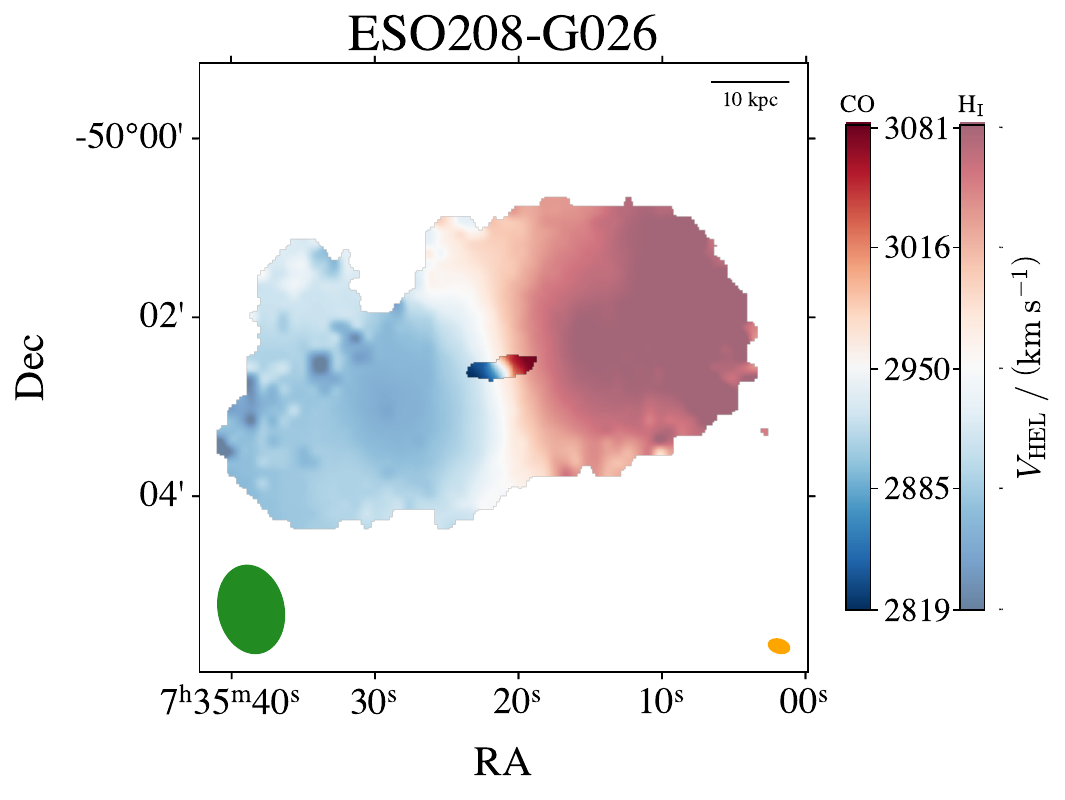}
         \includegraphics[width=8cm]{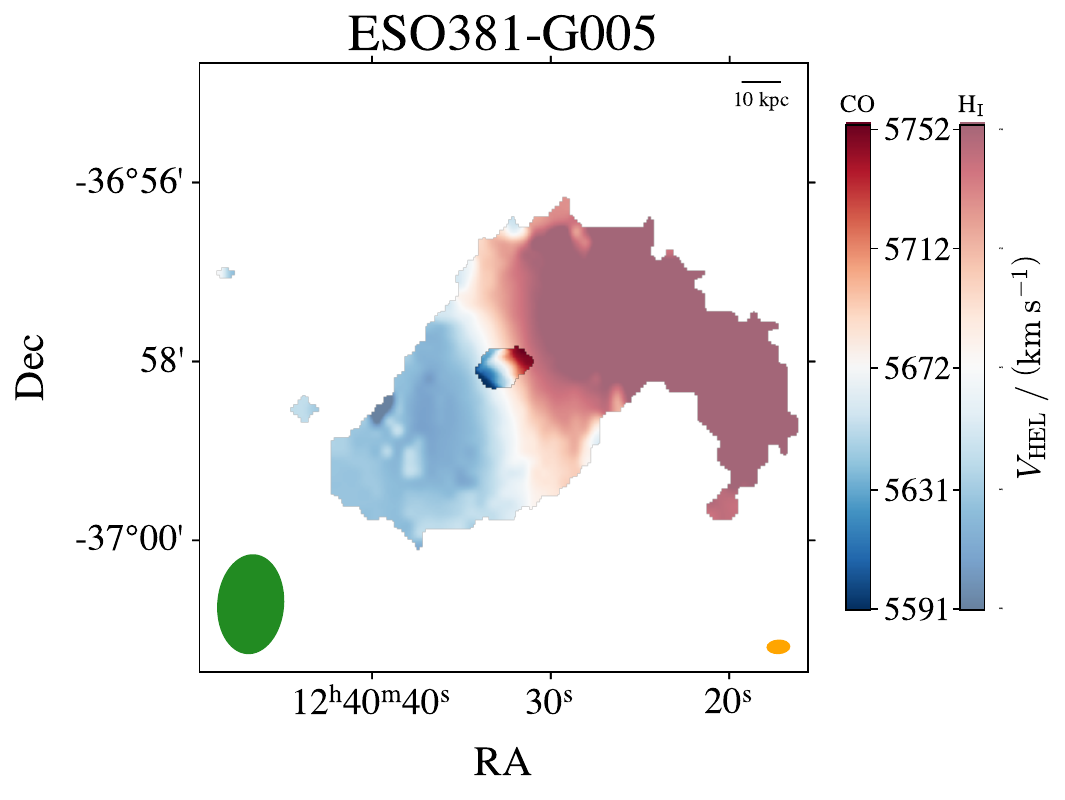}
         \includegraphics[width=8cm]{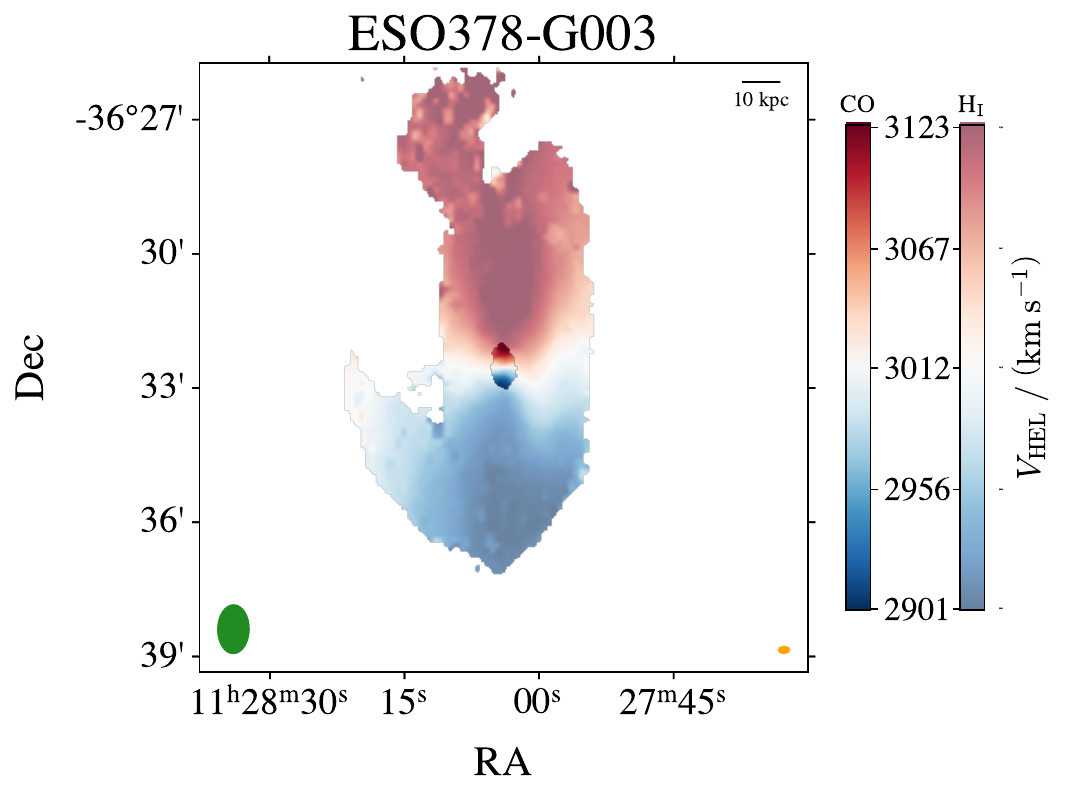}
         \includegraphics[width=8cm]{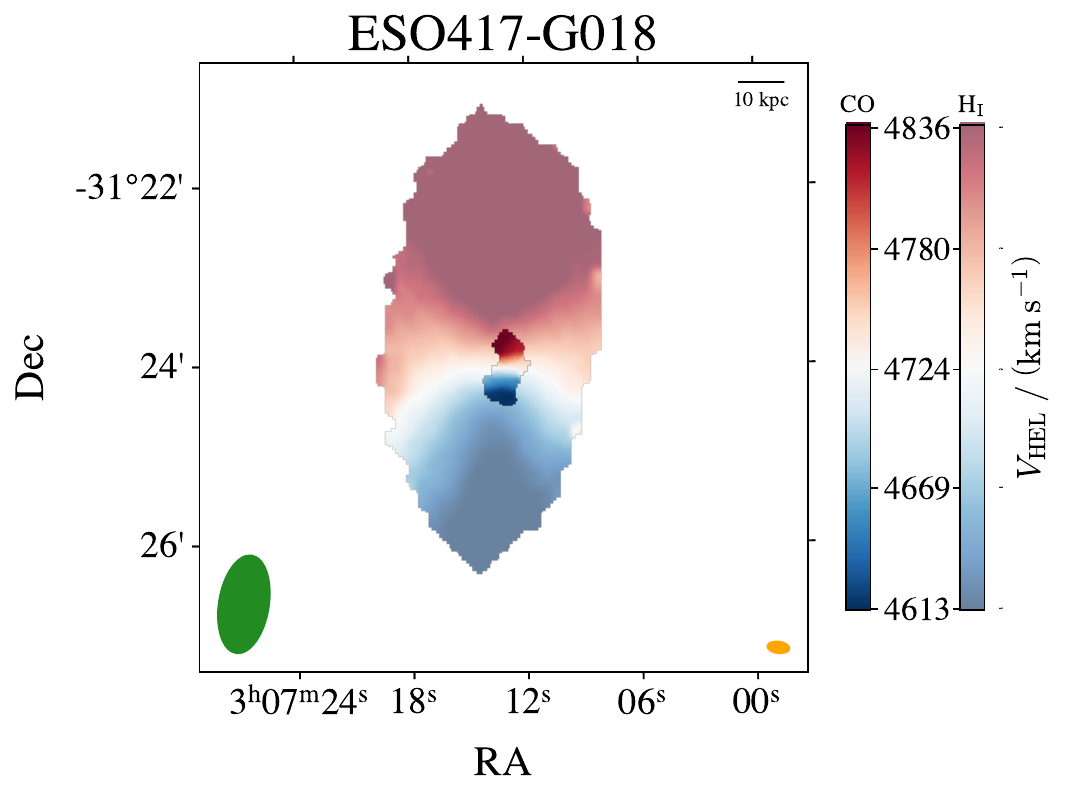}
         \includegraphics[width=8cm]{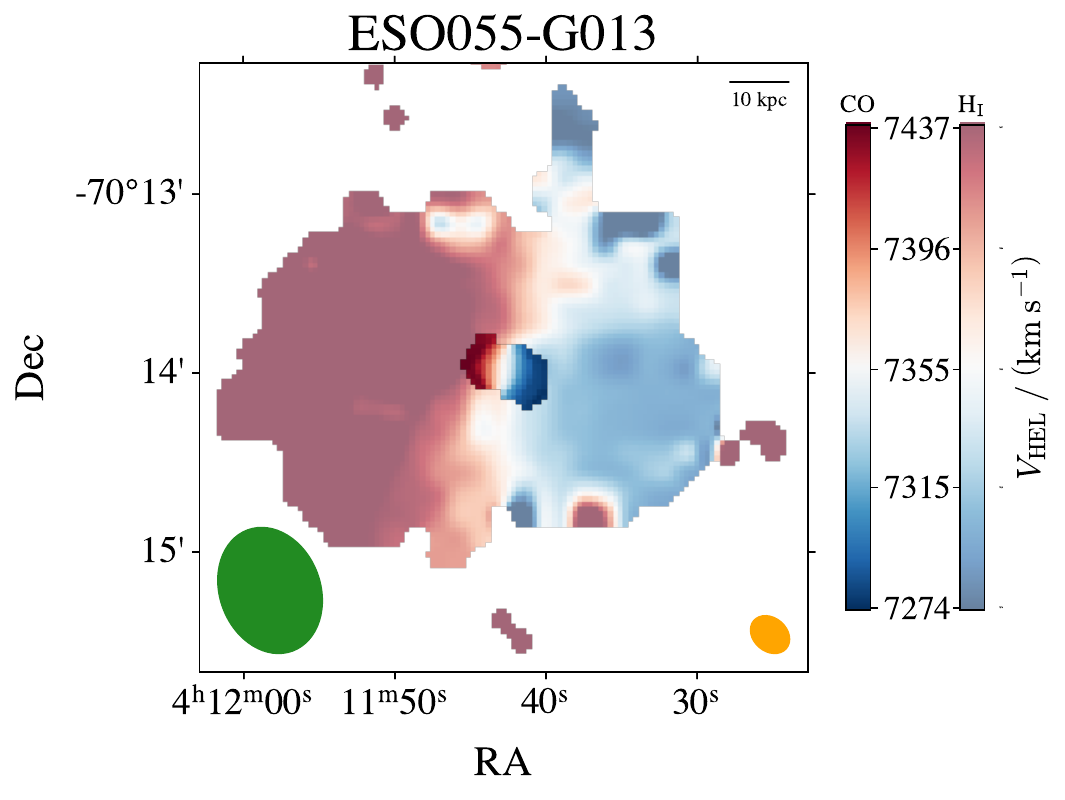}
        \caption{Intensity-weighted velocity maps of the \co\ and \hi\ gas discs of each galaxy in the sample, as labelled. To the right of each graph is its respective colour scale. The synthesised beams of the \hi\ (in green) and \co\ (in orange) data are shown respectively in the lower left and right corners of each plot.}
        \label{fig:hicooverlays}
\end{figure*}

To reliably detect the \co\ signal we use the \hi\ Source Finding Application (\textsc{SoFiA}, \citealt{Serra2015, Westmeier2021MNRAS}), a source finding software designed to detect and characterise extragalactic spectral line sources in 3D data cubes, where the flux of a spectral line is mapped as a function of position on the sky and line-of-sight velocity of the emitting matter. We obtain the following data products from \textsc{SoFiA}: the integrated flux intensity, velocity and velocity dispersion maps (0th, 1st, and 2nd moment maps), the integrated spectrum, and a 3D mask. We also obtain the position of the centroid, the redshift, the systemic velocity and the luminosity distance for each source, as summarised in Table \ref{tab:generalprop}. Figure~\ref{fig:overlays} shows Legacy Survey (DR10) $g$-band optical images of the sample galaxies with \co\ intensity contours from the \textsc{SoFiA} moment 0 maps and Figure~\ref{fig:spectra} shows the CO spectra for all 5 galaxies. Additionally, Figure~\ref{fig:hicooverlays} presents the intensity-weighted velocity maps (1st moment maps) for both the \co\ and \hi\ gas discs.

With the exception of ESO055-G013, the \co\ discs of all galaxies are resolved by several beams. ESO055-G013 is spatially unresolved. The molecular gas disc, indicated by the \co\ contours, is in all cases approximately the same size as the inner part of the stellar disc, which roughly corresponds to the inner star forming regions of the galaxies. We do not detect the more extended spiral arms in \co, which may be due to the sensitivity limit of the ALMA observations, although we have achieved the targeted $1\, \mathrm{M}_\odot\,\mathrm{pc}^{-2}$ sensitivity for all 5 galaxies. In the case of ESO417-G018 the \co\ emission only covers the bulge, the bar and the start of the spiral arms of the galaxy. While the ALMA observations may miss some of the low surface density gas in the other discs due to small beam filling factors, the dominant gas phase at these surface densities is expected to be \hi\ (e.g. \citealt{Leroy2008}). Therefore, the low surface density gas is a negligible fraction of the global molecular mass and our derived $M_{\mathrm{mol}}$ values represent robust measurements of the total molecular gas content.

\subsection{Auxiliary data for \hix\ galaxies}

We have previously compiled a rich data set for the \hix\ galaxies: For all sample galaxies deep \hi\ observations are available from the ATCA. The stellar content and the star formation activity can be measured from the combination of the 2 Micron All Sky Survey (2MASS) and the associated Extended Source Catalogue (2MASX; \citealt{Skrutskie2006}), the Wide-field Infrared Survey Explorer (WISE, \citealt{Wright2010}) and the Galaxy Evolution Explorer (GALEX, \citealt{Martin2005}) imaging. The stellar mass ($M_{\star}$) and the star formation rate (SFR) for our sample are adopted from \citet{Lutz2017} and \citet{Lutz2018}.

The stellar masses were derived from the total 2MASX $K$-band magnitudes using the empirical scaling relation by \citet{Wen2013}. The star formation rates were estimated using the GALEX NUV luminosity combined with the WISE $W3$-band luminosity to account for both unobscured and dust-obscured star formation \citep{Lutz2017}. GALEX photometry was unavailable for ESO208-G026, so its SFR was calculated using only the $W3$-band magnitude from WISE following the prescription by \citet{Cluver2014}.

We use optical images for overlays from the most recent data release 10\footnote{\url{https://www.legacysurvey.org/}} (DR10) of the DESI Legacy Imaging Surveys \citep{Dey2019}. This survey is a combination of three different surveys (the Dark Energy Camera Legacy Survey, the Beijing–Arizona Sky Survey, and the Mayall z-band Legacy Survey) carried out with two telescopes at the Kitt Peak National Observatory and the Cerro Tololo Inter-American Observatory, providing imaging in $g$, $r$, $i$ and $z$ bands over a region of $\sim 20\,000$ square degrees.  

In addition to the imaging data, we also use data from published catalogues, such as the xCOLD GASS \citep{Saintonge2017} \co\ data and xGASS \citep{Catinella2018} \hi\ data. We use these data sets to compare the properties of the \hix\ galaxies to global \hi\ and \Htwo\ scaling relations. For typical nearby galaxies the xGASS and xCOLD GASS \citep{Saintonge2017,Catinella2018} surveys provide insights into the interplay of the global \hi\ and \Htwo\ properties. For all comparisons, we restrict the comparison sample strictly to galaxies with \hi\ and \co\ detections. We do not include non-detections with upper limits to ensure a direct comparison with gas rich galaxies.

\section{Results}
\label{sec:Results}

\subsection{Molecular gas, scaling relations and atomic mass fraction}

\subsubsection{\co\ to \Htwo\ conversion}

\Htwo\ is the most abundant molecule in the universe and has a central role in the evolution of galaxies as the main ingredient for star formation. We calculate the  \Htwo\ mass of each galaxy from the \co\ data.
First we obtain the \co\ line luminosity ($L'_{\text{\co}}$) as a function of the velocity-integrated flux ($S_{\text{\co}}\Delta v$), the luminosity distance ($D_{\mathrm{L}}$) and the rest frequency of the emission ($\nu_{\mathrm{rest}}$) corrected for redshift, to obtain the intrinsic source brightness temperature \citep{Solomon1997},
\begin{equation}
    L'_{\text{\co}}=\left(\frac{c^2}{2k}\right)\frac{S_{\text{\co}}\Delta v D_{\mathrm{L}}^{2}}{\nu_\text{rest}(1+z)}.
    \label{eq:solomon}
\end{equation}

The molecular gas is directly proportional to this \co(1-0) line luminosity by a conversion factor,
\begin{equation}
    M_{\text{mol}}= \alpha_{\text{\co}}\, L'_{\text{\co}},
    \label{eq:mmol}
\end{equation}
where $\alpha_{\text{\co}}$ is the mass-to-light ratio. To convert directly from integrated flux into molecular gas, we can write equations \eqref{eq:solomon} and \eqref{eq:mmol} with constants in the corresponding units, in one equation (see \citealt{Bolatto2013}),
\begin{equation}
    M_\text{mol} = 1.05 \times 10^{4} \left(\frac{X_\text{CO}}{2\times 10^{20}\,\mathrm{cm^{-2}\,(K\,km\,s^{-1})^{-1}}}\right)\frac{S_\text{\co}\Delta v D_\text{L}^{2}}{1+z},
    \label{eq:bolatto}
\end{equation}
where $S_\text{\co}\Delta v$ is the integrated \co(1-0) line flux density in $\mathrm{Jy\,km\,s^{-1}}$, $D_\text{L}$ is the luminosity distance to the source in Mpc, $z$ is the redshift, and $M_\text{mol}$ is the mass of the molecular gas in units of $\mathrm{M}_\odot$. $X_{\text{\co}}$ is the factor for the relation between the \co(1-0) line emission intensity ($W_\text{\co}$) and the \Htwo\ column density ($N_\text{\Htwo}$),
\begin{equation}
    N_\text{\Htwo} = X_{\text{\co}}\, W_\text{\co}.
\end{equation}

$X_{\text{\co}}$ and $\alpha_{\text{\co}}$ are known as the ``\co-to-\Htwo\ conversion factors''. $\alpha_{\text{\co}}$ has units of $\mathrm{M}_\odot\mathrm{(K\,km\,s^{-1}\;pc^2)^{-1}}$, while $X_{\text{\co}}$ has units of $\mathrm{cm^{-2}(K\,km\,s^{-1})^{-1}}$.

Many observational studies have measured $X_\text{\co}$ by deriving the \Htwo\ mass independently of the \co(1-0) line emission. The \co-to-\Htwo\ conversion factor is known to depend on the ISM metallicity, with $\alpha_{\text{\co}}$ increasing by a factor of 2--5 at sub-solar metallicities \citep{Bolatto2013}. Our \hix\ sample galaxies exhibit slightly sub-solar metallicities \citep{Lutz2020}, but do not show unusual metallicity gradients, which could suggest past gas accretion events. We calculate the \Htwo\ mass using equation \eqref{eq:bolatto}, with constant $\alpha_\text{\co} = 4.35\,\mathrm{M}_\odot\mathrm{(K\,km\,s^{-1}\;pc^2)^{-1}}$, corresponding to $X_{\text{\co}} = 2.0 \times 10^{20}\,\mathrm{cm^{-2}(K\,km\,s^{-1})^{-1}}$ that is the recommended value for the Milky Way (see \citealt{Bolatto2013}, \citealt{Saintonge2017}, \citealt{Accurso2017}). We use this $\alpha_\text{\co}$ value to be consistent with the comparison sample from xCOLD GASS. 

While the \co-to-\Htwo\ conversion factor is known to vary with ISM metallicity and environmental properties, as extensively demonstrated by Herschel observations (e.g., \citealt{Sandstrom2013}) and implemented in large-scale surveys (e.g., \citealt{Saintonge2017}), our sample does not reach the extremely low-metallicity regime characteristic of dwarf galaxies where this effect becomes dominant. Furthermore, applying a metallicity-dependent conversion would generally result in higher molecular gas mass estimates. This would only reinforce our finding that these galaxies possess substantial gas reservoirs, without altering the conclusion that the \hi\ discs are anomalously stable. Thus, employing the standard value represents a conservative approach for calculating the \Htwo\ mass. 

Table~\ref{tab:masses} presents the calculated \Htwo\ masses together with the \hi\ masses and the stellar masses from \citet{Lutz2018}. 

\begin{table*}
\begin{tabular}{ccccccccc}
\hline
ID & $\log\left[\dfrac{M_{\star}}{\mathrm{M}_\odot}\right]$ & $\log\left[\dfrac{M_\text{\hi}}{\mathrm{M}_\odot}\right]$ & $\log\left[\dfrac{M_\text{mol}}{\mathrm{M}_\odot}\right]$ & $R_\text{mol}$ & $\log\left[\dfrac{M_\text{atm}}{\mathrm{M}_\odot}\right]$ & $\log\left[\dfrac{M_\text{bar}}{\mathrm{M}_\odot}\right]$ & $f_\text{atm}$\\ 
(1) & (2) & (3) & (4) & (5) & (6) & (7) & (8)\\
\hline
ESO055-G013 & $10.19 \pm 0.14$ & $10.28 \pm 0.06$ & $9.00 \pm 0.09$ & $-1.28 \pm 0.11$ & $10.41 \pm 0.06$ & $10.62 \pm 0.06$ & $0.61 \pm 0.08$\\
ESO208-G026 & $9.81 \pm 0.14$  & $9.91 \pm 0.06$  & $8.65 \pm 0.09$ & $-1.26 \pm 0.11$ & $10.04 \pm 0.06$ & $10.25 \pm 0.06$ & $0.62 \pm 0.08$\\
ESO378-G003 & $10.07 \pm 0.14$ & $10.27 \pm 0.06$ & $8.46 \pm 0.09$ & $-1.80 \pm 0.11$ & $10.40 \pm 0.06$ & $10.57 \pm 0.06$ & $0.68 \pm 0.06$\\
ESO381-G005 & $10.07 \pm 0.14$ & $10.43 \pm 0.06$ & $8.93 \pm 0.10$ & $-1.49 \pm 0.11$ & $10.56 \pm 0.06$ & $10.69 \pm 0.06$ & $0.74 \pm 0.08$\\
ESO417-G018 & $10.28 \pm 0.14$ & $10.43 \pm 0.06$ & $8.77 \pm 0.09$ & $-1.65 \pm 0.11$ & $10.56 \pm 0.06$ & $10.75 \pm 0.06$ & $0.65 \pm 0.08$\\
\hline
UGC 6168 & 10.37 & 10.35 & 8.96 & -1.39 & 10.48 & 10.74 & 0.55\\
UGC 7899 & 10.49 & 10.42 & 9.68 & -0.74 & 10.55 & 10.85 & 0.50\\
NGC 5230 & 10.89 & 10.53 & 10.02 & -0.51 & 10.66 & 11.13 & 0.34\\
UGC 2885 & $11.68 \pm 0.14$ & $10.57 \pm 0.04$ & $11.28 \pm 0.06$ & $0.70 \pm 0.07$ & $10.70 \pm 0.04$ & $11.86 \pm 0.09$ & $0.07 \pm 0.02$\\
\hline
\end{tabular}
\caption{Galaxy masses. (1): The galaxy ID. (2): Stellar mass derived via $K$-band scaling relation. (3): Neutral hydrogen mass from ATCA data (see \citealt{Lutz2018}). (4): Molecular mass from integrated \co(1-0) line flux density. (5): Logarithmic molecular-to-atomic gas ratio. (6): Atomic gas mass. (7): Baryonic mass. (8): Atomic gas fraction.}
\label{tab:masses}
\end{table*}

\subsubsection{Atomic gas fraction}
Atomic gas in galaxies has two main components: neutral atomic hydrogen and helium. We calculate the atomic gas mass, $M_\text{atm}$, for the sample as
\begin{equation}
    M_\text{atm} = 1.35 M_\text{\hi},
\end{equation}
where 1.35 is the correction factor to account for helium, based on cosmological abundances.

On the other hand, the total baryonic mass, $M_\text{bar}$, is the sum of the contributions from atomic gas, molecular gas and stars, so we have
\begin{equation}
    M_\text{bar} = M_\star + M_\text{atm} + M_\text{mol}.
\end{equation}

The atomic gas fraction, $f_\text{atm}$, is the ratio between atomic gas mass and total baryonic mass,
\begin{equation}
    f_\text{atm} =\frac{M_\text{atm}}{M_\text{bar}}.
\end{equation}
Table \ref{tab:masses} summarises these three properties for our sample. For comparison we have also included the values for UGC 6168, UGC 7899, NGC 5230 from \citet{Hallenbeck2016} and the values for UGC 2885 from \citet{Carvalho2024}.

\subsubsection{Scaling relations}

\begin{figure*}
    \centering
    \includegraphics[width=0.45\textwidth]{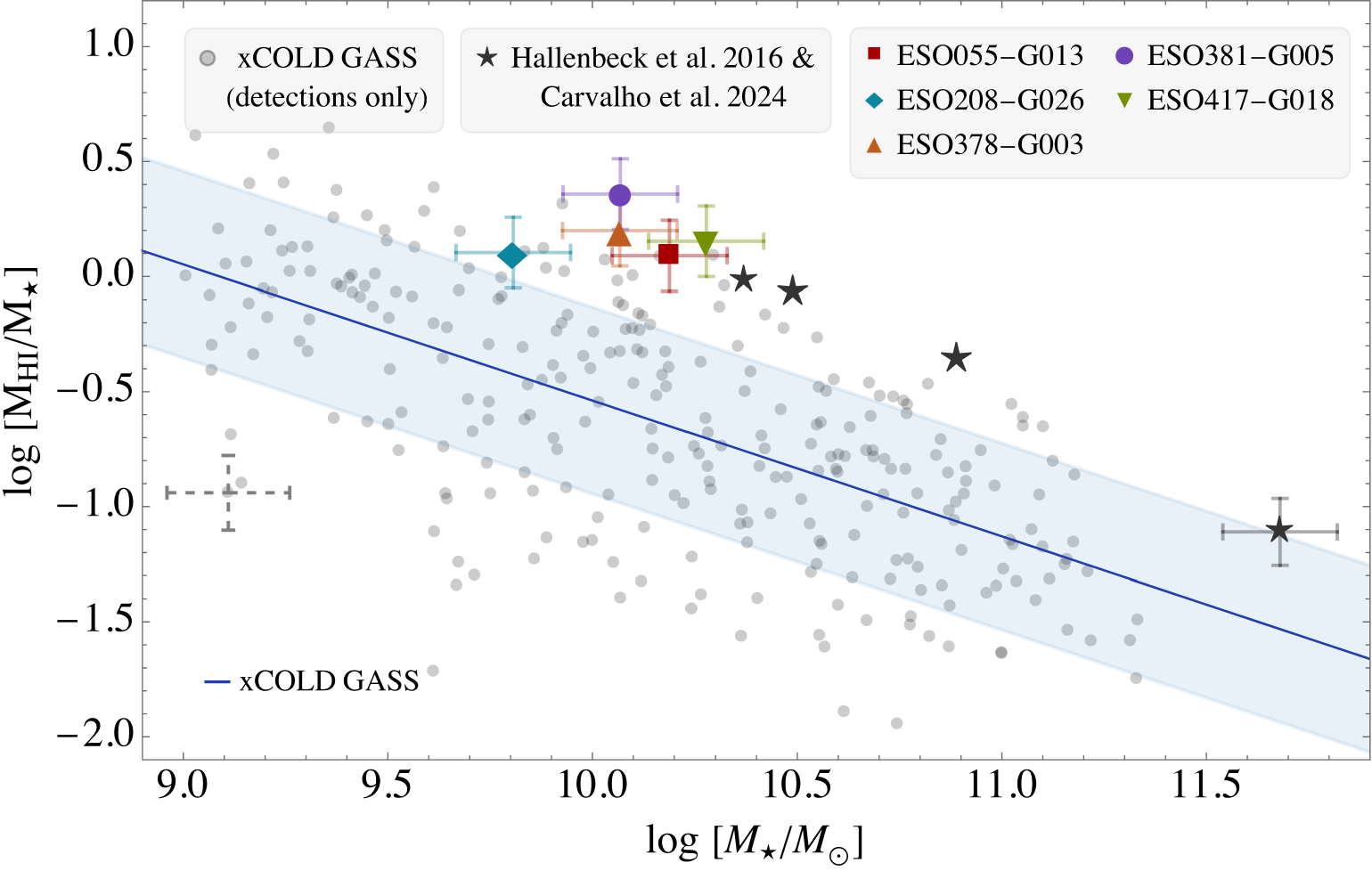}\qquad
    \includegraphics[width=0.45\textwidth]{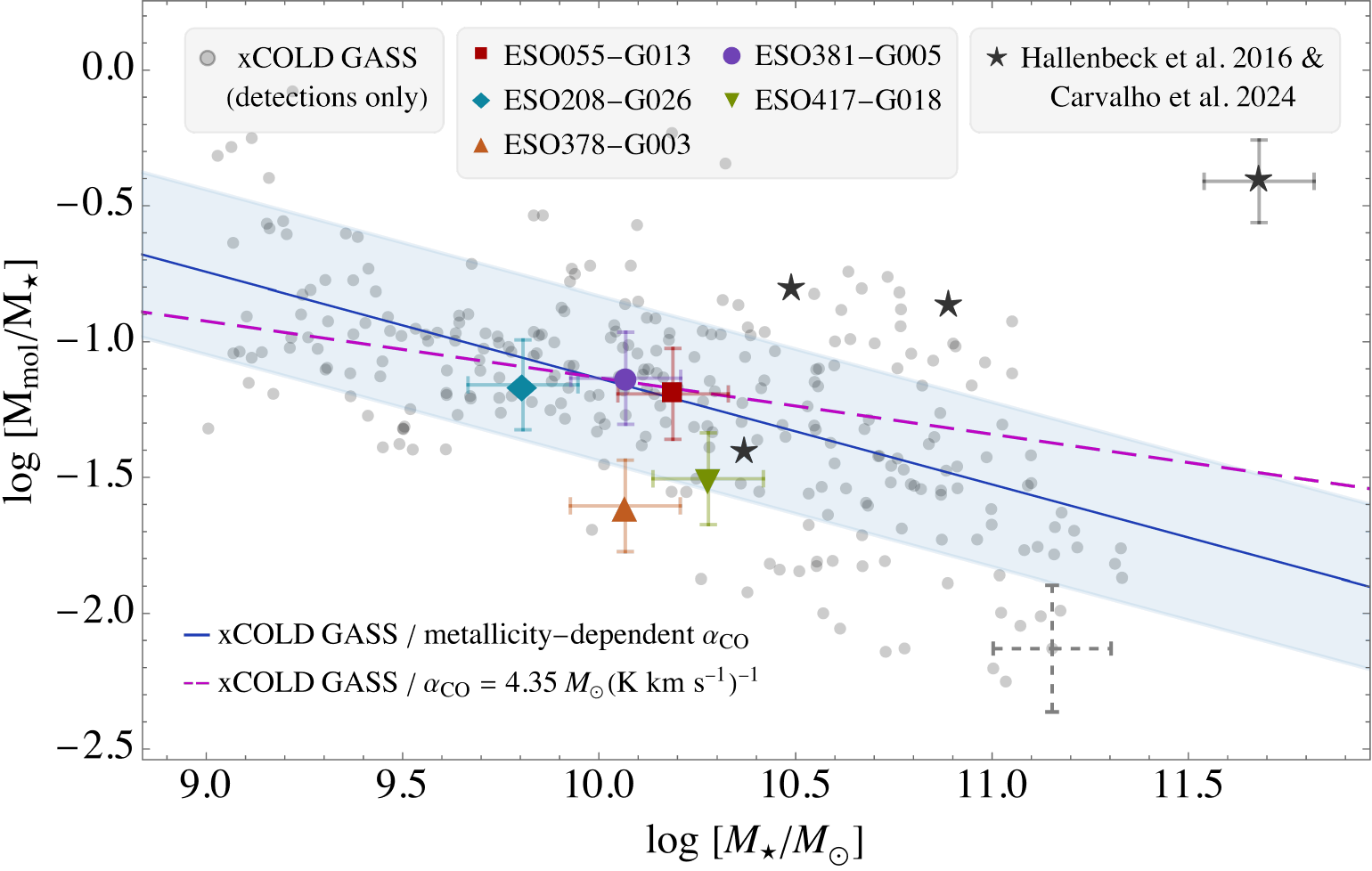}\\
    \;\\
    \includegraphics[width=0.45\textwidth]{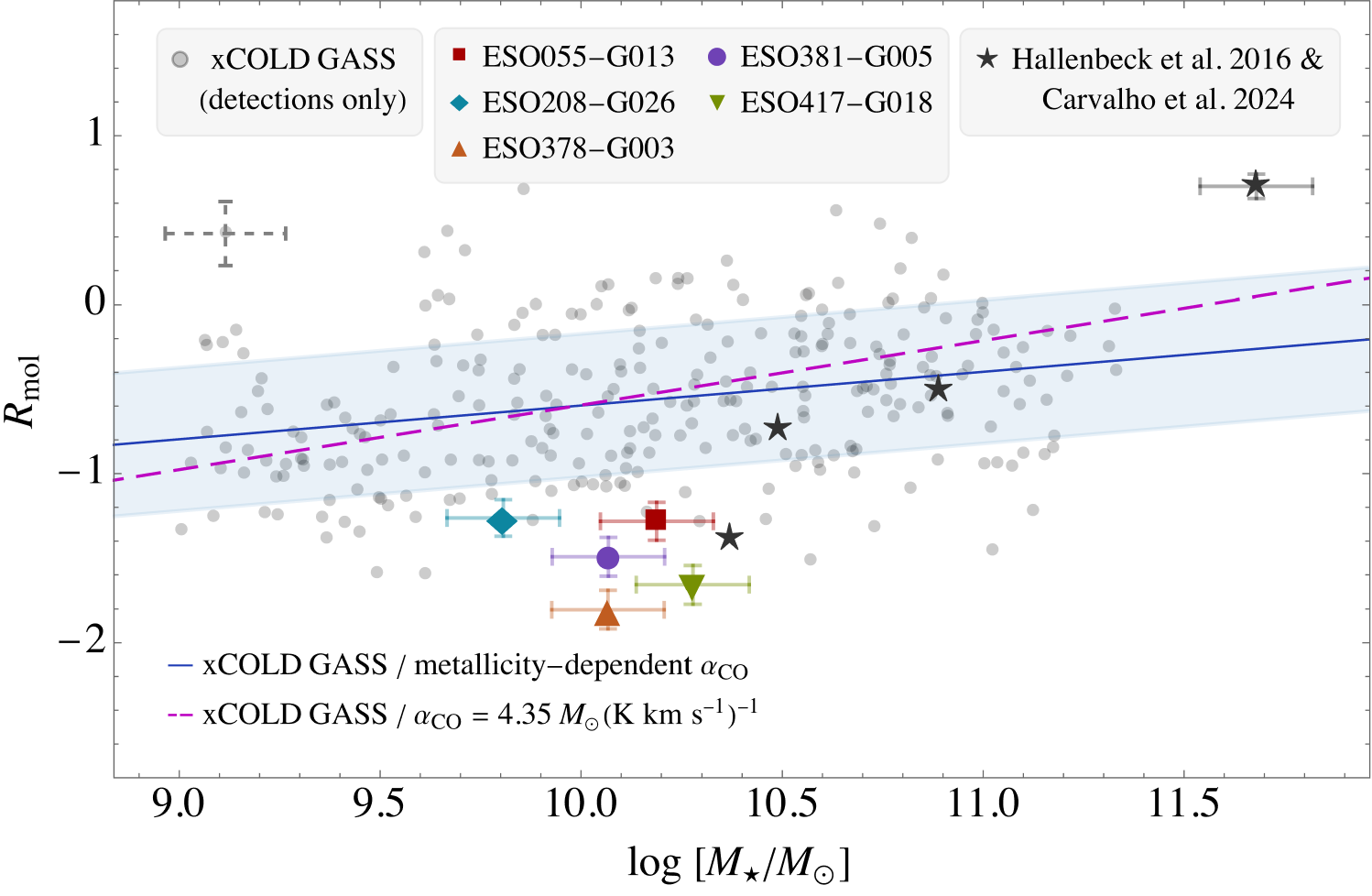}
    \caption{Comparison of (a) \hi\ gas mass, (b) \Htwo\ gas mass, and (c) \Htwo-\hi\ ratio of the five galaxies with xCOLD GASS survey data. The black stars represent UGC 6168, UGC 7899, NGC 5230, and the black star with error bars represents UGC 2885. The conversion factor between CO(1-0) radiation and \Htwo\ is $\alpha_\text{\co} = 4.35\, \mathrm{M}_\odot\mathrm{(K\,km\,s^{-1}\;pc^2)^{-1}}$.}
    \label{fig:scaling_relations}
\end{figure*}

We compare the amount of \hi, \Htwo\ and the molecular gas fraction ($R_\text{mol}$) with the data from the xCOLD GASS survey for 273 galaxies. Figure~\ref{fig:scaling_relations} shows the \hi\ and \Htwo\ scaling relations normalised by the stellar mass ($M_{\text{\hi}}/M_{\star}$ and $M_{\mathrm{mol}}/M_{\star}$ fractions), as well as the logarithmic molecular-to-atomic gas ratio ($R_{\mathrm{mol}} = \log[M_\text{mol}/M_\text{\hi}]$), as a function of stellar mass ($M_{\star}$). We show data from the xCOLD GASS survey with grey dots and measurements from our sample with coloured symbols. For comparison we have included the values for UGC 6168, UGC 7899, NGC 5230, and UGC 2885 as black stars. We also include the linear regression fit and the $1\sigma$ lines obtained from \citet{Saintonge2011}. The linear regression is fitted for molecular masses calculated with a metallicity-dependent \co-to-\Htwo\ conversion factor (see \citealt{Accurso2017} and \citealt{Saintonge2017}), and with a fixed \co-to-\Htwo\ conversion factor. To account for potential metallicity effects on the relations, we calculate scaling relation fits using both a fixed Milky Way conversion factor and a metallicity-dependent prescription following \citet{Accurso2017} and \citet{Saintonge2017}. The values of the fitted lines are summarised in Table \ref{tab:scalingrelations}. Our results confirm the relative excess of \hi\ in our sample galaxies. We find regular amounts of \Htwo\ in ESO055-G013, ESO208-G026, ESO381-G005, and ESO417-G018, and a deficit of \Htwo\ in ESO378-G003. As a result, all sample galaxies have an $R_\text{mol}$ below the expected range of values. From the additional \hi\ rich galaxies, UGC 6168 is the most similar to the \hix\ galaxies. This galaxy has the same low $R_\text{mol}$ fraction, while UGC 7899 and NGC 5230 show normal $R_\text{mol}$ fractions and UGC 2885 shows a very elevated $R_\text{mol}$ fraction.

\begin{table*}
\begin{tabular}{ccccccccc}
\hline
\multicolumn{3}{c}{} & \multicolumn{3}{c}{Metallicity-dependent $\alpha_\text{\co}$} & \multicolumn{3}{c}{Fixed $\alpha_\text{\co} = 4.35 \,\mathrm{M}_\odot\mathrm{(K\,km\,s^{-1}\;pc^2)^{-1}}$}\\
\multicolumn{3}{c}{} & \multicolumn{3}{c}{(1)} & \multicolumn{3}{c}{(2)}\\
$y$ & $x$ & $x_0$ & $m$ & $b$ & $\sigma$ & $m$ & $b$ & $\sigma$\\ \hline
$R_\text{mol}$ & $\log\left[M_\star/\mathrm{M}_\odot\right]$ & 10.70 & $0.200 \pm 0.043$ & $-0.457 \pm 0.034$ & $0.425$ & $0.382 \pm 0.045$ & $-0.326 \pm 0.036$ & $0.447$\\
$\log\left[M_\text{\Htwo}/M_\star\right]$ & $\log\left[M_\star/\mathrm{M}_\odot\right]$ & 10.70 & $-0.391 \pm 0.031$ & $-1.409 \pm 0.024$ & $0.304$ & $-0.208 \pm 0.031$ & $-1.279 \pm 0.024$ & $0.304$\\
$\log\left[M_\text{\hi}/M_\star\right]$ & $\log\left[M_\star/\mathrm{M}_\odot\right]$ & 10.70 & $-0.591 \pm 0.041$ & $-0.952 \pm 0.033$ & $0.406$ &  &  & \\ \hline
\end{tabular}
\caption{Molecular-to-atomic hydrogen gas mass relations from the xCOLDGASS survey. The relations are parametrised as $y = m(x - x_0)+b$. From the fit, we obtained the standard deviation, $\sigma$. Parameters are fitted with molecular masses calculated (1): with a metallicity-dependent \co-to-\Htwo\ conversion factor (see \citealt{Accurso2017} and \citealt{Saintonge2017}), and (2): with a fixed \co-to-\Htwo\ conversion factor.}
\label{tab:scalingrelations}
\end{table*}

\subsection{3D dynamical modelling}
\label{sec:dynamical_modeling}

The derivation of kinematic properties of galaxies is fundamental to understanding their dynamical state and evolution. For this reason, we performed 3D modelling of the CO disc of the sample galaxies using the 3D-Based Analysis of Rotating Objects via Line Observations \textsc{3DBarolo} \citep{DiTeodoro2015}\footnote{\url{https://editeodoro.github.io/Bbarolo/}}. The models for the gas discs consist of a number of concentric rings with non-zero thickness. The software compares the model to the real data ring by ring until the best match between the model and the observational data is found. To obtain the best fit we use the \textsc{SoFiA} mask together with the \co\ data cubes. We get the inclination, position angles (shown in Table \ref{tab:generalprop}), the rotation velocity and surface density profiles from the \textsc{3DBarolo} models.

For the fitting process with \textsc{3DBarolo}, rings are defined with a separation of 3 arcseconds and a list of initial guess parameters: rotation velocity, inclination and position angle based on the \hi\ data (adopted from \citealt{Lutz2018}). We define the centroid position, according to the flux distribution in the 0th moment map, and we estimate the radius of the disc based on the \textsc{SoFiA} results. After \textsc{3DBarolo} finds the most appropriate model, we then perform the fit again using the new initial parameters from the model to obtain the final fit parameters, which are presented in Table \ref{tab:generalprop}.

\citet{Lutz2018} has modelled the \hi\ discs using the \textit{Ti}lted \textit{Ri}ng \textit{Fi}tting Code \textsc{TiRiFiC} \citep{Jozsa2007}. Although their models provide a robust description of the \hi\ kinematics that is consistent with our findings, to directly compare with the \co\ disc models we model the \hi\ discs with \textsc{3DBarolo} as well, using the same method as with the \co\ data. We obtain the rotation velocity of the \hi\ disc and compare it with the \co\ models. We find that the kinematics of the \co\ and \hi\ gas discs are overall consistent. We find only small discrepancies in the inclination, position angle and the centroid of the galaxy.

\subsubsection{Surface density profiles}
\begin{figure*}
    \centering
    \includegraphics[width=0.42\textwidth]{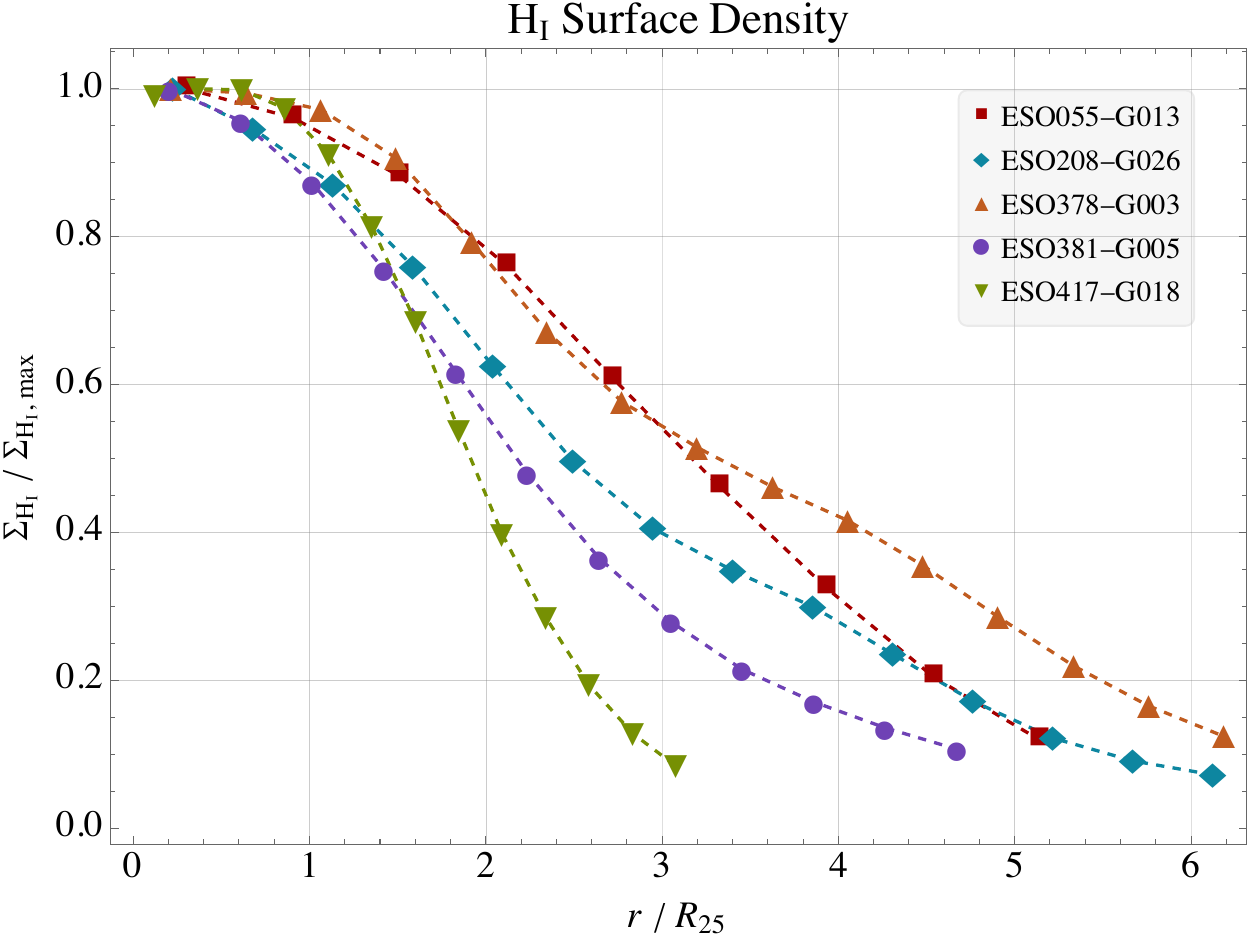}\qquad\includegraphics[width=0.42\textwidth]{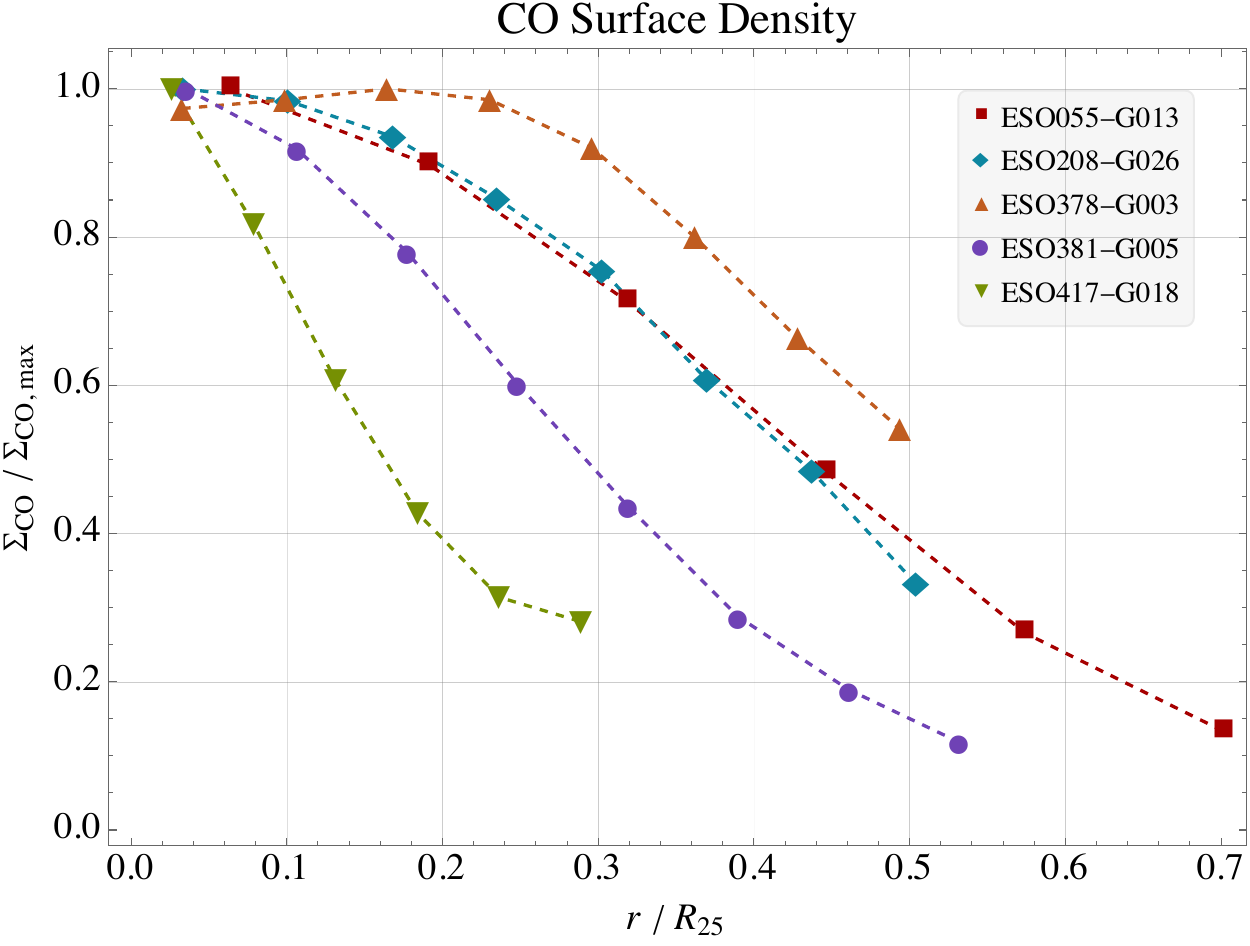}
    \caption{Peak-normalised radial surface density profiles of \hi\ and \co\ gas for the sample of galaxies from \textsc{3DBarolo}. Left: \hi\ surface density. Right: \co\ surface density. The galactocentric radius $r$ is normalised by the optical radius $R_{25}$ of each galaxy.}
    \label{fig:surfacedensity}
\end{figure*}

\textsc{3DBarolo} also provides the measured surface brightness densities in the fitted rings. The surface density profiles for both \hi\ and \co\ gas discs are shown in Figure \ref{fig:surfacedensity}. The \co\ gas discs have a surface density profile as expected. The surface density profile of \hi\ gas is also regular, except for ESO208-G026 and ESO378-G003, where the exponential decay is distorted by an unexpected flux increase around 30 kpc and 40 kpc from the centroid, respectively. Both of these galaxies show irregular extensions in their \hi\ morphology. 

\subsubsection{Radial velocity profiles}
\begin{figure*}
    \centering
    \includegraphics[width=0.42\textwidth]{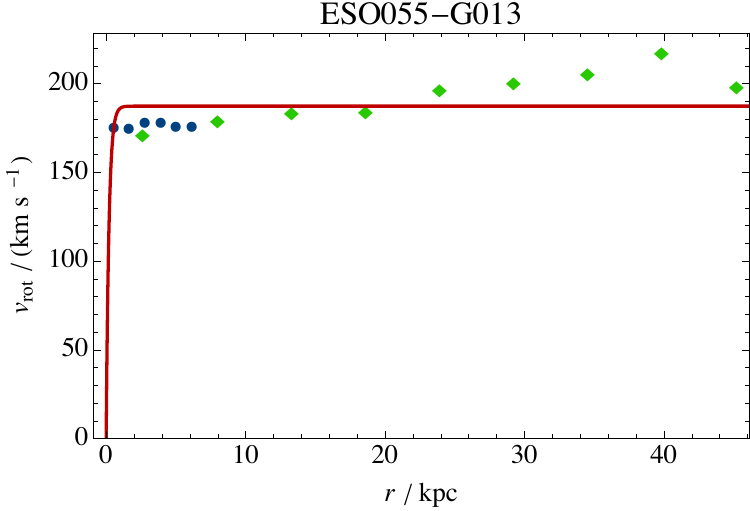}
    \includegraphics[width=0.42\textwidth]{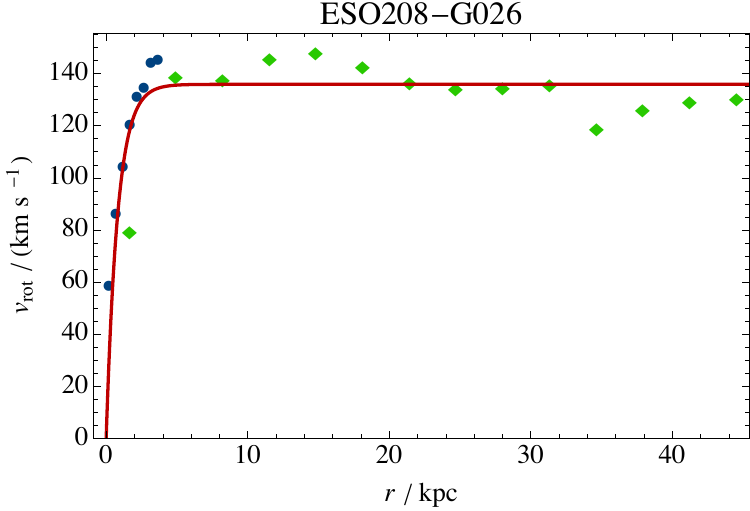}
    \includegraphics[width=0.42\textwidth]{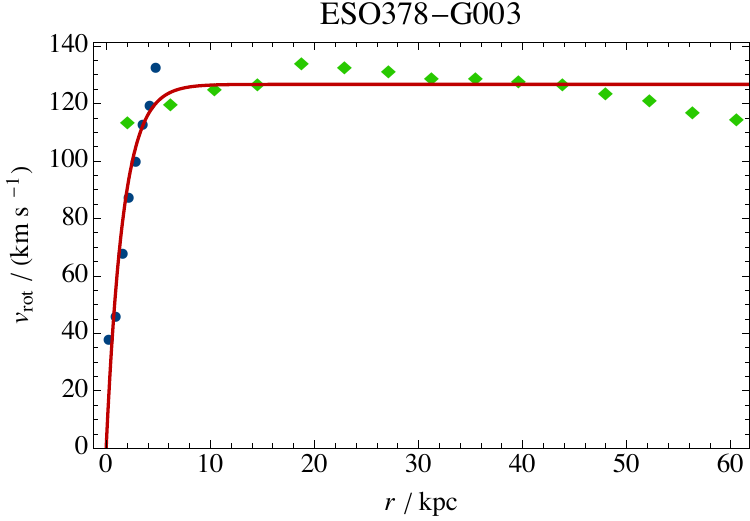}
    \includegraphics[width=0.42\textwidth]{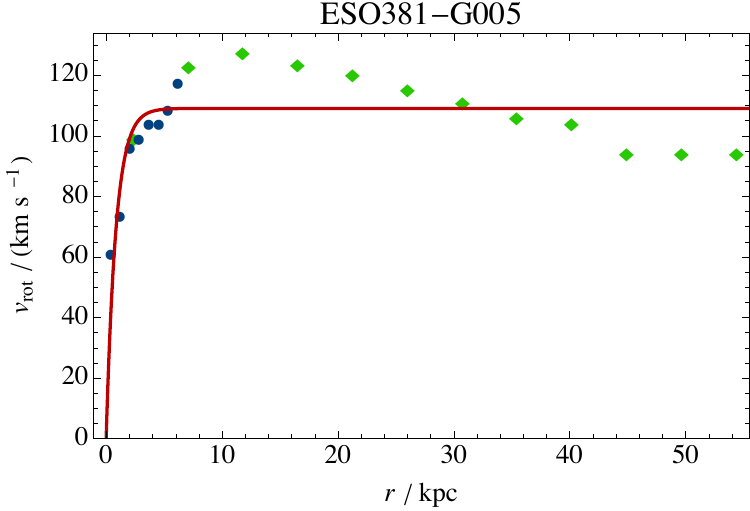}
    \includegraphics[width=0.42\textwidth]{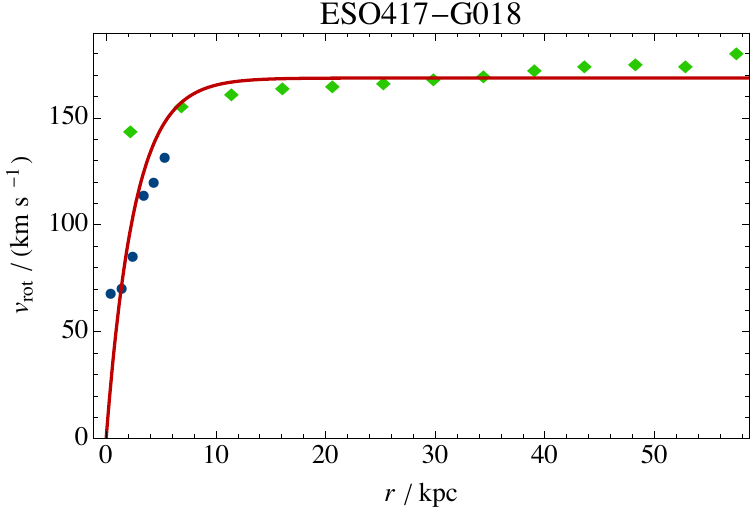}
    \caption{Rotation velocity calculated by rings with \textsc{3DBarolo} (green for \hi\ and blue for \co, each point, one ring) and rotation curves fitted with function \eqref{eq:vrot} in red, for each galaxy as labelled.}
    \label{fig:rotationcurves}
\end{figure*}

We use the rotation velocity values for each ring in the \co\ and the \hi\ gas discs to construct the rotation curves of the galaxies by fitting the data with the following function:
\begin{equation}
    v_\text{rot}(r) = v_\text{flat}\left[1-\exp\left(-\frac{r}{r_\text{flat}}\right)\right],
    \label{eq:vrot}
\end{equation}
where $v_\text{rot}(r)$ is the rotation velocity as a function of galactocentric radius $r$, $v_\text{flat}$ is the asymptotic flat rotation velocity reached in the outer disc, and $r_\text{flat}$ is the scale radius at which the rotation curve transitions from the rising inner region to the flat outer region. Due to the resolution and the size of the gas discs, the rising part of the curve near the centre corresponds to the velocities obtained from the \co\ data, while the flat part of the curve corresponds to the \hi\ data. The values for $r_{\mathrm{flat}}$ and $v_{\mathrm{flat}}$ obtained by fitting equation~\eqref{eq:vrot} are shown in Table \ref{tab:rot_param}, and the rotation curves and rotation velocity data are shown in Figure \ref{fig:rotationcurves}. Combining both gas phases allows us to improve the rotation curves presented in \citet{Lutz2018}. The \hi\ data alone often yields unphysical $r_{\text{flat}}$ values for the rotation curves. In addition, the good agreement between the \hi\ and \co\ data also confirms that our kinematic models are robust. These kinematic models are the essential foundation for calculating the specific angular momentum and the global stability parameter used in our subsequent analysis.

\begin{table}
\centering
\begin{tabular}{ccc}
\hline
\multirow{2}{*}{ID} & $v_\text{flat}$ & $r_\text{flat}$\\
 & (\kms) & (kpc)\\
(1) & (2) & (3)\\
\hline
ESO055-G013 & $187.28 \pm 3.70$ & $0.201 \pm 0.090$ \\
ESO208-G026 & $135.80 \pm 3.23$ & $0.814 \pm 0.121$ \\
ESO378-G003 & $126.61 \pm 2.32$ & $1.618 \pm 0.151$ \\
ESO381-G005 & $109.04 \pm 2.84$ & $0.843 \pm 0.174$ \\
ESO417-G018 & $168.53 \pm 5.03$ & $2.550 \pm 0.346$ \\
\hline
\end{tabular} 
\caption{Parameters of the fitted rotation curve function from \textsc{3DBarolo} results. (1): The galaxy ID. (2): Maximum velocity of the model with the standard error. (3): Kinematic radius of the model with the standard error.}
\label{tab:rot_param}
\end{table}

\subsubsection{Velocity dispersion}
\begin{figure*}
    \centering
    \includegraphics[width=0.32\textwidth]{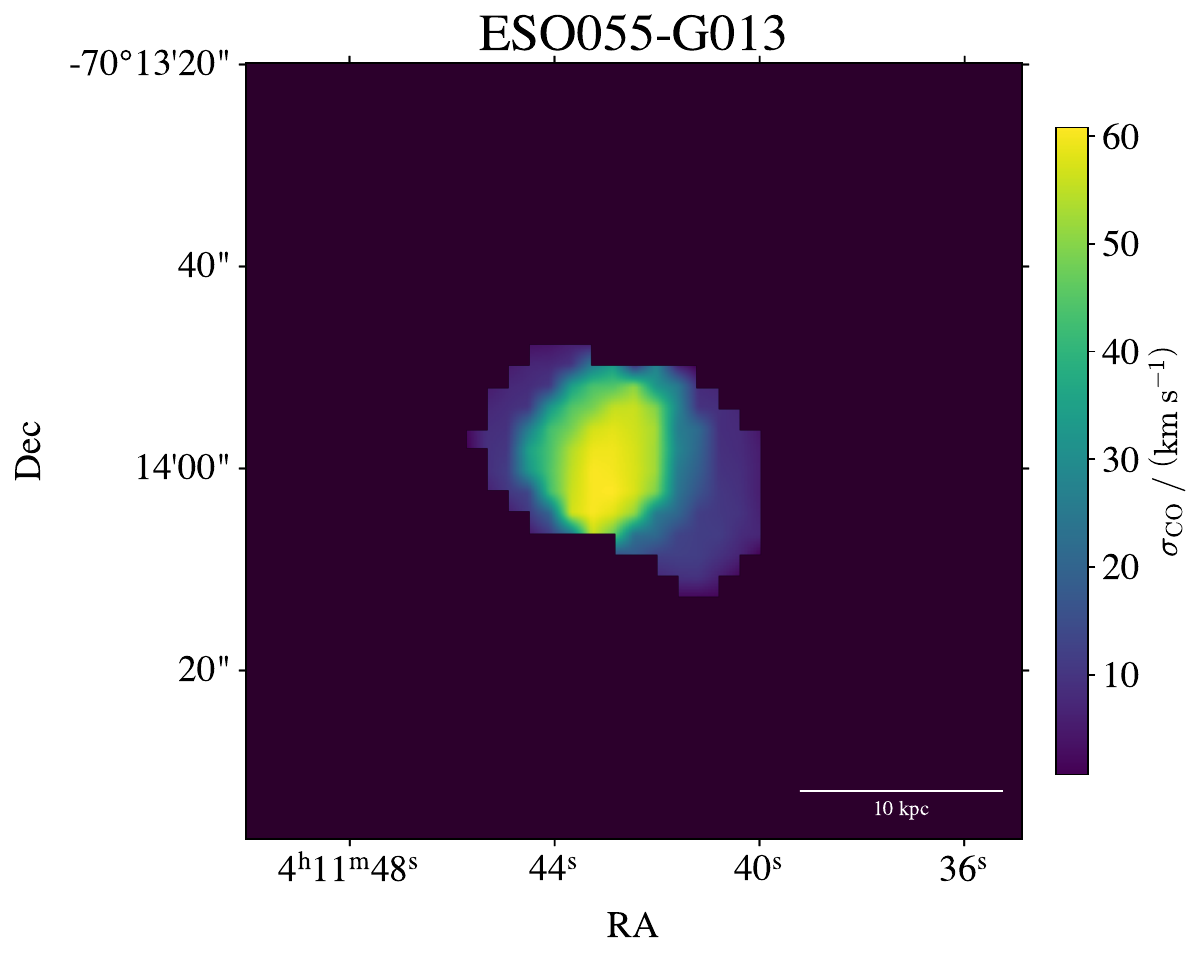}\quad\includegraphics[width=0.246\textwidth]{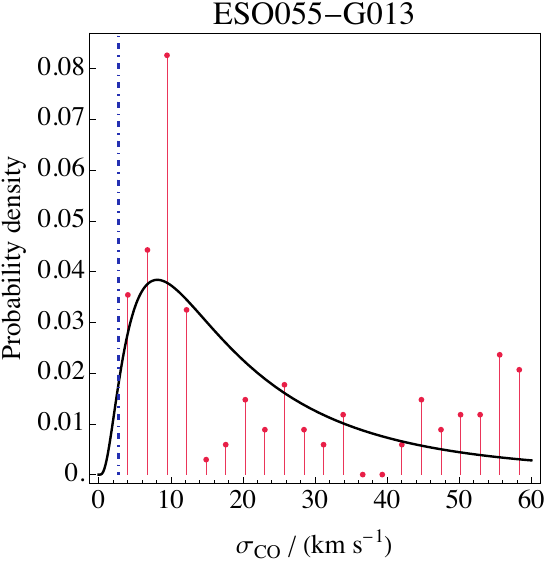}\quad\includegraphics[width=0.246\textwidth]{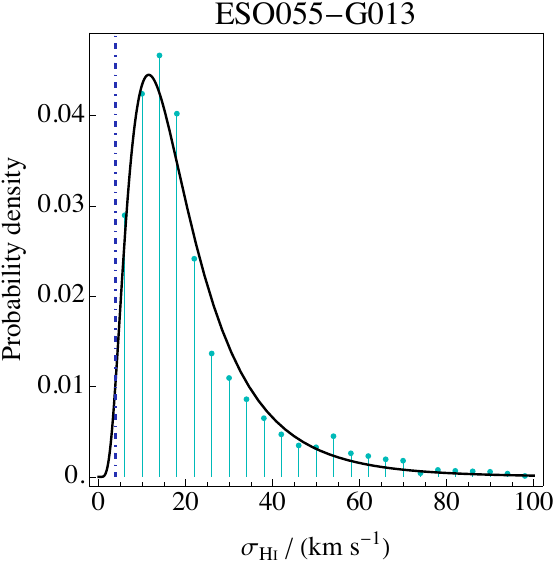}\\
    \includegraphics[width=0.32\textwidth]{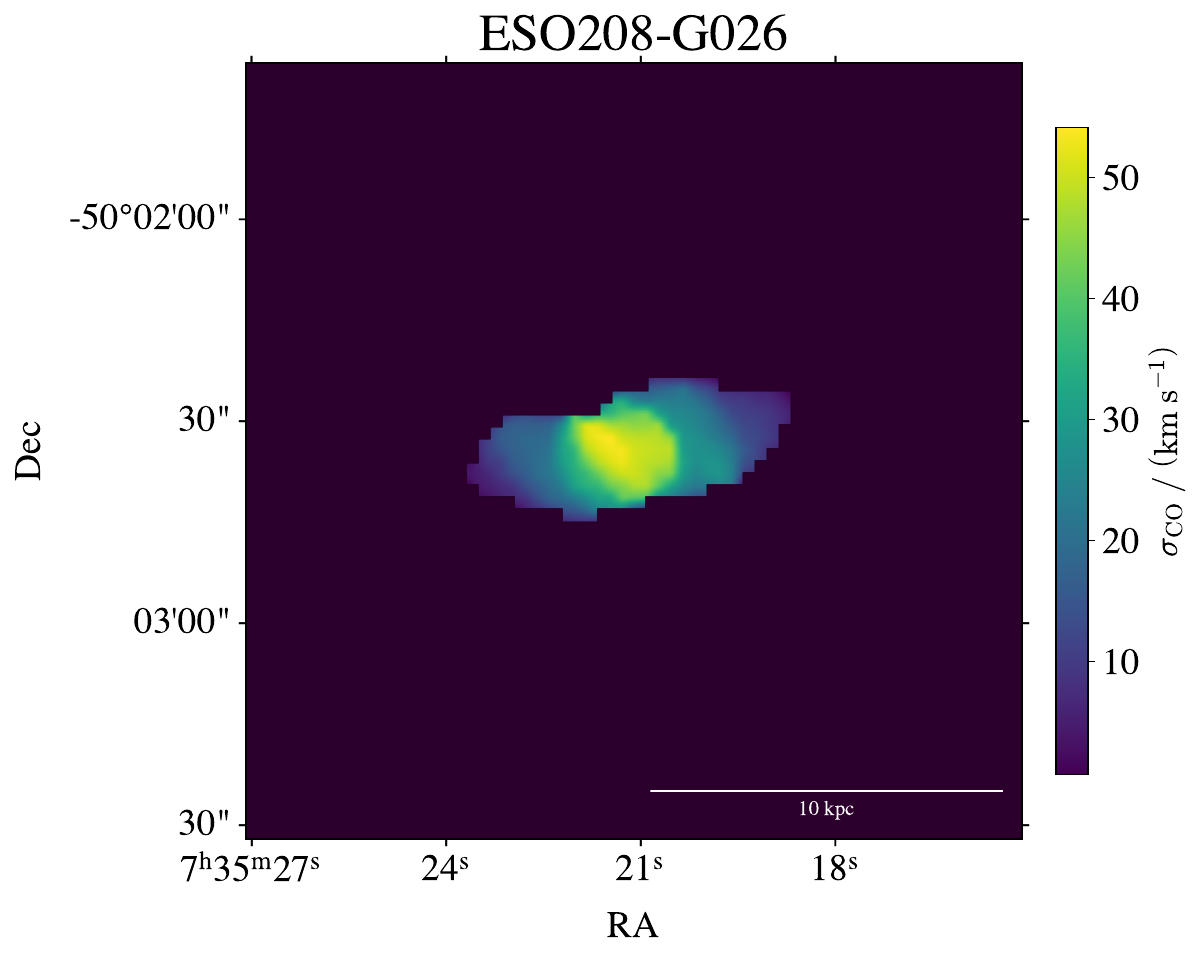}\quad\includegraphics[width=0.246\textwidth]{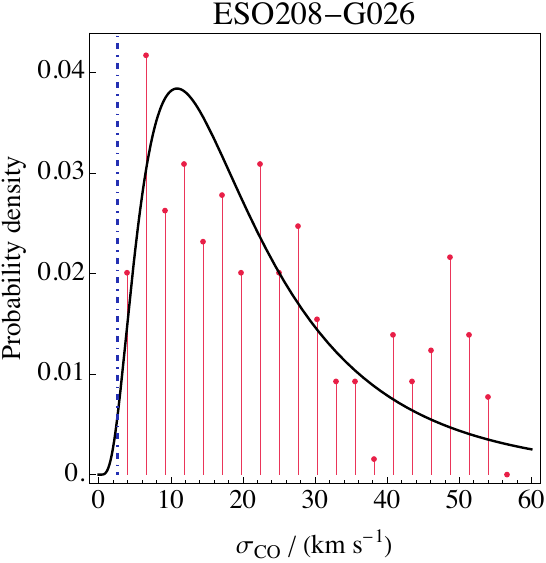}\quad\includegraphics[width=0.246\textwidth]{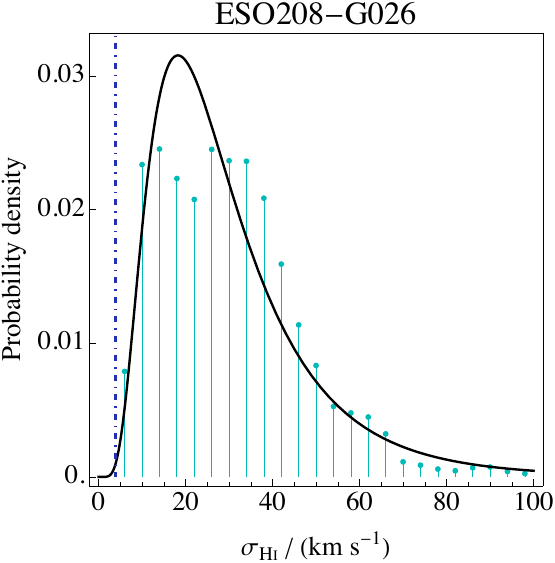}\\
    \includegraphics[width=0.32\textwidth]{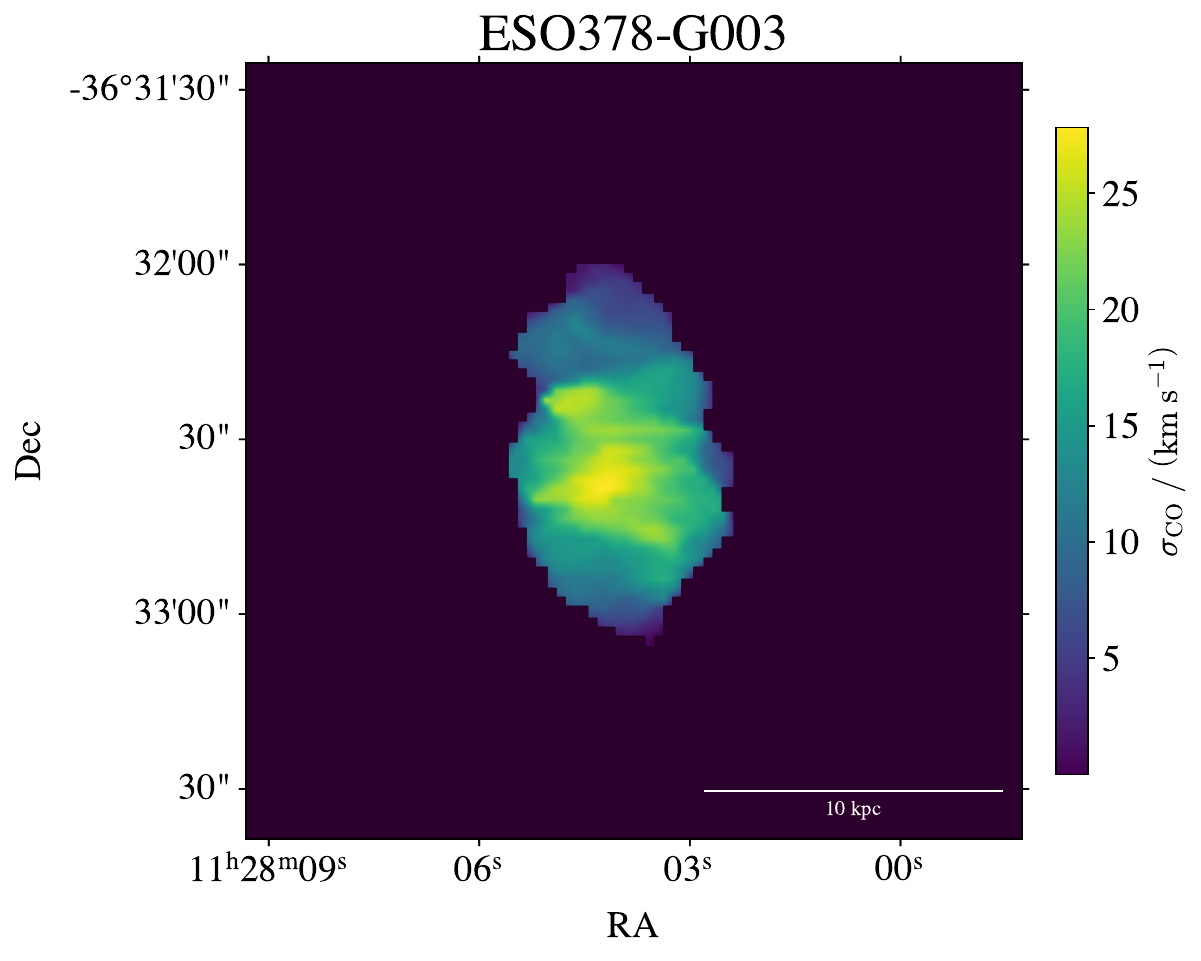}\quad\includegraphics[width=0.246\textwidth]{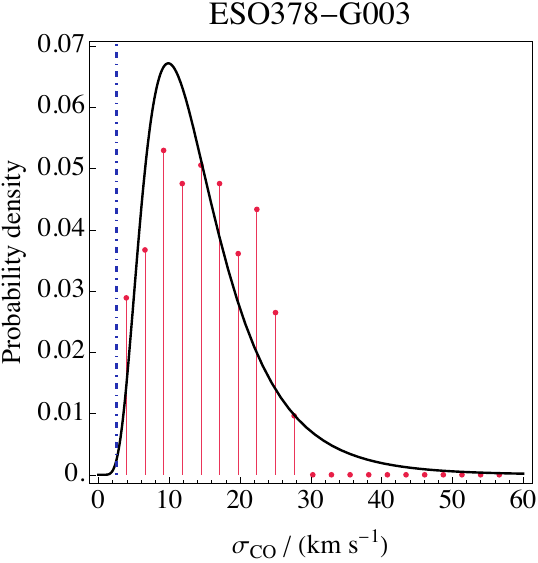}\quad\includegraphics[width=0.246\textwidth]{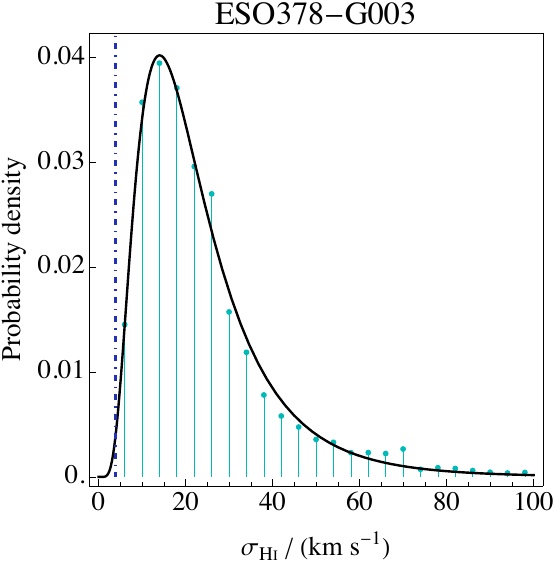}\\
    \includegraphics[width=0.32\textwidth]{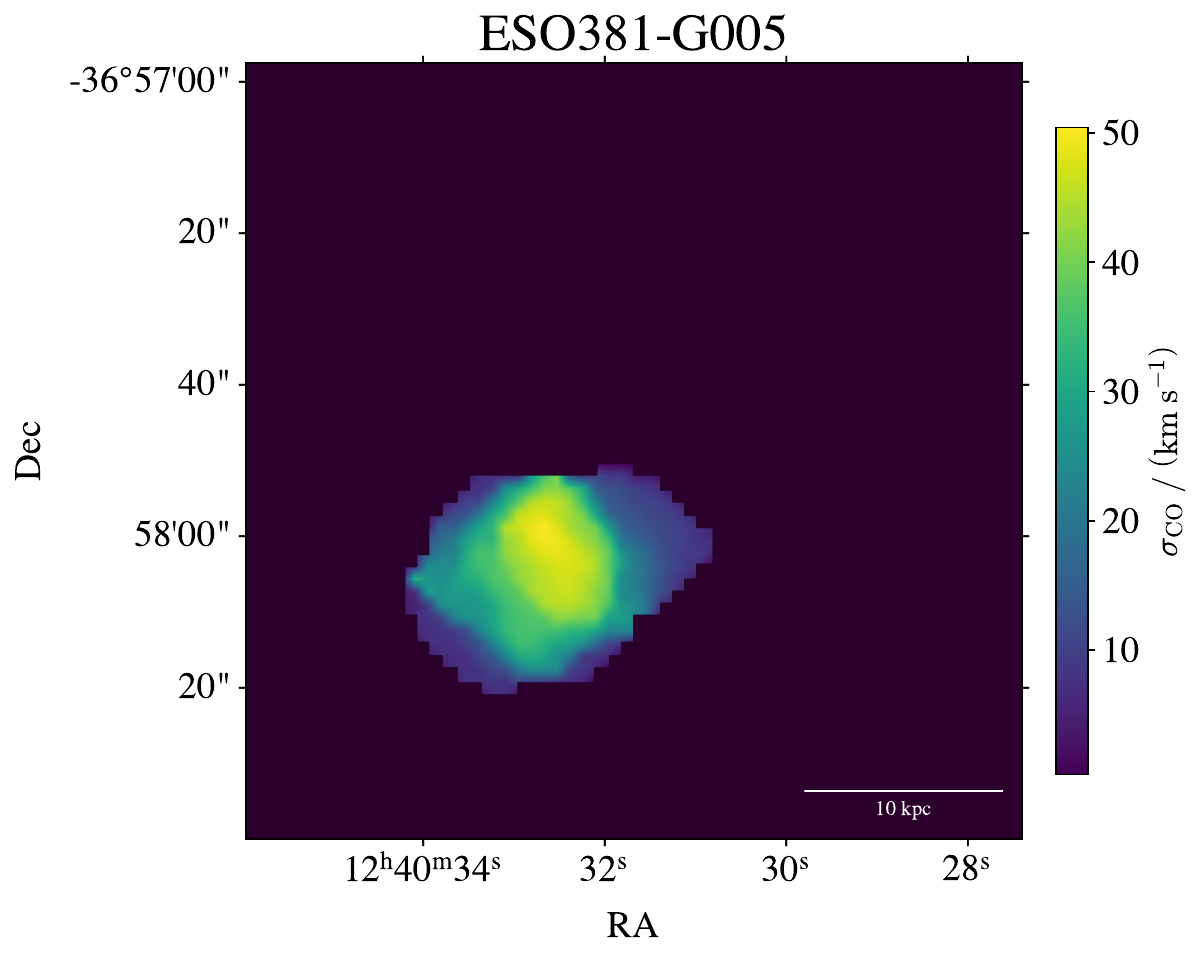}\quad\includegraphics[width=0.246\textwidth]{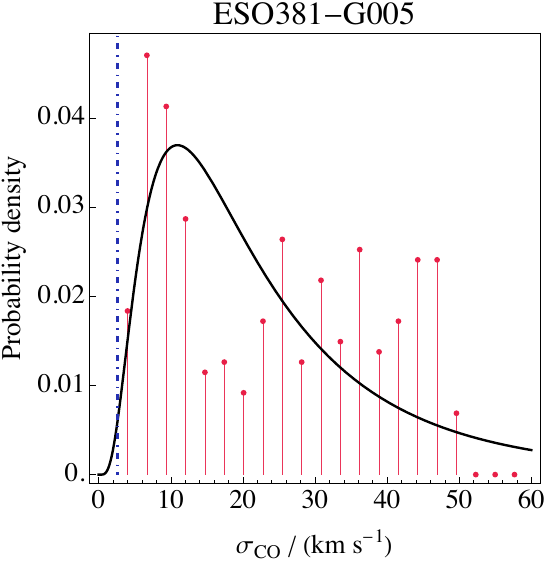}\quad\includegraphics[width=0.246\textwidth]{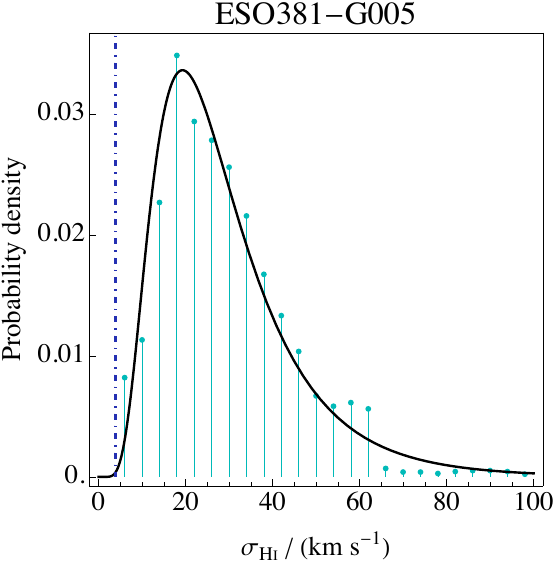}\\
    \includegraphics[width=0.32\textwidth]{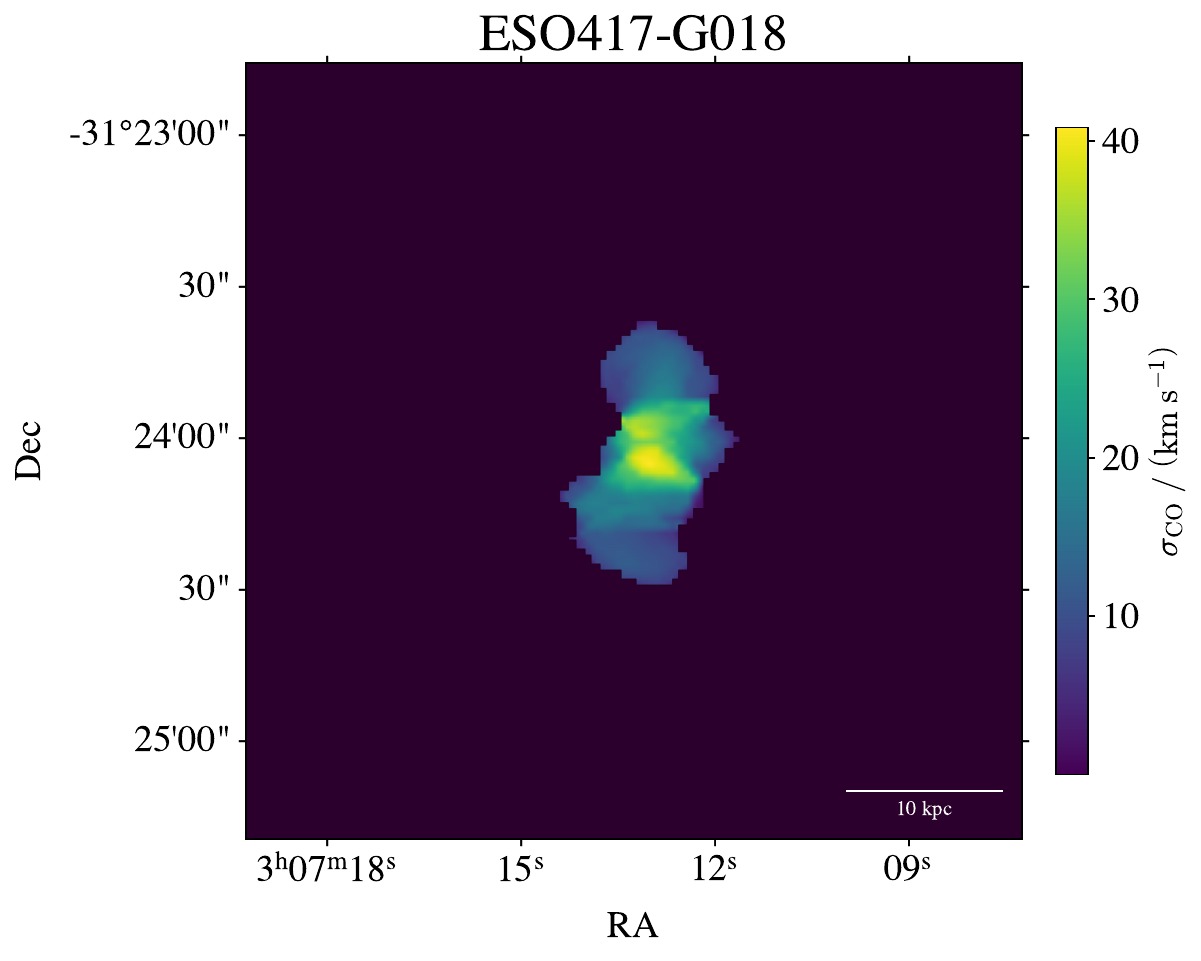}\quad\includegraphics[width=0.246\textwidth]{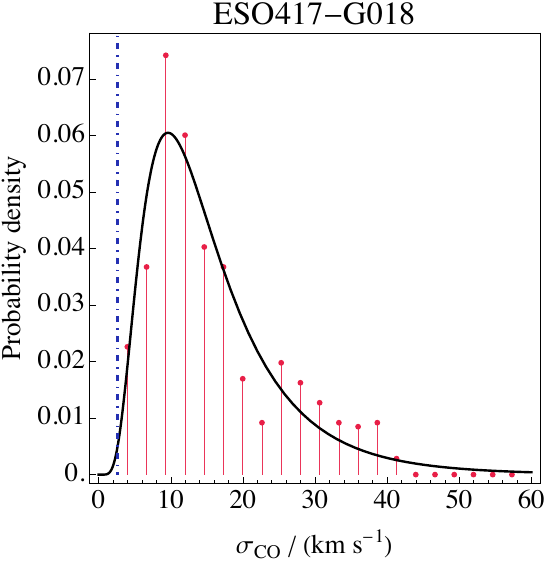}\quad\includegraphics[width=0.246\textwidth]{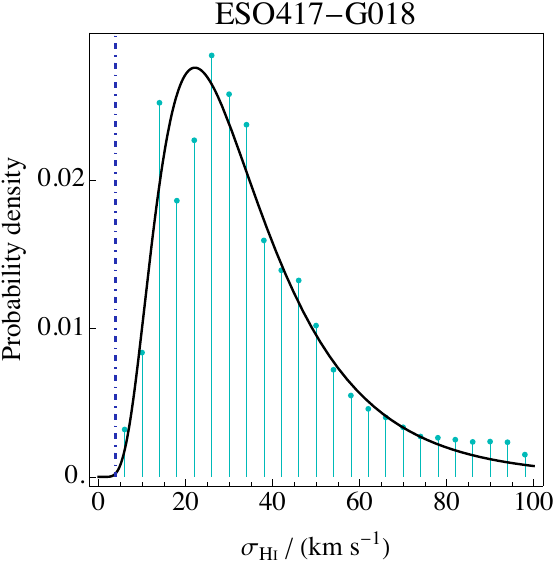}
    \caption{Left: the 2nd moment map for each galaxy (pbcor data or combined if available). Right: Velocity dispersion histograms of the \co\ (red) and \hi\ (cyan) data. Bin sizes are the same as the velocity resolution of the data. The black solid line shows the fitted probability density function to the data. The blue dashed line indicates the velocity resolution cut-off.}
    \label{fig:pbdisp_pbcor}
\end{figure*}

We obtain the distribution of the velocity dispersion from the 2nd moment maps created by \textsc{SoFiA}. We create density histograms using the velocity resolution for the bin sizes ($\sim 2.6$ \kms for \co\, and $\sim4.0$ \kms for \hi), establishing a lower limit at this value. Since the velocity dispersion has only positive numbers, the peak of the histograms is clearly after the velocity resolution cut-off, and to avoid beam smearing effects, we assume that the data follow a log-normal distribution, which is asymmetrical with a heavy tail on the right. We obtain the parameters of the probability density function by maximising the log-likelihood function. We calculate the mean, median, and mode for the velocity dispersion of the \co\ and \hi\ discs from the fitted probability density function. We show the histograms and the fitted density functions in Figure~\ref{fig:pbdisp_pbcor} for each galaxy. It is relevant to note that the velocity dispersion of the \hi\ disc follows the log-normal distribution more tightly than that of the \co, this is most likely because the \hi\ data is much better sampled due to the extended \hi\ disc compared to the compact \co\ disc. In addition, it is well known that the resolution of the data can strongly influence the measured velocity dispersion. Table~\ref{tab:veldisp} summarises the means, medians and modes of the fitted probability density function for the \co\ and \hi\ velocity dispersion. For subsequent calculations we only use the mode, which is quite robust as it is the peak of the distribution, and the mean and median are conditioned by the long tail of the distribution due to beam smearing.

\hi\ gas commonly exhibits a higher velocity dispersion than the molecular gas \citep{Mogotsi2016}. Even so, the measured values for this sample are higher than expected. For example, \citet{Tamburro2009} found a mean $\sigma_\text{\hi}$ of ${\sim}10 \text{ \kms}$, \citet{Stilp2013} found $\sigma_\text{\hi}$ of ${\sim}6\text{--}10 \text{ \kms}$, and \citet{Mogotsi2016} found $\sigma_\text{\hi}$ of ${\sim}12 \text{ \kms}$. Compared to this, all galaxies in our sample have $\sigma_\text{\hi} > 11 \text{ \kms}$, and an average mode of $\sigma_\text{\hi} {\sim} 17 \text{ \kms}$. According to \citet{Tamburro2009}, a velocity dispersion above the thermal value (${\sim}8 \text{ \kms}$) from warm \hi\ implies that turbulence drivers must be responsible for creating these high $\sigma_\text{\hi}$ values. On the other hand, \citet{CalduPrimo2013} found an average $\sigma_\text{\co}$ of ${\sim}12 \text{ \kms}$ similar to $\sigma_\text{\hi}$, attributing this to a diffuse and thick molecular disc. \citet{Mogotsi2016} found $\sigma_\text{\co} {\sim} 7 \text{ \kms}$ while our sample has an average mode of $\sigma_\text{\co} {\sim} 10 \text{ \kms}$. 

We also calculate the ratio between \hi\ and \co\ velocity dispersion, pixel by pixel, only where the two values are above the velocity resolution. The mean, median, and mode of this pixel-by-pixel ratio are presented in Table \ref{tab:veldisp}. These values are also calculated from the corresponding fitted log-normal distribution, and show values strictly greater than 1.4, clearly illustrating the higher velocity dispersion in the \hi\ gas. For comparison, \citet{CalduPrimo2013} found a ratio of $1.0 \pm 0.2$, while \citet{Mogotsi2016} found a ratio of $1.4 \pm 0.2$, for typical nearby late-type galaxies.

The velocity dispersion is an important parameter for global stability against gravitational collapse, and star formation. A high velocity dispersion on galactic scales indicates more turbulent motions in the gas, which can slow down the conversion of \hi\ into \Htwo\ (e.g. \citealt{Krumholz2009}; \citealt{Ostriker2010}) and consequently suppress star formation efficiency (e.g. \citealt{Tamburro2009}; \citealt{Mogotsi2016}).

\begin{table*}
\centering
\begin{tabular}{cccccccccc}
\hline
\multirow{2}{*}{ID} & \multicolumn{3}{c}{$\sigma_\text{\co}$} & \multicolumn{3}{c}{$\sigma_\text{\hi}$} & \multicolumn{3}{c}{\multirow{2}{*}{$\dfrac{\sigma_\text{\hi}}{\sigma_\text{\co}}$}} \\
 & \multicolumn{3}{c}{(\kms)} & \multicolumn{3}{c}{(\kms)} & \multicolumn{3}{c}{} \\
(1) & \multicolumn{3}{c}{(2)} & \multicolumn{3}{c}{(3)} & \multicolumn{3}{c}{(4)} \\
\hline
 & Mean & Median & Mode & Mean & Median & Mode & Mean & Median & Mode \\
\hline
ESO055-G013 & 25.5 & 17.4 & 8.2 & 21.1 & 17.3 & 11.5 & 3.4 & 2.5 & 1.4 \\
ESO208-G026 & 24.3 & 18.6 & 10.9 & 30.4 & 25.7 & 18.2 & 3.8 & 3.1 & 2.1 \\
ESO378-G003 & 14.9 & 13.0 & 9.9 & 23.8 & 20.0 & 14.0 & 4.5 & 4.2 & 3.5 \\
ESO381-G005 & 25.5 & 19.1 & 10.9 & 29.6 & 25.6 & 19.3 & 3.4 & 2.8 & 1.8 \\
ESO417-G018 & 15.9 & 13.5 & 9.6 & 35.4 & 30.3 & 22.1 & 5.7 & 4.9 & 3.7\\
\hline
\end{tabular}
\caption{Measures of central tendency of the velocity dispersion distribution function fitted from 2nd moment maps following a log-normal distribution. (1): Galaxy ID. (2): Velocity dispersion of \co\ disc. (3): Velocity dispersion of \hi\ disc. (4): Pixel by pixel ratio between \hi\ and \co\ velocity dispersion.}
\label{tab:veldisp}
\end{table*}

\subsection{Specific angular momentum and global stability factor}

\subsubsection{\emph{\hi}, molecular, stellar and baryonic specific angular momentum}
The specific angular momentum (the angular momentum per unit mass, $j = J/M$) is one of the fundamental parameters in the evolution of galaxies. Combining the \hi\ and molecular mass density and rotation velocity profiles from \textsc{3DBarolo} we can calculate the specific angular momentum of the \hi, molecular and stellar discs following \citet{Lutz2018}:
\begin{equation}
    j_\text{disc} = \frac{\sum_i M_i\;v_{\text{rot},i}\; r_i}{\sum_i M_i},
\end{equation}
where $M_i$, $v_{\text{rot},i}$ and $r_i$ are the mass, rotation velocity and radius of the $i$-th ring. Because direct stellar velocity information is not available, the stellar specific angular momentum ($j_\star$) is estimated by combining the stellar mass profiles with the \hi-\co\ rotation curve parameters in Table \ref{tab:rot_param}. This approach operates under the standard assumption that the kinematics of the stellar disc broadly behave as the kinematics of the \hi\ disc \citep{Lutz2018}.

The baryonic specific angular momentum is the mass-weighted average of the specific angular momentum of each baryonic component (e.g. \citealt{Hardwick2022b}), as follows:
\begin{equation}
    j_\text{bar} = \frac{j_\star M_\star + j_\text{mol} M_\text{mol} + 1.35 j_\text{\hi} M_\text{\hi}}{M_\star + M_\text{mol} + 1.35 M_\text{\hi}}.
\end{equation}
The baryonic specific angular momentum and the angular momentum of each baryonic component are summarised in Table \ref{tab:SAM}. Values for UGC 6168 and UGC 7899 are adopted from \citet{Lutz2018}.

\begin{table*}
\centering
\begin{tabular}{cccccc}
\hline
ID & $\log\left[\dfrac{j_\star}{\mathrm{kpc\,km\,s^{-1}}}\right]$ & $\log\left[\dfrac{j_\text{mol}}{\mathrm{kpc\,km\,s^{-1}}}\right]$ & $\log\left[\dfrac{j_\text{\hi}}{\mathrm{kpc\,km\,s^{-1}}}\right]$ & $\log\left[\dfrac{j_\text{bar}}{\mathrm{kpc\,km\,s^{-1}}}\right]$ & $\log\left[q\right]$ \\
(1) & (2) & (3) & (4) & (5) & (6) \\
\hline
ESO055-G013 & $2.39$ & $2.77$ & $3.69$ & $3.49$ & $-0.71$ \\
ESO208-G026 & $2.26$ & $2.49$ & $3.47$ & $3.28$ & $-0.34$ \\
ESO378-G003 & $2.86$ & $2.55$ & $3.61$ & $3.48$ & $-0.57$ \\
ESO381-G005 & $2.50$ & $2.54$ & $3.48$ & $3.36$ & $-0.67$ \\
ESO417-G018 & $2.94$ & $2.58$ & $3.70$ & $3.55$ & $-0.48$ \\
\hline
UGC 6168 & -- & -- & -- & $3.62$ & $-0.73$ \\
UGC 7899 & -- & -- & -- & $3.73$ & $-0.86$ \\
\hline
\end{tabular}
\caption{Specific angular momentum of the sample. (1): The galaxy ID. (2): Stellar, (3): molecular, (4): atomic, and (5): baryonic specific angular momentum. (6): Global stability factor.}
\label{tab:SAM}
\end{table*}

\subsubsection{Stellar and baryonic Fall relation}

The stellar mass and the stellar specific angular momentum follow a power-law relation, known as the Fall relation $j_\star \propto M_\star^{2/3}$ predicted from the $\Lambda$CDM model \citep{Fall1983}. In addition, there is also a plane defined by the stellar mass, the stellar specific angular momentum and the atomic mass fraction $f_{\star,\mathrm{atm}}$ \citep{Hardwick2022b}.

\begin{figure}
    \centering
    \includegraphics[width=0.4\textwidth]{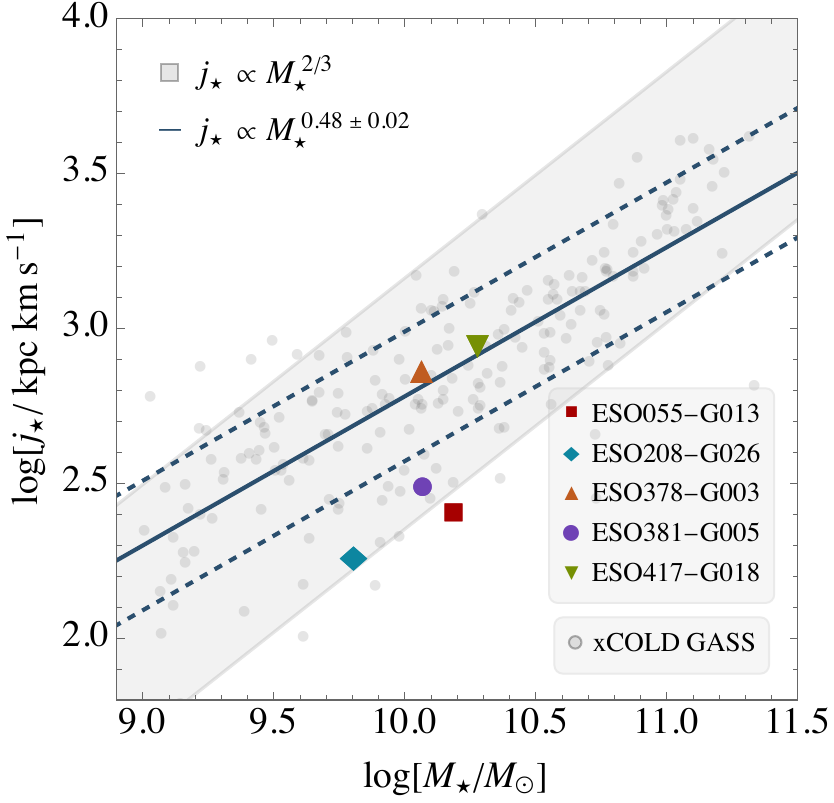}
    \caption{Stellar mass versus stellar specific angular momentum. In light grey, the projection of the plane $M_\star$-$j_\star$-$f_{\star,\mathrm{atm}}$ from the xCOLDGASS data. The dark line indicates the 2D best-fitting linear relation while the dashed lines indicate the $1\sigma$ scatter of the relation.}
    \label{fig:jsvsMs}
\end{figure}

In Figure \ref{fig:jsvsMs} we compare the stellar specific angular momentum measurements to the Fall relation and galaxies from the xCOLD GASS sample. ESO378-G003 and ESO417-G018 are in the range of expected values for this relation. However, ESO055-G013, ESO208-G026 and ESO381-G005 fall below the $1\sigma$ scatter of the xCOLD GASS sample. Specifically, their stellar specific angular momentum is so anomalously low that it aligns with the theoretical projection for gas-poor galaxies (where $f_{\star,\mathrm{atm}}=0.01$), despite the fact that their actual atomic mass fraction is highly gas-rich ($f_{\star,\mathrm{atm}}>1$) \citep{ManceraPina2021}.

\begin{figure}
    \centering
    \includegraphics[width=0.4\textwidth]{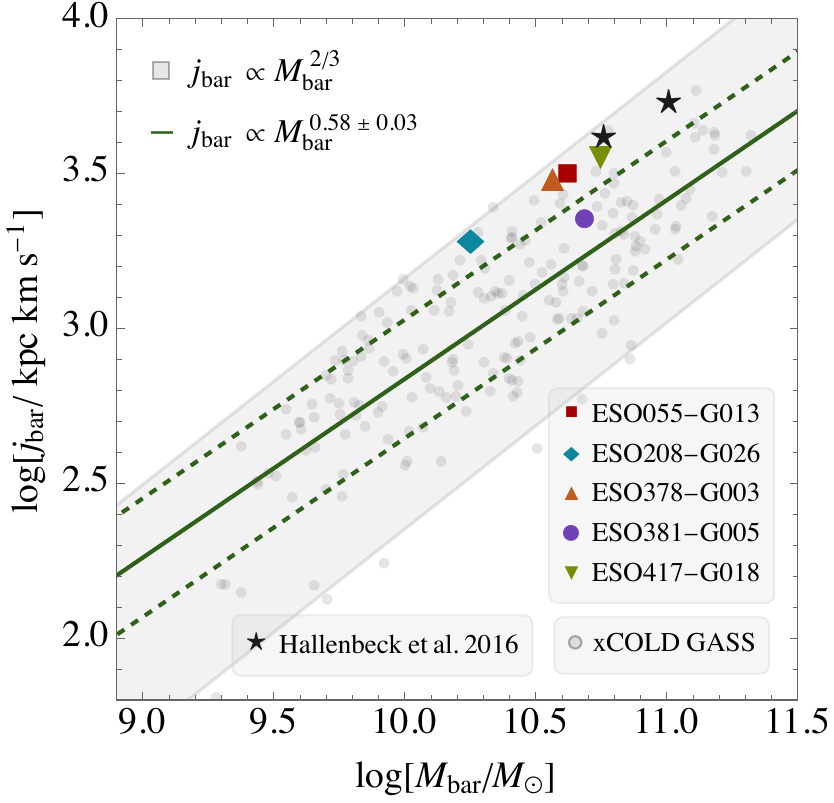}
    \caption{Baryonic mass versus baryonic specific angular momentum. In light grey, the projection of the plane $M_{\mathrm{bar}}$-$j_{\mathrm{bar}}$-$f_{\mathrm{atm}}$ from the xCOLDGASS data. The dark line indicates the 2D best-fitting linear relation while the dashed lines indicate the $1\sigma$ scatter of the relation. The black stars represent UGC 6168 and UGC 7899.}
    \label{fig:jbvsMb}
\end{figure}

Conversely, the Fall relation for baryonic mass and baryonic specific angular momentum shows a different trend (Figure \ref{fig:jbvsMb}). All of our sample galaxies—except ESO381-G005—lie above the expected 1$\sigma$ range of the xCOLD GASS control sample. UGC 6168 and UGC 7899 show the same trend as our sample. The baryonic specific angular momentum of these galaxies is remarkably high, aligning near the theoretical upper limit projection for galaxies composed entirely of gas ($f_\text{atm}=1$), even though their actual atomic gas fractions are between 0.6 and 0.7. \citet{ManceraPina2021} and \citet{Hardwick2022b} found that the distance from the main baryonic relation is correlated with the gas fraction and our results agree well with their findings for galaxies with high atomic gas fractions. 

This combination of low stellar specific angular momentum with elevated baryonic specific angular momentum indicates that the bulk of the galaxies' angular momentum resides in the gas component, particularly in the extended \hi\ disc.
 
\subsubsection{Global disc stability parameter}

\begin{figure}
    \centering
    \includegraphics[width=0.4\textwidth]{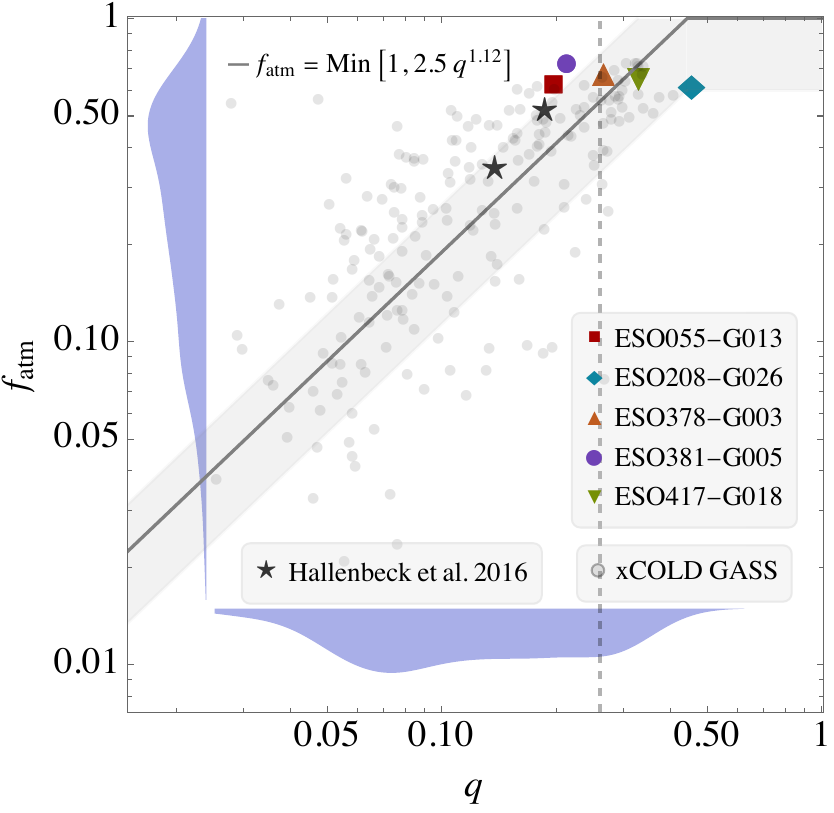}
    \caption[Atomic gas fraction versus global stability factor.]{Atomic gas fraction versus global stability factor in logarithmic scale. The solid black line is the relation found by \citet{Obreschkow2016}, and the shaded area represents 40\% of scatter. The dashed grey line is the limit for global stability, $q\approx 1/(\sqrt{2}\,e)$, where a higher $q$ parameter implies $f_\text{atm} = 1$. The black stars represent UGC 6168 and UGC 7899. The purple shaded histograms on the left and on the bottom show the distribution of the xCOLD GASS data points on the respective axes.}
    \label{fig:qvsfatm}
\end{figure}

The combination of atomic gas, baryonic specific angular momentum and baryonic mass can be expressed with a dimensionless stability parameter \citep{Obreschkow2016}, as follows:
\begin{equation}
    q = \frac{j_\text{bar}\,\sigma_\text{\hi}}{G\,M_\text{bar}},
\end{equation}
where $\sigma_\text{\hi}$ is the velocity dispersion of the \hi\ gas disc (we use the mode from Table \ref{tab:veldisp} in our calculations), and $G$ is the gravitational constant. As specified by \citet{Obreschkow2016}, the atomic gas fraction and the stability parameter $q$ are approximately related by
\begin{equation}
    f_\text{atm} = \min\left\{1,2.5q^{1.12}\right\}.
\end{equation}
Figure \ref{fig:qvsfatm} presents the theoretical relation in comparison with our sample and the xCOLDGASS sample. Our sample appears close to the upper limit of this relation with 3 galaxies inside the range of the 40 per cent scatter. The two galaxies outside the 40 per cent range are ESO381-G005 and ESO055-G013, with relatively high atomic gas fractions compared to their $q$ values. UGC 6168 has similarly high atomic gas fraction and $q$ parameter compared to the \hix\ galaxies, while UGC 7899 has lower values of both parameters. The specific logarithmic $q$ values for our galaxy sample are presented in Table \ref{tab:SAM}.

The dashed grey line in Figure \ref{fig:qvsfatm} represents the theoretical limit for global stability ($q = 1/(\sqrt{2}\,e)$). Three of our sample galaxies ESO208-G026, ESO378-G003, and ESO417-G018 have $q > 1/(\sqrt{2}\,e)$, indicating an entirely stable \hi\ disc against gravitational collapse \citep{Obreschkow2016}. This parameter space of high $q$ and high $f_\text{atm}$ is typical for gas rich dwarf galaxies. ESO381-G005 and ESO055-G013 exhibit lower $q$ values closer to or below this critical threshold, indicating that their \hi\ discs are not completely stable. Importantly, the $q$ parameter specifically describes the \hi\ disc stability \citep{Obreschkow2016}, not the molecular gas stability. The compact molecular gas discs, which are confined to the inner stellar regions, have already formed from \hi\ gas that successfully overcame the high angular momentum barrier, likely in regions where local instabilities, spiral density waves, or external perturbations (as tidal interactions, see Section 4.1) facilitated \hi-to-\Htwo\ conversion. 

\section{Discussion}
\label{sec:Discussion}

When comparing our sample galaxies to the xCOLD GASS sample we clearly find that their molecular gas mass is within the expected range, however the velocity dispersion for the \hix\ sample is clearly relatively high compared to xCOLD GASS. Based on the global stability factor, our main finding is the stability of the \hi\ gas disc against fragmentation or collapse. This stability is linked to a high specific angular momentum and also high velocity dispersion compared to other galaxies.

The fact that three out of the five sample galaxies show \hi\ extension and tails might be evidence of the external origin of the excess \hi, whose angular momentum differs from the rest of the disc components. Alternatively, the high angular momentum and the high-velocity dispersion, may also be due to gravitational interactions with nearby galaxies (see \citealt{Besla2007, Karanchentsev1999}). Because of this, we explore the possibility of ongoing tidal interactions in the next section.

\subsection{Tidal interactions}
We investigate the role of tidal interactions as a potential explanation for the observed \hi\ excess in the galaxies of our sample. Specifically, we calculate the tidal index for ESO378-G003, ESO381-G005, ESO055-G013, and ESO417-G018 which have neighbouring galaxies that could potentially tidally interact with them.

Due to the magnitude limits of large-scale environmental catalogues (such as the 2MRS group catalogues; e.g., \citealt{Lambert_2020, Lambert_2023}), pre-calculated macro-environmental densities are not available for our sample. Therefore, to quantify the local environment and interaction strength, we calculate the tidal index $\Theta$ (e.g. \citealt{Karanchentsev1999, Holwerda2011}). \citet{Karanchentsev1999} consider a tidal index to measure and rank the gravitational environment of a galaxy, and its interaction strength with their most significant neighbours. For a galaxy $i$, the tidal force magnitude $F_n$ by a neighbour $n$ is approximated as $F_n \propto M_n/D_{in}^3$, where $D_{in}$ is the distance between the galaxy and its disturber, and $M_n$ is the mass of the disturber proportional to its luminosity. This allows us to define the tidal index $\Theta$ as
\begin{equation}
    \Theta = \max_n\left\{\log\left(\frac{M_n}{D_{in}^3}\right)\right\}+C, \quad n = 1,2,3,\dots, N,\label{eq:tidal}
\end{equation}
where $n$ represents each of the $N$ neighbouring galaxies, and the constant $C=-11.75$ is chosen so that at $\Theta=0$ the galaxy $i$ is located on the zero velocity sphere relative to the main disturber. $\Theta>0$ means gravitational influence or member of a physical group. On the other hand, $\Theta<0$ means minimal current gravitational influence. The $\max$ function helps to determine the main disturber (MD) of the target galaxy.

We obtained the angular separation and the redshift from the NASA/IPAC Extragalactic Database \footnote{\url{https://ned.ipac.caltech.edu}} for all galaxies within one degree of angular separation and $\pm 500$ \kms of systemic velocity separation of our sample galaxies. From these, we define the potential disturbers as the most massive neighbour within the closest projected distance. We found potential disturbers near ESO055-G013, ESO378-G003, ESO381-G005 and ESO417-G018.
The nearest neighbours to ESO378-G003 are NGC 3706 and 2MASX J11262478-3615344. The nearest neighbour to ESO381-G005 is 2MASS J12402306-3702272 (PGC 629239). The potential disturbers for ESO055-G013 and ESO417-G018 are PGC 14802 and PGC 011830 (ESO417-G019), respectively. We computed the stellar masses of the neighbours using the $K_s$ (\SI{2.159}{\micro\metre}) band magnitude from the Two Micron All Sky Survey (2MASS) Point Source Catalog (PSC) and Extended Source Catalog (XSC) (see \citealt{Skrutskie2006}). To ensure consistency, we converted the $K_s(\mathrm{Vega})$ magnitudes to the AB system following \citet{Finlator2000} and derived the corresponding luminosities ($L_\nu(K_s)$) using the luminosity distances listed in Table \ref{tab:generalprop}. Finally, we calculated the stellar masses using the mass-to-light ratio relation provided by \citet{Wen2013}:
\begin{equation}
        \log\left(\frac{M_\star}{\mathrm{M}_\odot}\right)= (-0.498\pm 0.002) + (1.105\pm0.001)\log\left(\frac{\nu L_\nu(K_s)}{\mathrm{L}_\odot}\right).
\end{equation}

For these calculations, the distance to the main disturber, $D_{in} $ in \eqref{eq:tidal}, is determined using the projected distance. This is calculated from the observed angular separation between the two galaxies, assuming both lie at the same distance from Earth (using the average distance of the pair), to avoid the bias introduced by peculiar velocities if two galaxies are physically bound and interacting.

\begin{table*}
    \centering
    \begin{tabular}{llccc}
        \hline
        Galaxy & Main Disturber & \multirow{2}{*}{$\log\left[\dfrac{M_\star}{\mathrm{M}_\odot}\right]$} & $D_{in}$ &\multirow{2}{*}{$\Theta$}\\
        ID & ID & & (Mpc)& \\
        (1) & (2) & (3) & (4) & (5)\\
        \hline
        ESO378-G003 & NGC 3706 & $11.28$ & $0.268$ & $1.251$\\
        NGC 3706 & ESO378-G003 & $10.07$ & $0.268$ & $0.035$\\
        ESO381-G005 & PGC 629239 & $8.34$ & $0.112$ & $-0.552$\\
        PGC 629239 & ESO381-G005 & $10.07$ & $0.112$ & $1.172$\\
        ESO055-G013 & PGC 14802 & $10.89$ & $0.946$ & $-0.791$\\
        PGC 14802 & ESO055-G013 & $10.19$ & $0.946$ & $-1.489$\\
        ESO417-G018 & ESO417-G019 & $9.86$ & $0.757$ & $-1.526$\\
        ESO417-G019 & ESO417-G018 & $10.28$ & $0.757$ & $-1.109$\\
        \hline
    \end{tabular}
    \caption{Tidal index table. (1) ID of the target galaxy. (2) ID of the main disturber. (3) Log of the stellar mass of the main disturber. (4) Distance between galaxies using the angular separation. (5) Tidal index.}
    \label{tab:tidalin}
\end{table*}

Table \ref{tab:tidalin} shows the calculated stellar masses for the neighbouring galaxies and the calculated tidal indices. The tidal indices between ESO378-G003 and NGC 3706 indicate that while ESO378-G003 is strongly influenced by NGC 3706, the latter does not experience a strong tidal force from ESO378-G003. In other words, NGC 3706 is the dominant galaxy in this pair. This directional influence can alter the internal gas dynamics of ESO378-G003, leading to less efficient star formation and thus to a build-up of \hi\ gas.

In contrast, the tidal indices between ESO381-G005 and PGC 629239 indicate that ESO381-G005 is the dominant galaxy and the other is a gravitationally bound satellite. Figure~\ref{fig:381_HI_map} shows a map of the \hi\ distribution of the ESO381-G005 and PGC 629239 system. PGC 629239 shows an \hi\ extension in the direction of ESO381-G005. This suggests that ESO381-G005 might be accreting gas from PGC 629239 contributing to its \hi\ excess. As demonstrated by \citet{Bekki2008} with hydrodynamical simulations, the orbital angular momentum from gas-rich mergers or interactions can be efficiently transferred to the surrounding gas halo. Rather than collapsing towards the centre to drive a burst of star formation, the gas gains enough rotation to form an extended, stable \hi\ disc with elevated specific angular momentum similar to the kinematic properties we observe in the \hix\ sample.

For isolated systems such as ESO208-G026 and ESO417-G018 where the index indicates minimal current gravitational influence ($\Theta < 0$), their globally stable \hi\ discs ($q > 1/(\sqrt{2}\,e)$) are likely maintained by the intrinsically high spin of their dark matter halos \citep{Lutz2018}. This scenario is supported by high-resolution observations of similar isolated, \hi-rich galaxies (e.g., \citealt{Hallenbeck2014}), which demonstrate that halos with elevated spin parameters prevent massive atomic gas reservoirs from losing the angular momentum required to trigger local collapse. Furthermore, continuous cold gas accretion from cosmic filaments could be replenishing these stable discs (e.g. \citealt{Kleiner2017, Healy2024, Radburn-Smith_2014}).

In the case of ESO055-G013, which also lacks significant tidal influences, as $\Theta < 0$, this galaxy exhibits an atypically low stellar specific angular momentum whereas its baryonic specific angular momentum remains elevated, indicating that the angular momentum of the system is locked within the extended \hi\ disc. Consequently, even though the disc is not strictly globally stable, this massive angular momentum reservoir prevents the atomic gas from flowing inward to feed the compact molecular core.

\subsection{Star formation rates and gas depletion times}

External processes such as accretion or tidal interactions can supply galaxies with angular momentum and turbulence that suppress the conversion from atomic to molecular gas needed for star formation. A way to assess the star formation potential of a galaxy is through the gas depletion times.

\begin{figure*}
    \centering
    \includegraphics[width=0.42\linewidth]{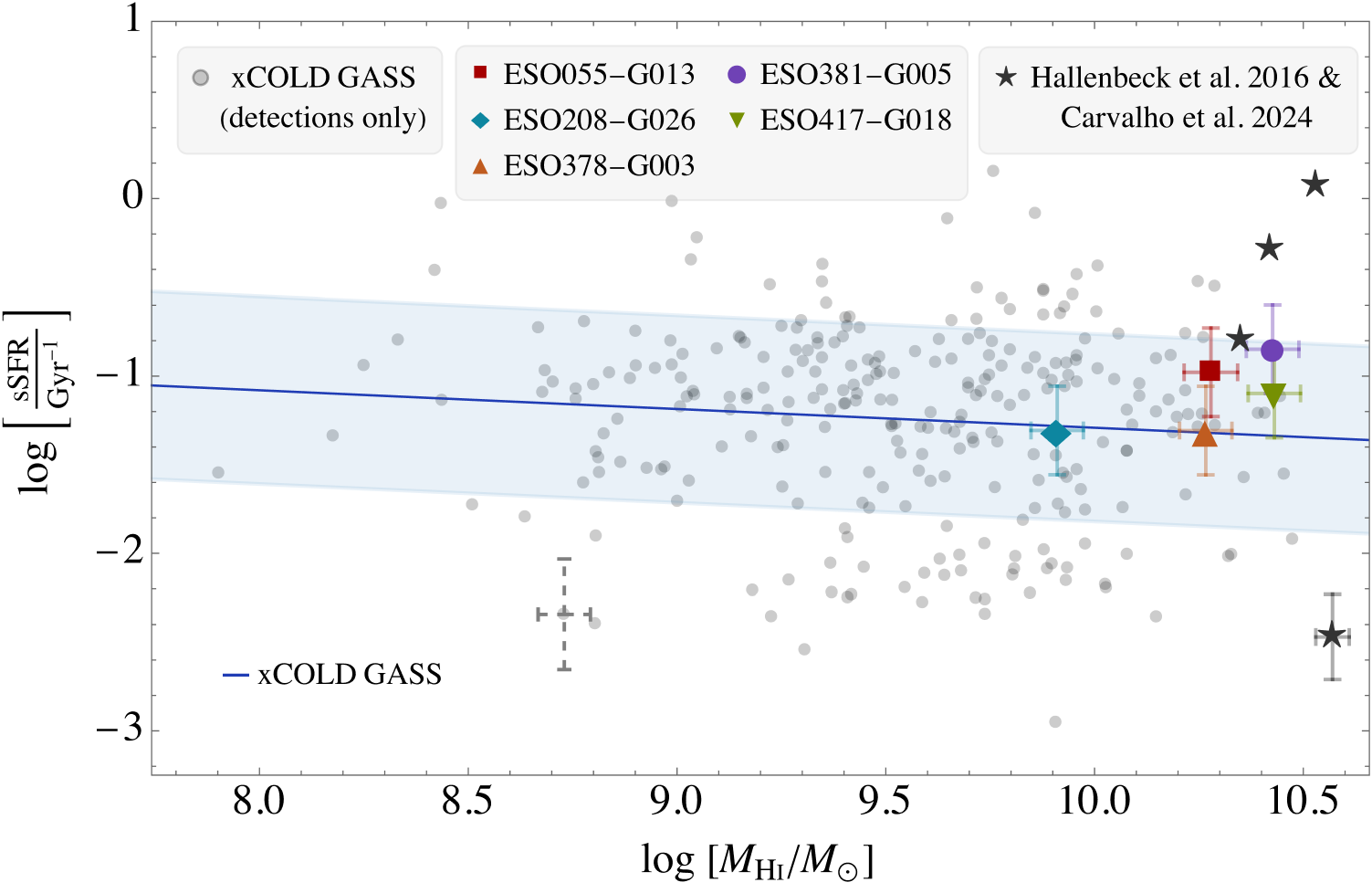}
    \includegraphics[width=0.42\linewidth]{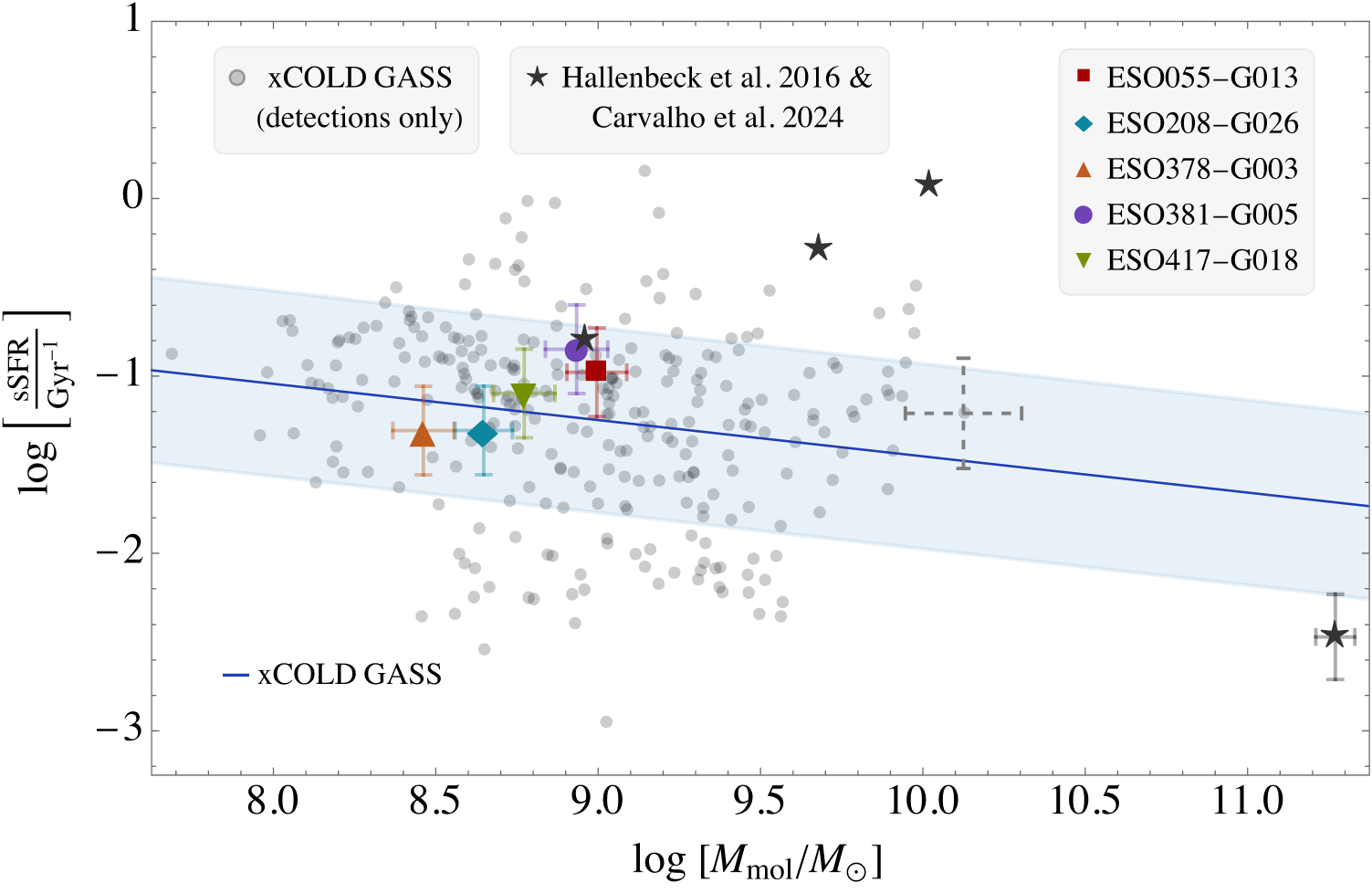}
    \includegraphics[width=0.42\linewidth]{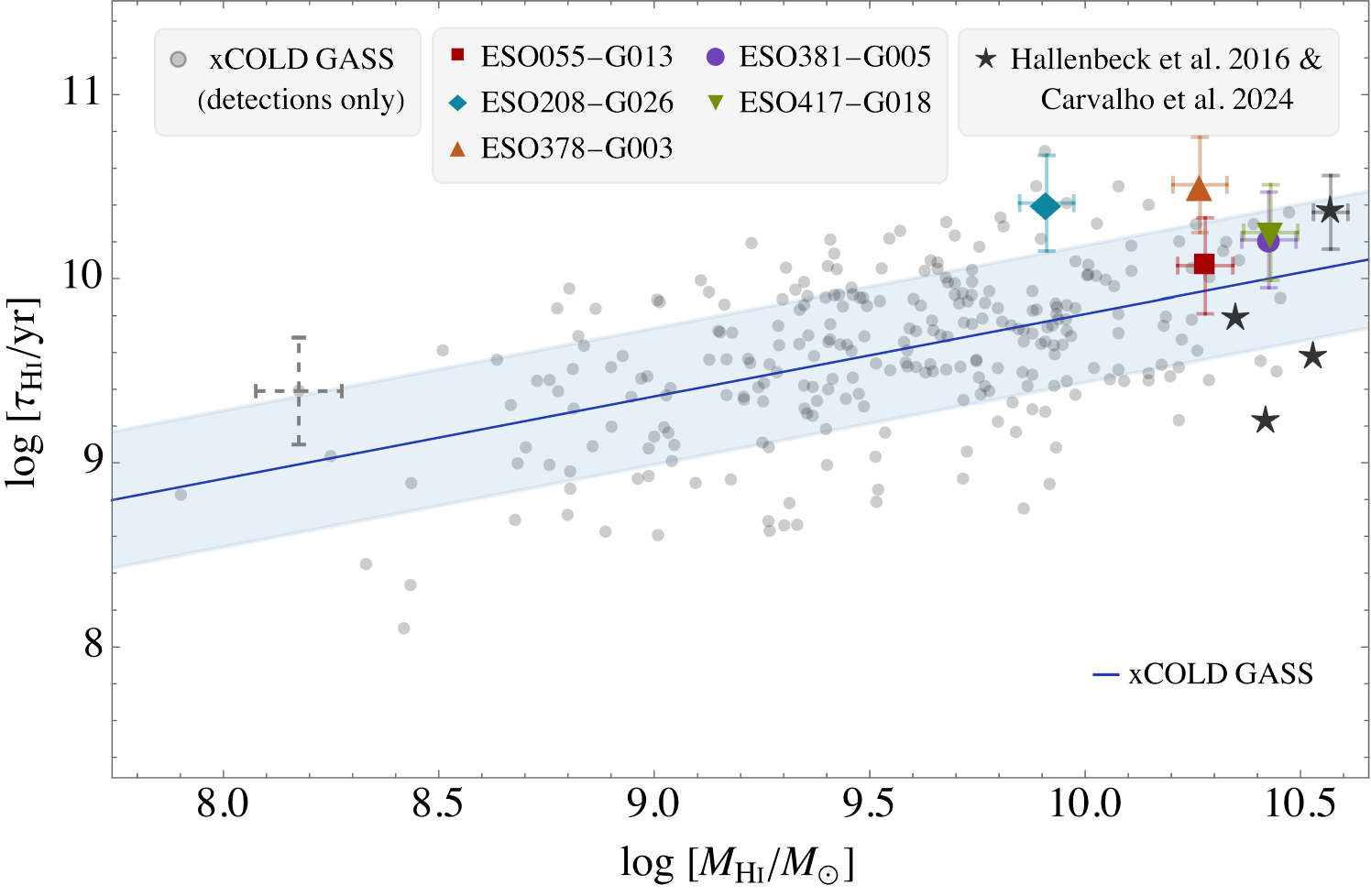}
    \includegraphics[width=0.42\linewidth]{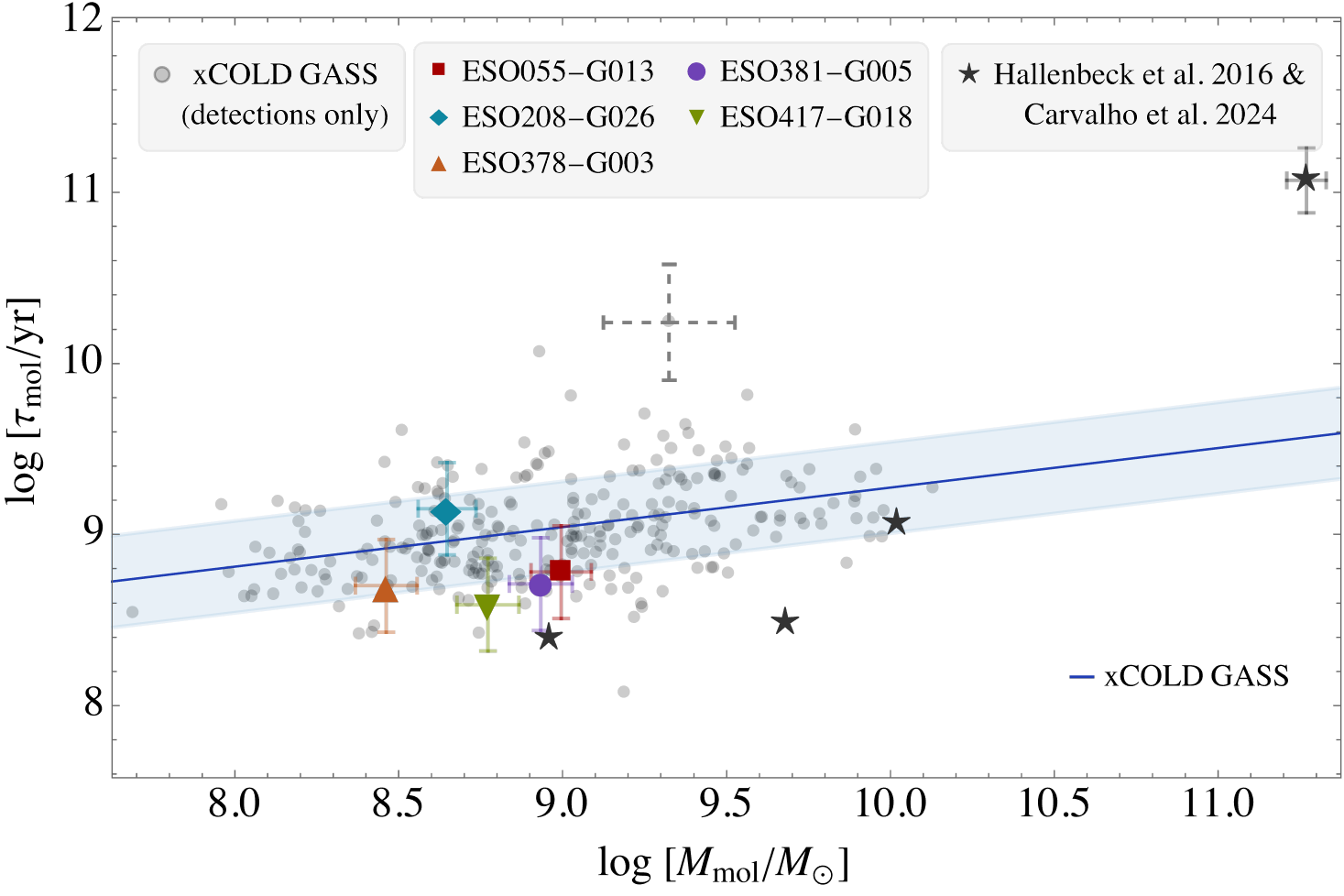}
    \caption{Star formation properties of our sample compared with xCOLD GASS. The black stars represent UGC 6168, UGC 7899, NGC 5230, and the black star with error bars represents UGC 2885. Top left: Specific star formation rate against \hi\ mass. Top right: Specific star formation rate against \Htwo\ mass. Bottom left: \hi\ depletion time against \hi\ mass. Bottom right: \Htwo\ depletion time against \Htwo\ mass. The solid blue lines represent the linear regression (best fit) for xCOLD GASS data. The shaded region indicates 1$\sigma$ interval.}
    \label{fig:SFRplots}
\end{figure*}

Depletion times, which indicate how long the current gas reservoir would last at the present SFR, are obtained using \hi\ and \Htwo\ masses presented earlier. Table \ref{tab:SFR} summarises the calculated depletion times based on the \hi\ and molecular gas masses. For comparison we have also included the values for UGC 6168, UGC 7899, NGC 5230 from \citet{Hallenbeck2016} and the values for UGC 2885 from \citet{Carvalho2024}. To visualise the divergence of the \hi-excess galaxies from the standard population, Figure \ref{fig:SFRplots} compares the specific star formation rates (sSFR) and gas depletion times against the xCOLD GASS scaling relations. The plots reveal a striking contrast: while some \hix\ galaxies appear as distinct outliers in relation to their \hi\ content (left panels), exhibiting very long depletion times, they fall back within the normal scatter of typical star-forming galaxies when analysed relative to their molecular gas mass (right panels). This reinforces our earlier findings that the discrepancy between \hi\ content and SFR is due to the inefficient conversion of \hi\ to molecular gas.

\begin{table*}
    \centering
    \begin{tabular}{lcccc}
    \hline
    ID  & $\log\left[\dfrac{\mathrm{SFR}}{\mathrm{M}_\odot\, \mathrm{yr^{-1}}}\right]$ & $\log\left[\dfrac{\mathrm{sSFR}}{\mathrm{Gyr^{-1}}}\right]$ & $\log \left[\dfrac{\tau_{\mathrm{mol}}}{\mathrm{yr}}\right]$ & $\log \left[\dfrac{\tau_{\text{\hi}}}{\mathrm{yr}}\right]$\\
    (1) & (2) & (3) & (4) & (5)\\
    \hline
    ESO055-G013 & $0.21 \pm 0.25$ & $-0.98 \pm 0.29$ & $8.78 \pm 0.27$ & $10.07 \pm 0.26$\\
    ESO208-G026 & $-0.50 \pm 0.25$ & $-1.31 \pm 0.29$ & $9.15 \pm 0.27$ & $10.41 \pm 0.26$\\
    ESO378-G003 & $-0.24 \pm 0.25$ & $-1.31 \pm 0.29$ & $8.70 \pm 0.27$ & $10.51 \pm 0.26$\\
    ESO381-G005 & $0.22 \pm 0.25$ & $-0.85 \pm 0.29$ & $8.71 \pm 0.27$ & $10.21 \pm 0.26$\\
    ESO417-G018 & $0.18 \pm 0.25$ & $-1.10 \pm 0.29$ & $8.59 \pm 0.27$ & $10.25 \pm 0.26$\\
    \hline
    UGC 6168 & $0.57$ & $-0.80$ & $8.39$ & $9.78$\\
    UGC 7899 & $1.20$ & $-0.29$ & $8.48$ & $9.22$\\
    NGC 5230 & $0.96$ & $0.07$ & $9.06$ & $9.57$\\
    UGC 2885 & $0.21 \pm 0.19$ & $-2.47 \pm 0.24$ & $11.07 \pm 0.19$ & $10.36 \pm 0.20$\\
    \hline
    \end{tabular}
    \caption{Star formation properties and depletion times for our sample and the literature comparison sample. (1) Galaxy ID. (2) Star formation rate from \citealt{Lutz2017} (for the main sample) and from the literature \citep{Hallenbeck2016, Carvalho2024}. (3) Specific star formation rate. (4) Molecular gas depletion time. (5) Atomic gas depletion time. Uncertainties for UGC 6168, UGC 7899, and NGC 5230 are absent as they are not reported in the source literature.}
    \label{tab:SFR}
\end{table*}

We find that the specific star formation rates are all in the normal range for \hix\ galaxies. However, the gas depletion times differ from the expected values. ESO208-G026 and ESO378-G003 have very long \hi\ gas depletion times, well above the 1$\sigma$ range of regular galaxies, providing a gas supply for the next Hubble time. These extreme depletion times are a hallmark of such massive, \hi-rich systems; for comparison, the giant spiral UGC 2885 exhibits an even more extreme molecular gas depletion time of $\sim 1.15 \times 10^{11}$\,yr and an \hi\ depletion time of $\sim 2.29 \times 10^{10}$\,yr \citep{Carvalho2024}. On the other hand, the molecular gas depletion times are all short, except for ESO208-G026. ESO378-G003 and ESO417-G018 have the shortest molecular gas depletion times below the 1$\sigma$ range. This highlights that ESO208-G026 is one of the most extreme cases of the \hi\ rich galaxies in terms of the high \hi\ content, low molecular gas content, regular star formation rate and the highest disc stability parameter.  

\section{Conclusions}
\label{sec:Conclusions}

The amount of molecular gas in this sample falls within the expected values compared to their stellar masses, in contrast to their large \hi\ reservoirs. 

Taking into account the mode or peak of their distributions (to avoid the distortion generated by very high values), the velocity dispersion in the \co\ disc, except for that of ESO055-G013, as well as in the \hi\ disc, is above the thermal value ($\sim 8$\,\kms), which is a sign of high turbulence. The ratio between the \hi\ and the \co\ dispersion is higher than expected, which indicates elevated dispersion and turbulence in the \hi\ gas disc compared to the molecular gas disc. 

The specific angular momentum of the stellar disc and the molecular disc typically have a ratio of around unity. Meanwhile, the ratio of the specific angular momentum of the stellar disc and the atomic gas is about 11, and between the molecular gas and the atomic gas is around 10. This is remarkably higher than the typical factor of 2--4 observed in normal spiral galaxies where the stellar and gaseous discs are dynamically coupled (e.g., \citealt{Obreschkow2014}). Such high ratios are rare but consistent with results found in other \hi-rich, high-spin galaxy populations (e.g. \citealt{ManceraPina2021}), reinforcing the scenario where the large \hi\ reservoir in \hix\ galaxies stores a large fraction of the angular momentum. 

The global stability factor $q$, analogous to the local Toomre parameter, of ESO208-G026, ESO378-G003 and ESO417-G018 is above the limit of $q\gtrsim 1/(\sqrt{2}\,e)$ indicating a globally stable disc, while ESO055-G013 and ESO381-G005 are below this limit for stability. 

The primary bottleneck in star formation of \hix\ galaxies is clearly the \hi-to-\Htwo\ conversion step. \hi\ depletion times are very long, in fact larger than the age of the universe, whereas \Htwo\ depletion times are near-normal, even low, indicating that molecular gas, once formed, is efficiently converted into stars. The likely cause for the \hi\ excess and the high angular momentum of the \hi\ disc might be past gas accretion or recent tidal interactions. Our analysis indicates that for ESO208-G026, ESO378-G003 and ESO417-G018 the low global star formation efficiency in \hix\ galaxies is not driven by an inability to form stars from molecular gas, but rather by the inefficiency of the production of molecular gas from the stable, turbulent \hi\ reservoir. Especially in the case of ESO378-G003, this stability against gravitational collapse may be caused by tidal interactions. Tidal interactions and recent gas accretion may explain the apparent excess of \hi\ gas in ESO381-G005. 

Comparing the \hix\ galaxies to other extremely \hi-rich galaxies from the literature implies that the same processes are likely responsible for the gas richness of UGC 6168, while NGC 5230 and UGC 7899 may have less extreme \hi-to-\Htwo\ conversion bottlenecks considering their more regular molecular gas fractions and the lower disc stability parameter of UGC 7899. UGC 2885 is the least similar galaxy to the \hix\ sample; its very high molecular gas fraction indicates that the gas content of this galaxy is likely controlled by different physical processes.

\hix\ galaxies represent a key phase in galaxy evolution, where environmental effects are involved. In these galaxies the excess \hi\ gas has not had time to condense into molecular gas yet. These systems are useful to study the effect of high-angular momentum discs and the long-term processes of gas accretion and tidal interactions on galaxy growth. Since this sample of nine \hi-rich galaxies is relatively small, further investigation is needed to establish a clear picture on the \hi\ excess in galaxies and the role of galaxy interactions in this process. An expansion of the \hix\ sample will be possible with the new high-resolution \hi\ surveys such as the Apertif imaging surveys \citep{Adams2022} and the WALLABY survey (e.g. \citealt{Koribalski2020, Westmeier2022}).


\section*{Acknowledgements}

We would like to thank B.W. Holwerda, the referee, for the useful comments and suggestions, which helped us to greatly improve the paper. This paper makes use of the following ALMA data: ADS/JAO.ALMA\#2019.2.00169.S. ALMA is a partnership of ESO (representing its member states), NSF (USA) and NINS (Japan), together with NRC (Canada), MOST and ASIAA (Taiwan), and KASI (Republic of Korea), in cooperation with the Republic of Chile. The Joint ALMA Observatory is operated by ESO, AUI/NRAO and NAOJ. In addition, publications from NA authors must include the standard NRAO acknowledgement: The National Radio Astronomy Observatory is a facility of the National Science Foundation operated under cooperative agreement by Associated Universities, Inc.

The Australia Telescope Compact Array is part of the Australia Telescope National Facility (grid.421683.a) which is funded by the Australian Government for operation as a National Facility managed by CSIRO. We acknowledge the Gomeroi people as the traditional owners of the Observatory site.

The Legacy Surveys consist of three individual and complementary projects: the Dark Energy Camera Legacy Survey (DECaLS; Proposal ID \#2014B-0404; PIs: David Schlegel and Arjun Dey), the Beijing-Arizona Sky Survey (BASS; NOAO Prop. ID \#2015A-0801; PIs: Zhou Xu and Xiaohui Fan), and the Mayall z-band Legacy Survey (MzLS; Prop. ID \#2016A-0453; PI: Arjun Dey). DECaLS, BASS and MzLS together include data obtained, respectively, at the Blanco telescope, Cerro Tololo Inter-American Observatory, NSF’s NOIRLab; the Bok telescope, Steward Observatory, University of Arizona; and the Mayall telescope, Kitt Peak National Observatory, NOIRLab. Pipeline processing and analyses of the data were supported by NOIRLab and the Lawrence Berkeley National Laboratory (LBNL). The Legacy Surveys project is honored to be permitted to conduct astronomical research on Iolkam Du’ag (Kitt Peak), a mountain with particular significance to the Tohono O’odham Nation.

NOIRLab is operated by the Association of Universities for Research in Astronomy (AURA) under a cooperative agreement with the National Science Foundation. LBNL is managed by the Regents of the University of California under contract to the U.S. Department of Energy.

This project used data obtained with the Dark Energy Camera (DECam), which was constructed by the Dark Energy Survey (DES) collaboration. Funding for the DES Projects has been provided by the U.S. Department of Energy, the U.S. National Science Foundation, the Ministry of Science and Education of Spain, the Science and Technology Facilities Council of the United Kingdom, the Higher Education Funding Council for England, the National Center for Supercomputing Applications at the University of Illinois at Urbana-Champaign, the Kavli Institute of Cosmological Physics at the University of Chicago, Center for Cosmology and Astro-Particle Physics at the Ohio State University, the Mitchell Institute for Fundamental Physics and Astronomy at Texas A\&M University, Financiadora de Estudos e Projetos, Fundacao Carlos Chagas Filho de Amparo, Financiadora de Estudos e Projetos, Fundacao Carlos Chagas Filho de Amparo a Pesquisa do Estado do Rio de Janeiro, Conselho Nacional de Desenvolvimento Cientifico e Tecnologico and the Ministerio da Ciencia, Tecnologia e Inovacao, the Deutsche Forschungsgemeinschaft and the Collaborating Institutions in the Dark Energy Survey. The Collaborating Institutions are Argonne National Laboratory, the University of California at Santa Cruz, the University of Cambridge, Centro de Investigaciones Energeticas, Medioambientales y Tecnologicas-Madrid, the University of Chicago, University College London, the DES-Brazil Consortium, the University of Edinburgh, the Eidgenossische Technische Hochschule (ETH) Zurich, Fermi National Accelerator Laboratory, the University of Illinois at Urbana-Champaign, the Institut de Ciencies de l’Espai (IEEC/CSIC), the Institut de Fisica d’Altes Energies, Lawrence Berkeley National Laboratory, the Ludwig Maximilians Universitat Munchen and the associated Excellence Cluster Universe, the University of Michigan, NSF’s NOIRLab, the University of Nottingham, the Ohio State University, the University of Pennsylvania, the University of Portsmouth, SLAC National Accelerator Laboratory, Stanford University, the University of Sussex, and Texas A\&M University. BASS is a key project of the Telescope Access Program (TAP), which has been funded by the National Astronomical Observatories of China, the Chinese Academy of Sciences (the Strategic Priority Research Program “The Emergence of Cosmological Structures” Grant \# XDB09000000), and the Special Fund for Astronomy from the Ministry of Finance. The BASS is also supported by the External Cooperation Program of Chinese Academy of Sciences (Grant \# 114A11KYSB20160057), and Chinese National Natural Science Foundation (Grant \# 12120101003, \# 11433005). The Legacy Survey team makes use of data products from the Near-Earth Object Wide-field Infrared Survey Explorer (NEOWISE), which is a project of the Jet Propulsion Laboratory/California Institute of Technology. NEOWISE is funded by the National Aeronautics and Space Administration. The Legacy Surveys imaging of the DESI footprint is supported by the Director, Office of Science, Office of High Energy Physics of the U.S. Department of Energy under Contract No. DE-AC02-05CH1123, by the National Energy Research Scientific Computing Center, a DOE Office of Science User Facility under the same contract; and by the U.S. National Science Foundation, Division of Astronomical Sciences under Contract No. AST-0950945 to NOAO.

This research has made use of {\sc NumPy} \citep{vanderWalt2011}, {\sc scipy} \citep{2020SciPy}, {\sc Astropy}, a community-developed core {\sc Python} package for Astronomy (\citealt{astropy:2013, astropy:2018}; http://www.astropy.org), and astroquery \citep{astroquery}.

\section*{Data Availability} 

All \hix\ \hi\ data cubes and tabular information from \citet{Lutz2017} and \citet{Lutz2018} are available from the online CDS VizieR data base (\url{https://vizier.cds.unistra.fr/viz-bin/VizieR}). The CO(1-0) data presented in this paper are available from the ALMA Science Archive (\url{https://almascience.nrao.edu/aq}).


\bibliographystyle{mnras}
\bibliography{hix4} 

@ARTICLE{Lutz2017,
       author = {{Lutz}, K.~A. and {Kilborn}, V.~A. and {Catinella}, B. and {Koribalski}, B.~S. and {Brown}, T.~H. and {Cortese}, L. and {D{\'e}nes}, H. and {J{\'o}zsa}, G.~I.~G. and {Wong}, O.~I.},
        title = "{The HIX galaxy survey I: Study of the most gas rich galaxies from HIPASS}",
      journal = {\mnras},
         year = 2017,
        month = {may},
       volume = {467},
       number = {1},
        pages = {1083-1097},
          doi = {10.1093/mnras/stx053},
archivePrefix = {arXiv},
       eprint = {1701.02447},
 primaryClass = {astro-ph.GA},
       adsurl = {https://ui.adsabs.harvard.edu/abs/2017MNRAS.467.1083L}
}

@ARTICLE{Lutz2018,
       author = {{Lutz}, K.~A. and {Kilborn}, V.~A. and {Koribalski}, B.~S. and {Catinella}, B. and {J{\'o}zsa}, G.~I.~G. and {Wong}, O.~I. and {Stevens}, A.~R.~H. and {Obreschkow}, D. and {D{\'e}nes}, H.},
        title = "{The H IX galaxy survey - II. H I kinematics of H I eXtreme galaxies}",
      journal = {\mnras},
         year = 2018,
        month = may,
       volume = {476},
       number = {3},
        pages = {3744-3780},
          doi = {10.1093/mnras/sty387},
archivePrefix = {arXiv},
       eprint = {1802.04043},
 primaryClass = {astro-ph.GA},
       adsurl = {https://ui.adsabs.harvard.edu/abs/2018MNRAS.476.3744L}
}

@ARTICLE{Lutz2020,
       author = {{Lutz}, K.~A. and {Kilborn}, V. and {Catinella}, B. and {Cortese}, L. and {Brown}, T.~H. and {Koribalski}, B.},
        title = "{The H IX galaxy survey. III. The gas-phase metallicity in H I eXtreme galaxies}",
      journal = {\aap},
         year = 2020,
        month = mar,
       volume = {635},
          eid = {A69},
        pages = {A69},
          doi = {10.1051/0004-6361/201936556},
archivePrefix = {arXiv},
       eprint = {2001.07516},
 primaryClass = {astro-ph.GA},
       adsurl = {https://ui.adsabs.harvard.edu/abs/2020A&A...635A..69L}
}

@ARTICLE{Denes2014,
       author = {{D{\'e}nes}, H. and {Kilborn}, V.~A. and {Koribalski}, B.~S.},
        title = "{New H I scaling relations to probe the H I content of galaxies via global H I-deficiency maps}",
      journal = {\mnras},
         year = 2014,
        month = oct,
       volume = {444},
       number = {1},
        pages = {667-681},
          doi = {10.1093/mnras/stu1337},
archivePrefix = {arXiv},
       eprint = {1407.1619},
 primaryClass = {astro-ph.GA},
       adsurl = {https://ui.adsabs.harvard.edu/abs/2014MNRAS.444..667D}
}

@ARTICLE{DiTeodoro2015,
       author = {{Di Teodoro}, E.~M. and {Fraternali}, F.},
        title = "{$^{3D}$ BAROLO: a new 3D algorithm to derive rotation curves of galaxies}",
      journal = {\mnras},
         year = 2015,
        month = aug,
       volume = {451},
       number = {3},
        pages = {3021-3033},
          doi = {10.1093/mnras/stv1213},
archivePrefix = {arXiv},
       eprint = {1505.07834},
 primaryClass = {astro-ph.GA},
       adsurl = {https://ui.adsabs.harvard.edu/abs/2015MNRAS.451.3021D}
}

@ARTICLE{astroquery,
       author = {{Ginsburg}, Adam and {Sip{\H{o}}cz}, Brigitta M. and {Brasseur}, C.~E. and {Cowperthwaite}, Philip S. and {Craig}, Matthew W. and {Deil}, Christoph and {Guillochon}, James and {Guzman}, Giannina and {Liedtke}, Simon and {Lian Lim}, Pey and {Lockhart}, Kelly E. and {Mommert}, Michael and {Morris}, Brett M. and {Norman}, Henrik and {Parikh}, Madhura and {Persson}, Magnus V. and {Robitaille}, Thomas P. and {Segovia}, Juan-Carlos and {Singer}, Leo P. and {Tollerud}, Erik J. and {de Val-Borro}, Miguel and {Valtchanov}, Ivan and {Woillez}, Julien and {Astroquery Collaboration} and {a subset of astropy Collaboration}},
        title = "{astroquery: An Astronomical Web-querying Package in Python}",
      journal = {\aj},
         year = 2019,
        month = mar,
       volume = {157},
       number = {3},
          eid = {98},
        pages = {98},
          doi = {10.3847/1538-3881/aafc33},
archivePrefix = {arXiv},
       eprint = {1901.04520},
 primaryClass = {astro-ph.IM},
       adsurl = {https://ui.adsabs.harvard.edu/abs/2019AJ....157...98G}
}

@ARTICLE{2020SciPy,
  author  = {Virtanen, Pauli and Gommers, Ralf and Oliphant, Travis E. and
            Haberland, Matt and Reddy, Tyler and Cournapeau, David and
            Burovski, Evgeni and Peterson, Pearu and Weckesser, Warren and
            Bright, Jonathan and {van der Walt}, St{\'e}fan J. and
            Brett, Matthew and Wilson, Joshua and Millman, K. Jarrod and
            Mayorov, Nikolay and Nelson, Andrew R. J. and Jones, Eric and
            Kern, Robert and Larson, Eric and Carey, C J and
            Polat, {\.I}lhan and Feng, Yu and Moore, Eric W. and
            {VanderPlas}, Jake and Laxalde, Denis and Perktold, Josef and
            Cimrman, Robert and Henriksen, Ian and Quintero, E. A. and
            Harris, Charles R. and Archibald, Anne M. and
            Ribeiro, Ant{\^o}nio H. and Pedregosa, Fabian and
            {van Mulbregt}, Paul and {SciPy 1.0 Contributors}},
  title   = {{{SciPy} 1.0: Fundamental Algorithms for Scientific
            Computing in Python}},
  journal = {Nature Methods},
  year    = {2020},
  volume  = {17},
  pages   = {261--272},
  adsurl  = {https://rdcu.be/b08Wh},
  doi     = {10.1038/s41592-019-0686-2},
}

@ARTICLE{vanderWalt2011,
       author = {{Van Der Walt}, St{\'e}fan and {Colbert}, S. Chris and {Varoquaux}, Ga{\"e}l},
        title = "{The NumPy Array: A Structure for Efficient Numerical Computation}",
      journal = {Computing in Science and Engineering},
         year = 2011,
        month = mar,
       volume = {13},
       number = {2},
        pages = {22-30},
          doi = {10.1109/MCSE.2011.37},
archivePrefix = {arXiv},
       eprint = {1102.1523},
 primaryClass = {cs.MS},
       adsurl = {https://ui.adsabs.harvard.edu/abs/2011CSE....13b..22V}
}

@article{astropy:2013,
Adsurl = {http://adsabs.harvard.edu/abs/2013A%26A...558A..33A},
Archiveprefix = {arXiv},
Author = {{Astropy Collaboration} and {Robitaille}, T.~P. and {Tollerud}, E.~J. and {Greenfield}, P. and {Droettboom}, M. and {Bray}, E. and {Aldcroft}, T. and {Davis}, M. and {Ginsburg}, A. and {Price-Whelan}, A.~M. and {Kerzendorf}, W.~E. and {Conley}, A. and {Crighton}, N. and {Barbary}, K. and {Muna}, D. and {Ferguson}, H. and {Grollier}, F. and {Parikh}, M.~M. and {Nair}, P.~H. and {Unther}, H.~M. and {Deil}, C. and {Woillez}, J. and {Conseil}, S. and {Kramer}, R. and {Turner}, J.~E.~H. and {Singer}, L. and {Fox}, R. and {Weaver}, B.~A. and {Zabalza}, V. and {Edwards}, Z.~I. and {Azalee Bostroem}, K. and {Burke}, D.~J. and {Casey}, A.~R. and {Crawford}, S.~M. and {Dencheva}, N. and {Ely}, J. and {Jenness}, T. and {Labrie}, K. and {Lim}, P.~L. and {Pierfederici}, F. and {Pontzen}, A. and {Ptak}, A. and {Refsdal}, B. and {Servillat}, M. and {Streicher}, O.},
Doi = {10.1051/0004-6361/201322068},
Eid = {A33},
Eprint = {1307.6212},
Journal = {\aap},
Month = oct,
Pages = {A33},
Primaryclass = {astro-ph.IM},
Title = {{Astropy: A community Python package for astronomy}},
Volume = 558,
Year = 2013}

@ARTICLE{astropy:2018,
       author = {{Astropy Collaboration} and {Price-Whelan}, A.~M. and
         {Sip{\H{o}}cz}, B.~M. and {G{\"u}nther}, H.~M. and {Lim}, P.~L. and
         {Crawford}, S.~M. and {Conseil}, S. and {Shupe}, D.~L. and
         {Craig}, M.~W. and {Dencheva}, N. and {Ginsburg}, A. and {Vand
        erPlas}, J.~T. and {Bradley}, L.~D. and {P{\'e}rez-Su{\'a}rez}, D. and
         {de Val-Borro}, M. and {Aldcroft}, T.~L. and {Cruz}, K.~L. and
         {Robitaille}, T.~P. and {Tollerud}, E.~J. and {Ardelean}, C. and
         {Babej}, T. and {Bach}, Y.~P. and {Bachetti}, M. and {Bakanov}, A.~V. and
         {Bamford}, S.~P. and {Barentsen}, G. and {Barmby}, P. and
         {Baumbach}, A. and {Berry}, K.~L. and {Biscani}, F. and {Boquien}, M. and
         {Bostroem}, K.~A. and {Bouma}, L.~G. and {Brammer}, G.~B. and
         {Bray}, E.~M. and {Breytenbach}, H. and {Buddelmeijer}, H. and
         {Burke}, D.~J. and {Calderone}, G. and {Cano Rodr{\'\i}guez}, J.~L. and
         {Cara}, M. and {Cardoso}, J.~V.~M. and {Cheedella}, S. and {Copin}, Y. and
         {Corrales}, L. and {Crichton}, D. and {D'Avella}, D. and {Deil}, C. and
         {Depagne}, {\'E}. and {Dietrich}, J.~P. and {Donath}, A. and
         {Droettboom}, M. and {Earl}, N. and {Erben}, T. and {Fabbro}, S. and
         {Ferreira}, L.~A. and {Finethy}, T. and {Fox}, R.~T. and
         {Garrison}, L.~H. and {Gibbons}, S.~L.~J. and {Goldstein}, D.~A. and
         {Gommers}, R. and {Greco}, J.~P. and {Greenfield}, P. and
         {Groener}, A.~M. and {Grollier}, F. and {Hagen}, A. and {Hirst}, P. and
         {Homeier}, D. and {Horton}, A.~J. and {Hosseinzadeh}, G. and {Hu}, L. and
         {Hunkeler}, J.~S. and {Ivezi{\'c}}, {\v{Z}}. and {Jain}, A. and
         {Jenness}, T. and {Kanarek}, G. and {Kendrew}, S. and {Kern}, N.~S. and
         {Kerzendorf}, W.~E. and {Khvalko}, A. and {King}, J. and {Kirkby}, D. and
         {Kulkarni}, A.~M. and {Kumar}, A. and {Lee}, A. and {Lenz}, D. and
         {Littlefair}, S.~P. and {Ma}, Z. and {Macleod}, D.~M. and
         {Mastropietro}, M. and {McCully}, C. and {Montagnac}, S. and
         {Morris}, B.~M. and {Mueller}, M. and {Mumford}, S.~J. and {Muna}, D. and
         {Murphy}, N.~A. and {Nelson}, S. and {Nguyen}, G.~H. and
         {Ninan}, J.~P. and {N{\"o}the}, M. and {Ogaz}, S. and {Oh}, S. and
         {Parejko}, J.~K. and {Parley}, N. and {Pascual}, S. and {Patil}, R. and
         {Patil}, A.~A. and {Plunkett}, A.~L. and {Prochaska}, J.~X. and
         {Rastogi}, T. and {Reddy Janga}, V. and {Sabater}, J. and
         {Sakurikar}, P. and {Seifert}, M. and {Sherbert}, L.~E. and
         {Sherwood-Taylor}, H. and {Shih}, A.~Y. and {Sick}, J. and
         {Silbiger}, M.~T. and {Singanamalla}, S. and {Singer}, L.~P. and
         {Sladen}, P.~H. and {Sooley}, K.~A. and {Sornarajah}, S. and
         {Streicher}, O. and {Teuben}, P. and {Thomas}, S.~W. and
         {Tremblay}, G.~R. and {Turner}, J.~E.~H. and {Terr{\'o}n}, V. and
         {van Kerkwijk}, M.~H. and {de la Vega}, A. and {Watkins}, L.~L. and
         {Weaver}, B.~A. and {Whitmore}, J.~B. and {Woillez}, J. and
         {Zabalza}, V. and {Astropy Contributors}},
        title = "{The Astropy Project: Building an Open-science Project and Status of the v2.0 Core Package}",
      journal = {\aj},
         year = 2018,
        month = sep,
       volume = {156},
       number = {3},
          eid = {123},
        pages = {123},
          doi = {10.3847/1538-3881/aabc4f},
archivePrefix = {arXiv},
       eprint = {1801.02634},
 primaryClass = {astro-ph.IM},
       adsurl = {https://ui.adsabs.harvard.edu/abs/2018AJ....156..123A}
}

@ARTICLE{Besla2007,
       author = {{Besla}, Gurtina and {Kallivayalil}, Nitya and {Hernquist}, Lars and {Robertson}, Brant and {Cox}, T.~J. and {van der Marel}, Roeland P. and {Alcock}, Charles},
        title = "{Are the Magellanic Clouds on Their First Passage about the Milky Way?}",
      journal = {\apj},
         year = 2007,
        month = oct,
       volume = {668},
       number = {2},
        pages = {949-967},
          doi = {10.1086/521385},
archivePrefix = {arXiv},
       eprint = {astro-ph/0703196},
 primaryClass = {astro-ph},
       adsurl = {https://ui.adsabs.harvard.edu/abs/2007ApJ...668..949B}
}

@ARTICLE{DiTeodoro2019,
       author = {{Di Teodoro}, E.~M. and {McClure-Griffiths}, N.~M. and {Jameson}, K.~E. and {D{\'e}nes}, H. and {Dickey}, John M. and {Stanimirovi{\'c}}, S. and {Staveley-Smith}, L. and {Anderson}, C. and {Bunton}, J.~D. and {Chippendale}, A. and {Lee-Waddell}, K. and {MacLeod}, A. and {Voronkov}, M.~A.},
        title = "{On the dynamics of the Small Magellanic Cloud through high-resolution ASKAP H I observations}",
      journal = {\mnras},
         year = 2019,
        month = feb,
       volume = {483},
       number = {1},
        pages = {392-406},
          doi = {10.1093/mnras/sty3095},
archivePrefix = {arXiv},
       eprint = {1811.09627},
 primaryClass = {astro-ph.GA},
       adsurl = {https://ui.adsabs.harvard.edu/abs/2019MNRAS.483..392D}
}

@ARTICLE{Saintonge2011,
       author = {{Saintonge}, Am{\'e}lie and {Kauffmann}, Guinevere and {Kramer}, Carsten and {Tacconi}, Linda J. and {Buchbender}, Christof and {Catinella}, Barbara and {Fabello}, Silvia and {Graci{\'a}-Carpio}, Javier and {Wang}, Jing and {Cortese}, Luca and {Fu}, Jian and {Genzel}, Reinhard and {Giovanelli}, Riccardo and {Guo}, Qi and {Haynes}, Martha P. and {Heckman}, Timothy M. and {Krumholz}, Mark R. and {Lemonias}, Jenna and {Li}, Cheng and {Moran}, Sean and {Rodriguez-Fernandez}, Nemesio and {Schiminovich}, David and {Schuster}, Karl and {Sievers}, Albrecht},
        title = "{COLD GASS, an IRAM legacy survey of molecular gas in massive galaxies - I. Relations between H$_{2}$, H I, stellar content and structural properties}",
      journal = {\mnras},
         year = 2011,
        month = jul,
       volume = {415},
       number = {1},
        pages = {32-60},
          doi = {10.1111/j.1365-2966.2011.18677.x},
archivePrefix = {arXiv},
       eprint = {1103.1642},
 primaryClass = {astro-ph.CO},
       adsurl = {https://ui.adsabs.harvard.edu/abs/2011MNRAS.415...32S}
}

@ARTICLE{Saintonge2017,
       author = {{Saintonge}, Am{\'e}lie and {Catinella}, Barbara and {Tacconi}, Linda J. and {Kauffmann}, Guinevere and {Genzel}, Reinhard and {Cortese}, Luca and {Dav{\'e}}, Romeel and {Fletcher}, Thomas J. and {Graci{\'a}-Carpio}, Javier and {Kramer}, Carsten and {Heckman}, Timothy M. and {Janowiecki}, Steven and {Lutz}, Katharina and {Rosario}, David and {Schiminovich}, David and {Schuster}, Karl and {Wang}, Jing and {Wuyts}, Stijn and {Borthakur}, Sanchayeeta and {Lamperti}, Isabella and {Roberts-Borsani}, Guido W.},
        title = "{xCOLD GASS: The Complete IRAM 30 m Legacy Survey of Molecular Gas for Galaxy Evolution Studies}",
      journal = {\apjs},
         year = 2017,
        month = dec,
       volume = {233},
       number = {2},
          eid = {22},
        pages = {22},
          doi = {10.3847/1538-4365/aa97e0},
archivePrefix = {arXiv},
       eprint = {1710.02157},
 primaryClass = {astro-ph.GA},
       adsurl = {https://ui.adsabs.harvard.edu/abs/2017ApJS..233...22S}
}

@ARTICLE{Catinella2018,
       author = {{Catinella}, Barbara and {Saintonge}, Am{\'e}lie and {Janowiecki}, Steven and {Cortese}, Luca and {Dav{\'e}}, Romeel and {Lemonias}, Jenna J. and {Cooper}, Andrew P. and {Schiminovich}, David and {Hummels}, Cameron B. and {Fabello}, Silvia and {Ger{\'e}b}, Katinka and {Kilborn}, Virginia and {Wang}, Jing},
        title = "{xGASS: total cold gas scaling relations and molecular-to-atomic gas ratios of galaxies in the local Universe}",
      journal = {\mnras},
         year = 2018,
        month = may,
       volume = {476},
       number = {1},
        pages = {875-895},
          doi = {10.1093/mnras/sty089},
archivePrefix = {arXiv},
       eprint = {1802.02373},
 primaryClass = {astro-ph.GA},
       adsurl = {https://ui.adsabs.harvard.edu/abs/2018MNRAS.476..875C}
}

@BOOK{Lauberts1982,
       author = {{Lauberts}, A.},
        title = "{ESO/Uppsala survey of the ESO(B) atlas}",
         year = 1982,
       adsurl = {https://ui.adsabs.harvard.edu/abs/1982euse.book.....L}
}

@ARTICLE{Weilbacher2003,
       author = {{Weilbacher}, P.~M. and {Duc}, P. -A. and {Fritze-v. Alvensleben}, U.},
        title = "{Tidal dwarf candidates in a sample of interacting galaxies. II. Properties and kinematics of the ionized gas}",
      journal = {\aap},
         year = 2003,
        month = jan,
       volume = {397},
        pages = {545-555},
          doi = {10.1051/0004-6361:20021522},
archivePrefix = {arXiv},
       eprint = {astro-ph/0210393},
 primaryClass = {astro-ph},
       adsurl = {https://ui.adsabs.harvard.edu/abs/2003A&A...397..545W}
}

@ARTICLE{Dey2019,
       author = {{Dey}, Arjun and {Schlegel}, David J. and {Lang}, Dustin and {Blum}, Robert and {Burleigh}, Kaylan and {Fan}, Xiaohui and {Findlay}, Joseph R. and {Finkbeiner}, Doug and {Herrera}, David and {Juneau}, St{\'e}phanie and {Landriau}, Martin and {Levi}, Michael and {McGreer}, Ian and {Meisner}, Aaron and {Myers}, Adam D. and {Moustakas}, John and {Nugent}, Peter and {Patej}, Anna and {Schlafly}, Edward F. and {Walker}, Alistair R. and {Valdes}, Francisco and {Weaver}, Benjamin A. and {Y{\`e}che}, Christophe and {Zou}, Hu and {Zhou}, Xu and {Abareshi}, Behzad and {Abbott}, T.~M.~C. and {Abolfathi}, Bela and {Aguilera}, C. and {Alam}, Shadab and {Allen}, Lori and {Alvarez}, A. and {Annis}, James and {Ansarinejad}, Behzad and {Aubert}, Marie and {Beechert}, Jacqueline and {Bell}, Eric F. and {BenZvi}, Segev Y. and {Beutler}, Florian and {Bielby}, Richard M. and {Bolton}, Adam S. and {Brice{\~n}o}, C{\'e}sar and {Buckley-Geer}, Elizabeth J. and {Butler}, Karen and {Calamida}, Annalisa and {Carlberg}, Raymond G. and {Carter}, Paul and {Casas}, Ricard and {Castander}, Francisco J. and {Choi}, Yumi and {Comparat}, Johan and {Cukanovaite}, Elena and {Delubac}, Timoth{\'e}e and {DeVries}, Kaitlin and {Dey}, Sharmila and {Dhungana}, Govinda and {Dickinson}, Mark and {Ding}, Zhejie and {Donaldson}, John B. and {Duan}, Yutong and {Duckworth}, Christopher J. and {Eftekharzadeh}, Sarah and {Eisenstein}, Daniel J. and {Etourneau}, Thomas and {Fagrelius}, Parker A. and {Farihi}, Jay and {Fitzpatrick}, Mike and {Font-Ribera}, Andreu and {Fulmer}, Leah and {G{\"a}nsicke}, Boris T. and {Gaztanaga}, Enrique and {George}, Koshy and {Gerdes}, David W. and {Gontcho}, Satya Gontcho A. and {Gorgoni}, Claudio and {Green}, Gregory and {Guy}, Julien and {Harmer}, Diane and {Hernandez}, M. and {Honscheid}, Klaus and {Huang}, Lijuan Wendy and {James}, David J. and {Jannuzi}, Buell T. and {Jiang}, Linhua and {Joyce}, Richard and {Karcher}, Armin and {Karkar}, Sonia and {Kehoe}, Robert and {Kneib}, Jean-Paul and {Kueter-Young}, Andrea and {Lan}, Ting-Wen and {Lauer}, Tod R. and {Le Guillou}, Laurent and {Le Van Suu}, Auguste and {Lee}, Jae Hyeon and {Lesser}, Michael and {Perreault Levasseur}, Laurence and {Li}, Ting S. and {Mann}, Justin L. and {Marshall}, Robert and {Mart{\'\i}nez-V{\'a}zquez}, C.~E. and {Martini}, Paul and {du Mas des Bourboux}, H{\'e}lion and {McManus}, Sean and {Meier}, Tobias Gabriel and {M{\'e}nard}, Brice and {Metcalfe}, Nigel and {Mu{\~n}oz-Guti{\'e}rrez}, Andrea and {Najita}, Joan and {Napier}, Kevin and {Narayan}, Gautham and {Newman}, Jeffrey A. and {Nie}, Jundan and {Nord}, Brian and {Norman}, Dara J. and {Olsen}, Knut A.~G. and {Paat}, Anthony and {Palanque-Delabrouille}, Nathalie and {Peng}, Xiyan and {Poppett}, Claire L. and {Poremba}, Megan R. and {Prakash}, Abhishek and {Rabinowitz}, David and {Raichoor}, Anand and {Rezaie}, Mehdi and {Robertson}, A.~N. and {Roe}, Natalie A. and {Ross}, Ashley J. and {Ross}, Nicholas P. and {Rudnick}, Gregory and {Safonova}, Sasha and {Saha}, Abhijit and {S{\'a}nchez}, F. Javier and {Savary}, Elodie and {Schweiker}, Heidi and {Scott}, Adam and {Seo}, Hee-Jong and {Shan}, Huanyuan and {Silva}, David R. and {Slepian}, Zachary and {Soto}, Christian and {Sprayberry}, David and {Staten}, Ryan and {Stillman}, Coley M. and {Stupak}, Robert J. and {Summers}, David L. and {Sien Tie}, Suk and {Tirado}, H. and {Vargas-Maga{\~n}a}, Mariana and {Vivas}, A. Katherina and {Wechsler}, Risa H. and {Williams}, Doug and {Yang}, Jinyi and {Yang}, Qian and {Yapici}, Tolga and {Zaritsky}, Dennis and {Zenteno}, A. and {Zhang}, Kai and {Zhang}, Tianmeng and {Zhou}, Rongpu and {Zhou}, Zhimin},
        title = "{Overview of the DESI Legacy Imaging Surveys}",
      journal = {\aj},
         year = 2019,
        month = may,
       volume = {157},
       number = {5},
          eid = {168},
        pages = {168},
          doi = {10.3847/1538-3881/ab089d},
archivePrefix = {arXiv},
       eprint = {1804.08657},
 primaryClass = {astro-ph.IM},
       adsurl = {https://ui.adsabs.harvard.edu/abs/2019AJ....157..168D}
}

@ARTICLE{Lada2010,
	author = {{Lada}, C.~J. and {Lombardi}, M. and {Alves}, J.~F.},
	title = "{On the Star Formation Rates in Molecular Clouds}",
	journal = {\apj},
	archivePrefix = "arXiv",
	eprint = {1009.2985},
	year = 2010,
	month = nov,
	volume = 724,
	pages = {687-693},
	doi = {10.1088/0004-637X/724/1/687},
	adsurl = {http://adsabs.harvard.edu/abs/2010ApJ...724..687L}
}

@ARTICLE{Leroy2009,
	author = {{Leroy}, A.~K. and {Walter}, F. and {Bigiel}, F. and {et al.}},
	title = "{Heracles: The HERA CO Line Extragalactic Survey}",
	journal = {\aj},
	archivePrefix = "arXiv",
	eprint = {0905.4742},
	primaryClass = "astro-ph.CO",
	year = 2009,
	month = jun,
	volume = 137,
	pages = {4670-4696},
	doi = {10.1088/0004-6256/137/6/4670},
	adsurl = {http://adsabs.harvard.edu/abs/2009AJ....137.4670L}
}

@ARTICLE{Bigiel2008,
	author = {{Bigiel}, F. and {Leroy}, A. and {Walter}, F. and {et al.}},
	title = "{The Star Formation Law in Nearby Galaxies on Sub-Kpc Scales}",
	journal = {\aj},
	archivePrefix = "arXiv",
	eprint = {0810.2541},
	year = 2008,
	month = dec,
	volume = 136,
	pages = {2846-2871},
	doi = {10.1088/0004-6256/136/6/2846},
	adsurl = {http://adsabs.harvard.edu/abs/2008AJ....136.2846B}
}

@ARTICLE{Bigiel2016,
       author = {{Bigiel}, Frank and {Leroy}, Adam K. and {Jim{\'e}nez-Donaire}, Maria J. and {Pety}, J{\'e}r{\^o}me and {Usero}, Antonio and {Cormier}, Diane and {Bolatto}, Alberto and {Garcia-Burillo}, Santiago and {Colombo}, Dario and {Gonz{\'a}lez-Garc{\'\i}a}, Manuel and {Hughes}, Annie and {Kepley}, Amanda A. and {Kramer}, Carsten and {Sandstrom}, Karin and {Schinnerer}, Eva and {Schruba}, Andreas and {Schuster}, Karl and {Tomicic}, Neven and {Zschaechner}, Laura},
        title = "{The EMPIRE Survey: Systematic Variations in the Dense Gas Fraction and Star Formation Efficiency from Full-disk Mapping of M51}",
      journal = {\apjl},
         year = 2016,
        month = may,
       volume = {822},
       number = {2},
          eid = {L26},
        pages = {L26},
          doi = {10.3847/2041-8205/822/2/L26},
archivePrefix = {arXiv},
       eprint = {1604.05316},
 primaryClass = {astro-ph.GA},
       adsurl = {https://ui.adsabs.harvard.edu/abs/2016ApJ...822L..26B}
}

@ARTICLE{Toribio2011,
	author = {{Toribio}, M.~C. and {Solanes}, J.~M. and {Giovanelli}, R. and 
	{Haynes}, M.~P. and {Masters}, K.~L.},
	title = "{H I Content and Optical Properties of Field Galaxies from the ALFALFA Survey. I. Selection of a Control Sample}",
	journal = {\apj},
	archivePrefix = "arXiv",
	eprint = {1103.0900},
	primaryClass = "astro-ph.CO",
	year = 2011,
	month = may,
	volume = 732,
	eid = {92},
	pages = {92},
	doi = {10.1088/0004-637X/732/2/92},
	adsurl = {http://adsabs.harvard.edu/abs/2011ApJ...732...92T}
}

@ARTICLE{Catinella2012,
       author = {{Catinella}, B. and {Schiminovich}, D. and {Kauffmann}, G. and {Fabello}, S. and {Hummels}, C. and {Lemonias}, J. and {Moran}, S.~M. and {Wu}, R. and {Cooper}, A. and {Wang}, J.},
        title = "{The GALEX Arecibo SDSS Survey. VI. Second data release and updated gas fraction scaling relations}",
      journal = {\aap},
         year = 2012,
        month = aug,
       volume = {544},
          eid = {A65},
        pages = {A65},
          doi = {10.1051/0004-6361/201219261},
archivePrefix = {arXiv},
       eprint = {1206.3059},
 primaryClass = {astro-ph.CO},
       adsurl = {https://ui.adsabs.harvard.edu/abs/2012A&A...544A..65C}
}

@ARTICLE{Chung2007,
	author = {{Chung}, A. and {van Gorkom}, J.~H. and {Kenney}, J.~D.~P. and 
	{Vollmer}, B.},
	title = "{Virgo Galaxies with Long One-sided H I Tails}",
	journal = {\apjl},
	eprint = {arXiv:astro-ph/0703338},
	year = 2007,
	month = apr,
	volume = 659,
	pages = {L115-L119},
	doi = {10.1086/518034},
	adsurl = {http://adsabs.harvard.edu/abs/2007ApJ...659L.115C}
}

@ARTICLE{Hibbard1996,
	author = {{Hibbard}, J.~E. and {van Gorkom}, J.~H.},
	title = "{HI, HII, and R-Band Observations of a Galactic Merger Sequence}",
	journal = {\aj},
	eprint = {arXiv:astro-ph/9512035},
	year = 1996,
	month = feb,
	volume = 111,
	pages = {655},
	doi = {10.1086/117815},
	adsurl = {http://adsabs.harvard.edu/abs/1996AJ....111..655H}
}

@ARTICLE{Leroy2008,
       author = {{Leroy}, Adam K. and {Walter}, Fabian and {Brinks}, Elias and {Bigiel}, Frank and {de Blok}, W.~J.~G. and {Madore}, Barry and {Thornley}, M.~D.},
        title = "{The Star Formation Efficiency in Nearby Galaxies: Measuring Where Gas Forms Stars Effectively}",
      journal = {\aj},
         year = 2008,
        month = dec,
       volume = {136},
       number = {6},
        pages = {2782-2845},
          doi = {10.1088/0004-6256/136/6/2782},
archivePrefix = {arXiv},
       eprint = {0810.2556},
 primaryClass = {astro-ph},
       adsurl = {https://ui.adsabs.harvard.edu/abs/2008AJ....136.2782L}
}

@ARTICLE{Hallenbeck2016,
       author = {{Hallenbeck}, Gregory and {Huang}, Shan and {Spekkens}, Kristine and {Haynes}, Martha P. and {Giovanelli}, Riccardo and {Adams}, Elizabeth A.~K. and {Brinchmann}, Jarle and {Carpenter}, John and {Chengalur}, Jayaram and {Hunt}, Leslie K. and {Masters}, Karen L. and {Saintonge}, Am{\'e}lie},
        title = "{HIghMass{\textemdash}High H I Mass, H I-rich Galaxies at z {\ensuremath{\sim}} 0: Combined H I and H2 Observations}",
      journal = {\aj},
         year = 2016,
        month = dec,
       volume = {152},
       number = {6},
          eid = {225},
        pages = {225},
          doi = {10.3847/1538-3881/152/6/225},
archivePrefix = {arXiv},
       eprint = {1610.03859},
 primaryClass = {astro-ph.GA},
       adsurl = {https://ui.adsabs.harvard.edu/abs/2016AJ....152..225H}
}

@ARTICLE{Hallenbeck2014,
       author = {{Hallenbeck}, Gregory and {Huang}, Shan and {Spekkens}, Kristine and {Haynes}, Martha P. and {Giovanelli}, Riccardo and {Adams}, Elizabeth A.~K. and {Brinchmann}, Jarle and {Chengalur}, Jayaram and {Hunt}, Leslie K. and {Masters}, Karen L. and {Saintonge}, Am{\'e}lie},
        title = "{HIghMass{\textemdash}High H I Mass, H I-rich Galaxies at z \raisebox{-0.5ex}\textasciitilde 0 High-resolution VLA Imaging of UGC 9037 and UGC 12506}",
      journal = {\aj},
         year = 2014,
        month = oct,
       volume = {148},
       number = {4},
          eid = {69},
        pages = {69},
          doi = {10.1088/0004-6256/148/4/69},
archivePrefix = {arXiv},
       eprint = {1407.1744},
 primaryClass = {astro-ph.GA},
       adsurl = {https://ui.adsabs.harvard.edu/abs/2014AJ....148...69H}
}

@ARTICLE{Mogotsi2016,
       author = {{Mogotsi}, K.~M. and {de Blok}, W.~J.~G. and {Cald{\'u}-Primo}, A. and {Walter}, F. and {Ianjamasimanana}, R. and {Leroy}, A.~K.},
        title = "{H I and CO Velocity Dispersions in Nearby Galaxies}",
      journal = {\aj},
         year = 2016,
        month = jan,
       volume = {151},
       number = {1},
          eid = {15},
        pages = {15},
          doi = {10.3847/0004-6256/151/1/15},
archivePrefix = {arXiv},
       eprint = {1511.06006},
 primaryClass = {astro-ph.GA},
       adsurl = {https://ui.adsabs.harvard.edu/abs/2016AJ....151...15M}
}

@ARTICLE{Martin2005,
       author = {{Martin}, D. Christopher and {Fanson}, James and {Schiminovich}, David and {Morrissey}, Patrick and {Friedman}, Peter G. and {Barlow}, Tom A. and {Conrow}, Tim and {Grange}, Robert and {Jelinsky}, Patrick N. and {Milliard}, Bruno and {Siegmund}, Oswald H.~W. and {Bianchi}, Luciana and {Byun}, Yong-Ik and {Donas}, Jose and {Forster}, Karl and {Heckman}, Timothy M. and {Lee}, Young-Wook and {Madore}, Barry F. and {Malina}, Roger F. and {Neff}, Susan G. and {Rich}, R. Michael and {Small}, Todd and {Surber}, Frank and {Szalay}, Alex S. and {Welsh}, Barry and {Wyder}, Ted K.},
        title = "{The Galaxy Evolution Explorer: A Space Ultraviolet Survey Mission}",
      journal = {\apjl},
         year = 2005,
        month = jan,
       volume = {619},
       number = {1},
        pages = {L1-L6},
          doi = {10.1086/426387},
archivePrefix = {arXiv},
       eprint = {astro-ph/0411302},
 primaryClass = {astro-ph},
       adsurl = {https://ui.adsabs.harvard.edu/abs/2005ApJ...619L...1M}
}

@ARTICLE{Wright2010,
       author = {{Wright}, Edward L. and {Eisenhardt}, Peter R.~M. and {Mainzer}, Amy K. and {Ressler}, Michael E. and {Cutri}, Roc M. and {Jarrett}, Thomas and {Kirkpatrick}, J. Davy and {Padgett}, Deborah and {McMillan}, Robert S. and {Skrutskie}, Michael and {Stanford}, S.~A. and {Cohen}, Martin and {Walker}, Russell G. and {Mather}, John C. and {Leisawitz}, David and {Gautier}, Thomas N., III and {McLean}, Ian and {Benford}, Dominic and {Lonsdale}, Carol J. and {Blain}, Andrew and {Mendez}, Bryan and {Irace}, William R. and {Duval}, Valerie and {Liu}, Fengchuan and {Royer}, Don and {Heinrichsen}, Ingolf and {Howard}, Joan and {Shannon}, Mark and {Kendall}, Martha and {Walsh}, Amy L. and {Larsen}, Mark and {Cardon}, Joel G. and {Schick}, Scott and {Schwalm}, Mark and {Abid}, Mohamed and {Fabinsky}, Beth and {Naes}, Larry and {Tsai}, Chao-Wei},
        title = "{The Wide-field Infrared Survey Explorer (WISE): Mission Description and Initial On-orbit Performance}",
      journal = {\aj},
         year = 2010,
        month = dec,
       volume = {140},
       number = {6},
        pages = {1868-1881},
          doi = {10.1088/0004-6256/140/6/1868},
archivePrefix = {arXiv},
       eprint = {1008.0031},
 primaryClass = {astro-ph.IM},
       adsurl = {https://ui.adsabs.harvard.edu/abs/2010AJ....140.1868W}
}

@article{Bolatto2013,
   author = {Alberto D. Bolatto and Mark Wolfire and Adam K. Leroy},
   doi = {10.1146/annurev-astro-082812-140944},
   journal = {\araa},
   month = {1},
   title = {The CO-to-H2 Conversion Factor},
   year = {2013},
}

@BOOK{Arp1987,
       author = {{Arp}, Halton C. and {Madore}, B.},
        title = "{A catalogue of southern peculiar galaxies and associations}",
         year = 1987,
       adsurl = {https://ui.adsabs.harvard.edu/abs/1987cspg.book.....A}
}

@ARTICLE{Mathewson1992,
       author = {{Mathewson}, D.~S. and {Ford}, V.~L. and {Buchhorn}, M.},
        title = "{A Southern Sky Survey of the Peculiar Velocities of 1355 Spiral Galaxies}",
      journal = {\apjs},
         year = 1992,
        month = aug,
       volume = {81},
        pages = {413},
          doi = {10.1086/191700},
       adsurl = {https://ui.adsabs.harvard.edu/abs/1992ApJS...81..413M}
}

@ARTICLE{Gereb2016,
       author = {{Ger{\'e}b}, K. and {Catinella}, B. and {Cortese}, L. and {Bekki}, K. and {Moran}, S.~M. and {Schiminovich}, D.},
        title = "{GASS 3505: the prototype of H I-excess, passive galaxies}",
      journal = {\mnras},
         year = 2016,
        month = oct,
       volume = {462},
       number = {1},
        pages = {382-394},
          doi = {10.1093/mnras/stw1675},
archivePrefix = {arXiv},
       eprint = {1607.01446},
 primaryClass = {astro-ph.GA},
       adsurl = {https://ui.adsabs.harvard.edu/abs/2016MNRAS.462..382G}
}

@ARTICLE{Accurso2017,
       author = {{Accurso}, G. and {Saintonge}, A. and {Catinella}, B. and {Cortese}, L. and {Dav{\'e}}, R. and {Dunsheath}, S.~H. and {Genzel}, R. and {Gracia-Carpio}, J. and {Heckman}, T.~M. and {Jimmy} and {Kramer}, C. and {Li}, Cheng and {Lutz}, K. and {Schiminovich}, D. and {Schuster}, K. and {Sternberg}, A. and {Sturm}, E. and {Tacconi}, L.~J. and {Tran}, K.~V. and {Wang}, J.},
        title = "{Deriving a multivariate {\ensuremath{\alpha}}$_{CO}$ conversion function using the [C II]/CO (1-0) ratio and its application to molecular gas scaling relations}",
      journal = {\mnras},
         year = 2017,
        month = oct,
       volume = {470},
       number = {4},
        pages = {4750-4766},
          doi = {10.1093/mnras/stx1556},
archivePrefix = {arXiv},
       eprint = {1702.03888},
 primaryClass = {astro-ph.GA},
       adsurl = {https://ui.adsabs.harvard.edu/abs/2017MNRAS.470.4750A}
}

@ARTICLE{Solomon1997,
       author = {{Solomon}, P.~M. and {Downes}, D. and {Radford}, S.~J.~E. and {Barrett}, J.~W.},
        title = "{The Molecular Interstellar Medium in Ultraluminous Infrared Galaxies}",
      journal = {\apj},
         year = 1997,
        month = mar,
       volume = {478},
       number = {1},
        pages = {144-161},
          doi = {10.1086/303765},
archivePrefix = {arXiv},
       eprint = {astro-ph/9610166},
 primaryClass = {astro-ph},
       adsurl = {https://ui.adsabs.harvard.edu/abs/1997ApJ...478..144S}
}

@ARTICLE{Tamburro2009,
       author = {{Tamburro}, D. and {Rix}, H.-W. and {Leroy}, A.~K. and {Mac Low}, M.-M. and {Walter}, F. and {Kennicutt}, R.~C. and {Brinks}, E. and {de Blok}, W.~J.~G.},
        title = "{What is Driving the H I Velocity Dispersion?}",
      journal = {\aj},
         year = 2009,
        month = may,
       volume = {137},
       number = {5},
        pages = {4424-4435},
          doi = {10.1088/0004-6256/137/5/4424},
archivePrefix = {arXiv},
       eprint = {0903.0183},
 primaryClass = {astro-ph.GA},
       adsurl = {https://ui.adsabs.harvard.edu/abs/2009AJ....137.4424T}
}

@ARTICLE{CalduPrimo2013,
       author = {{Cald{\'u}-Primo}, Anahi and {Schruba}, Andreas and {Walter}, Fabian and {Leroy}, Adam and {Sandstrom}, Karin and {de Blok}, W.~J.~G. and {Ianjamasimanana}, R. and {Mogotsi}, K.~M.},
        title = "{A High-dispersion Molecular Gas Component in Nearby Galaxies}",
      journal = {\aj},
         year = 2013,
        month = dec,
       volume = {146},
       number = {6},
          eid = {150},
        pages = {150},
          doi = {10.1088/0004-6256/146/6/150},
archivePrefix = {arXiv},
       eprint = {1309.6324},
 primaryClass = {astro-ph.CO},
       adsurl = {https://ui.adsabs.harvard.edu/abs/2013AJ....146..150C}
}

@ARTICLE{Hardwick2022b,
       author = {{Hardwick}, Jennifer A. and {Cortese}, Luca and {Obreschkow}, Danail and {Catinella}, Barbara},
        title = "{xGASS: the connection between angular momentum, mass, and atomic gas fraction in nearby galaxies}",
      journal = {\mnras},
         year = 2022,
        month = nov,
       volume = {516},
       number = {3},
        pages = {4043-4051},
          doi = {10.1093/mnras/stac2476},
archivePrefix = {arXiv},
       eprint = {2209.01720},
 primaryClass = {astro-ph.GA},
       adsurl = {https://ui.adsabs.harvard.edu/abs/2022MNRAS.516.4043H}
}

@ARTICLE{Obreschkow2016,
       author = {{Obreschkow}, D. and {Glazebrook}, K. and {Kilborn}, V. and {Lutz}, K.},
        title = "{Angular Momentum Regulates Atomic Gas Fractions of Galactic Disks}",
      journal = {\apjl},
         year = 2016,
        month = jun,
       volume = {824},
       number = {2},
          eid = {L26},
        pages = {L26},
          doi = {10.3847/2041-8205/824/2/L26},
archivePrefix = {arXiv},
       eprint = {1605.04927},
 primaryClass = {astro-ph.GA},
       adsurl = {https://ui.adsabs.harvard.edu/abs/2016ApJ...824L..26O}
}

@INPROCEEDINGS{Karanchentsev1999,
       author = {{Karachentsev}, I.~D. and {Makarov}, D.~I.},
        title = "{Galaxy Interactions in the Local Volume}",
    booktitle = {Galaxy Interactions at Low and High Redshift},
         year = 1999,
       editor = {{Barnes}, J.~E. and {Sanders}, D.~B.},
       series = {IAU Symposium},
       volume = {186},
        month = jan,
        pages = {109},
       adsurl = {https://ui.adsabs.harvard.edu/abs/1999IAUS..186..109K}
}

@ARTICLE{Skrutskie2006,
       author = {{Skrutskie}, M.~F. and {Cutri}, R.~M. and {Stiening}, R. and {Weinberg}, M.~D. and {Schneider}, S. and {Carpenter}, J.~M. and {Beichman}, C. and {Capps}, R. and {Chester}, T. and {Elias}, J. and {Huchra}, J. and {Liebert}, J. and {Lonsdale}, C. and {Monet}, D.~G. and {Price}, S. and {Seitzer}, P. and {Jarrett}, T. and {Kirkpatrick}, J.~D. and {Gizis}, J.~E. and {Howard}, E. and {Evans}, T. and {Fowler}, J. and {Fullmer}, L. and {Hurt}, R. and {Light}, R. and {Kopan}, E.~L. and {Marsh}, K.~A. and {McCallon}, H.~L. and {Tam}, R. and {Van Dyk}, S. and {Wheelock}, S.},
        title = "{The Two Micron All Sky Survey (2MASS)}",
      journal = {\aj},
         year = 2006,
        month = feb,
       volume = {131},
       number = {2},
        pages = {1163-1183},
          doi = {10.1086/498708},
       adsurl = {https://ui.adsabs.harvard.edu/abs/2006AJ....131.1163S}
}

@ARTICLE{Finlator2000,
       author = {{Finlator}, Kristian and {Ivezi{\'c}}, {\v{Z}}eljko and {Fan}, Xiaohui and {Strauss}, Michael A. and {Knapp}, Gillian R. and {Lupton}, Robert H. and {Gunn}, James E. and {Rockosi}, Constance M. and {Anderson}, John E. and {Csabai}, Istv{\'a}n and {Hennessy}, Gregory S. and {Hindsley}, Robert B. and {McKay}, Timothy A. and {Nichol}, Robert C. and {Schneider}, Donald P. and {Smith}, J. Allyn and {York}, Donald G. and {SDSS Collaboration}},
        title = "{Optical and Infrared Colors of Stars Observed by the Two Micron All Sky Survey and the Sloan Digital Sky Survey}",
      journal = {\aj},
         year = 2000,
        month = nov,
       volume = {120},
       number = {5},
        pages = {2615-2626},
          doi = {10.1086/316824},
archivePrefix = {arXiv},
       eprint = {astro-ph/0010052},
 primaryClass = {astro-ph},
       adsurl = {https://ui.adsabs.harvard.edu/abs/2000AJ....120.2615F}
}

@ARTICLE{Wen2013,
       author = {{Wen}, Xiao-Qing and {Wu}, Hong and {Zhu}, Yi-Nan and {Lam}, Man I. and {Wu}, Chao-Jian and {Wicker}, James and {Zhao}, Yong-Heng},
        title = "{The stellar masses of galaxies from the 3.4 {\ensuremath{\mu}}m band of the WISE All-Sky Survey}",
      journal = {\mnras},
         year = 2013,
        month = aug,
       volume = {433},
       number = {4},
        pages = {2946-2957},
          doi = {10.1093/mnras/stt939},
       adsurl = {https://ui.adsabs.harvard.edu/abs/2013MNRAS.433.2946W}
}

@article{Jozsa2007,
	author = {{Józsa, G. I. G.} and {Kenn, F.} and {Klein, U.} and {Oosterloo, T. A.}},
	title = {Kinematic modelling of disk galaxies - I. A new method to fit tilted rings to data cubes},
	DOI= "10.1051/0004-6361:20066164",
	url= "https://doi.org/10.1051/0004-6361:20066164",
	journal = {A&A},
	year = 2007,
	volume = 468,
	number = 2,
	pages = "731-774",
}

@INPROCEEDINGS{Fall1983,
       author = {{Fall}, S.~M.},
        title = "{Galaxy formation - Some comparisons between theory and observation}",
    booktitle = {Internal Kinematics and Dynamics of Galaxies},
         year = 1983,
       editor = {{Athanassoula}, E.},
       series = {IAU Symposium},
       volume = {100},
        month = jan,
        pages = {391-398},
       adsurl = {https://ui.adsabs.harvard.edu/abs/1983IAUS..100..391F}
}

@ARTICLE{Obreschkow2014,
       author = {{Obreschkow}, D. and {Glazebrook}, K.},
        title = "{Fundamental Mass-Spin-Morphology Relation Of Spiral Galaxies}",
      journal = {\apj},
         year = 2014,
        month = mar,
       volume = {784},
       number = {1},
          eid = {26},
        pages = {26},
          doi = {10.1088/0004-637X/784/1/26},
archivePrefix = {arXiv},
       eprint = {1312.4543},
 primaryClass = {astro-ph.GA},
       adsurl = {https://ui.adsabs.harvard.edu/abs/2014ApJ...784...26O}
}

@ARTICLE{ManceraPina2021,
       author = {{Mancera Pi{\~n}a}, Pavel E. and {Posti}, Lorenzo and {Fraternali}, Filippo and {Adams}, Elizabeth A.~K. and {Oosterloo}, Tom},
        title = "{The baryonic specific angular momentum of disc galaxies}",
      journal = {\aap},
         year = 2021,
        month = mar,
       volume = {647},
          eid = {A76},
        pages = {A76},
          doi = {10.1051/0004-6361/202039340},
archivePrefix = {arXiv},
       eprint = {2009.06645},
 primaryClass = {astro-ph.GA},
       adsurl = {https://ui.adsabs.harvard.edu/abs/2021A&A...647A..76M}
}

@ARTICLE{Haynes1984,
       author = {{Haynes}, Martha P. and {Giovanelli}, Ricardo and {Chincarini}, Guido L.},
        title = "{The Influence of Envirionment on the H I Content of Galaxies}",
      journal = {\araa},
         year = 1984,
        month = jan,
       volume = {22},
        pages = {445-470},
          doi = {10.1146/annurev.aa.22.090184.002305},
       adsurl = {https://ui.adsabs.harvard.edu/abs/1984ARA&A..22..445H}
}

@ARTICLE{Cortese2011,
       author = {{Cortese}, L. and {Catinella}, B. and {Boissier}, S. and {Boselli}, A. and {Heinis}, S.},
        title = "{The effect of the environment on the H I scaling relations}",
      journal = {\mnras},
         year = 2011,
        month = aug,
       volume = {415},
       number = {2},
        pages = {1797-1806},
          doi = {10.1111/j.1365-2966.2011.18822.x},
archivePrefix = {arXiv},
       eprint = {1103.5889},
 primaryClass = {astro-ph.CO},
       adsurl = {https://ui.adsabs.harvard.edu/abs/2011MNRAS.415.1797C}
}

@ARTICLE{Hess2013,
       author = {{Hess}, Kelley M. and {Wilcots}, Eric M.},
        title = "{Evolution in the H I Gas Content of Galaxy Groups: Pre-processing and Mass Assembly in the Current Epoch}",
      journal = {\aj},
         year = 2013,
        month = nov,
       volume = {146},
       number = {5},
          eid = {124},
        pages = {124},
          doi = {10.1088/0004-6256/146/5/124},
archivePrefix = {arXiv},
       eprint = {1308.4646},
 primaryClass = {astro-ph.CO},
       adsurl = {https://ui.adsabs.harvard.edu/abs/2013AJ....146..124H}
}

@ARTICLE{Stevens2018,
       author = {{Stevens}, Adam R.~H. and {Lagos}, Claudia del P. and {Obreschkow}, Danail and {Sinha}, Manodeep},
        title = "{Connecting and dissecting galaxies' angular momenta and neutral gas in a hierarchical universe: cue DARK SAGE}",
      journal = {\mnras},
         year = 2018,
        month = dec,
       volume = {481},
       number = {4},
        pages = {5543-5559},
          doi = {10.1093/mnras/sty2650},
archivePrefix = {arXiv},
       eprint = {1806.07402},
 primaryClass = {astro-ph.GA},
       adsurl = {https://ui.adsabs.harvard.edu/abs/2018MNRAS.481.5543S}
}

@ARTICLE{Westmeier2022,
       author = {{Westmeier}, T. and {Deg}, N. and {Spekkens}, K. and {Reynolds}, T.~N. and {Shen}, A.~X. and {Gaudet}, S. and {Goliath}, S. and {Huynh}, M.~T. and {Venkataraman}, P. and {Lin}, X. and {O'Beirne}, T. and {Catinella}, B. and {Cortese}, L. and {D{\'e}nes}, H. and {Elagali}, A. and {For}, B.-Q. and {J{\'o}zsa}, G.~I.~G. and {Howlett}, C. and {van der Hulst}, J.~M. and {Jurek}, R.~J. and {Kamphuis}, P. and {Kilborn}, V.~A. and {Kleiner}, D. and {Koribalski}, B.~S. and {Lee-Waddell}, K. and {Murugeshan}, C. and {Rhee}, J. and {Serra}, P. and {Shao}, L. and {Staveley-Smith}, L. and {Wang}, J. and {Wong}, O.~I. and {Zwaan}, M.~A. and {Allison}, J.~R. and {Anderson}, C.~S. and {Ball}, Lewis and {Bock}, D.~C.-J. and {Brodrick}, D. and {Bunton}, J.~D. and {Cooray}, F.~R. and {Gupta}, N. and {Hayman}, D.~B. and {Mahony}, E.~K. and {Moss}, V.~A. and {Ng}, A. and {Pearce}, S.~E. and {Raja}, W. and {Roxby}, D.~N. and {Voronkov}, M.~A. and {Warhurst}, K.~A. and {Courtois}, H.~M. and {Said}, K.},
        title = "{WALLABY pilot survey: Public release of H I data for almost 600 galaxies from phase 1 of ASKAP pilot observations}",
      journal = {\pasa},
         year = 2022,
        month = nov,
       volume = {39},
          eid = {e058},
        pages = {e058},
          doi = {10.1017/pasa.2022.50},
archivePrefix = {arXiv},
       eprint = {2211.07094},
 primaryClass = {astro-ph.GA},
       adsurl = {https://ui.adsabs.harvard.edu/abs/2022PASA...39...58W}
}

@ARTICLE{Koribalski2020,
       author = {{Koribalski}, B{\"a}rbel S. and {Staveley-Smith}, L. and {Westmeier}, T. and {Serra}, P. and {Spekkens}, K. and {Wong}, O.~I. and {Lee-Waddell}, K. and {Lagos}, C.~D.~P. and {Obreschkow}, D. and {Ryan-Weber}, E.~V. and {Zwaan}, M. and {Kilborn}, V. and {Bekiaris}, G. and {Bekki}, K. and {Bigiel}, F. and {Boselli}, A. and {Bosma}, A. and {Catinella}, B. and {Chauhan}, G. and {Cluver}, M.~E. and {Colless}, M. and {Courtois}, H.~M. and {Crain}, R.~A. and {de Blok}, W.~J.~G. and {D{\'e}nes}, H. and {Duffy}, A.~R. and {Elagali}, A. and {Fluke}, C.~J. and {For}, B.-Q. and {Heald}, G. and {Henning}, P.~A. and {Hess}, K.~M. and {Holwerda}, B.~W. and {Howlett}, C. and {Jarrett}, T. and {Jones}, D.~H. and {Jones}, M.~G. and {J{\'o}zsa}, G.~I.~G. and {Jurek}, R. and {J{\"u}tte}, E. and {Kamphuis}, P. and {Karachentsev}, I. and {Kerp}, J. and {Kleiner}, D. and {Kraan-Korteweg}, R.~C. and {L{\'o}pez-S{\'a}nchez}, {\'A}. R. and {Madrid}, J. and {Meyer}, M. and {Mould}, J. and {Murugeshan}, C. and {Norris}, R.~P. and {Oh}, S.-H. and {Oosterloo}, T.~A. and {Popping}, A. and {Putman}, M. and {Reynolds}, T.~N. and {Rhee}, J. and {Robotham}, A.~S.~G. and {Ryder}, S. and {Schr{\"o}der}, A.~C. and {Shao}, Li and {Stevens}, A.~R.~H. and {Taylor}, E.~N. and {van{\^A} der Hulst}, J.~M. and {Verdes-Montenegro}, L. and {Wakker}, B.~P. and {Wang}, J. and {Whiting}, M. and {Winkel}, B. and {Wolf}, C.},
        title = "{WALLABY ─ an SKA Pathfinder H I survey}",
      journal = {\apss},
         year = 2020,
        month = jul,
       volume = {365},
       number = {7},
          eid = {118},
        pages = {118},
          doi = {10.1007/s10509-020-03831-4},
archivePrefix = {arXiv},
       eprint = {2002.07311},
 primaryClass = {astro-ph.GA},
       adsurl = {https://ui.adsabs.harvard.edu/abs/2020Ap&SS.365..118K}
}

@ARTICLE{Adams2022,
       author = {{Adams}, E.~A.~K. and {Adebahr}, B. and {de Blok}, W.~J.~G. and {D{\'e}nes}, H. and {Hess}, K.~M. and {van der Hulst}, J.~M. and {Kutkin}, A. and {Lucero}, D.~M. and {Morganti}, R. and {Moss}, V.~A. and {Oosterloo}, T.~A. and {Orr{\'u}}, E. and {Schulz}, R. and {van Amesfoort}, A.~S. and {Berger}, A. and {Boersma}, O.~M. and {Bouwhuis}, M. and {van den Brink}, R. and {van Cappellen}, W.~A. and {Connor}, L. and {Coolen}, A.~H.~W.~M. and {Damstra}, S. and {van Diepen}, G.~N.~J. and {Dijkema}, T.~J. and {Ebbendorf}, N. and {Grange}, Y.~G. and {de Goei}, R. and {Gunst}, A.~W. and {Holties}, H.~A. and {Hut}, B. and {Ivashina}, M.~V. and {J{\'o}zsa}, G.~I.~G. and {van Leeuwen}, J. and {Loose}, G.~M. and {Maan}, Y. and {Mancini}, M. and {Mika}, {\'A}. and {Mulder}, H. and {Norden}, M.~J. and {Offringa}, A.~R. and {Oostrum}, L.~C. and {Pastor-Marazuela}, I. and {Pisano}, D.~J. and {Ponomareva}, A.~A. and {Romein}, J.~W. and {Ruiter}, M. and {Schoenmakers}, A.~P. and {van der Schuur}, D. and {Sluman}, J.~J. and {Smits}, R. and {Stuurwold}, K.~J.~C. and {Verstappen}, J. and {Vilchez}, N.~P.~E. and {Vohl}, D. and {Wierenga}, K.~J. and {Wijnholds}, S.~J. and {Woestenburg}, E.~E.~M. and {Zanting}, A.~W. and {Ziemke}, J.},
        title = "{First release of Apertif imaging survey data}",
      journal = {\aap},
         year = 2022,
        month = nov,
       volume = {667},
          eid = {A38},
        pages = {A38},
          doi = {10.1051/0004-6361/202244007},
archivePrefix = {arXiv},
       eprint = {2208.05348},
 primaryClass = {astro-ph.IM},
       adsurl = {https://ui.adsabs.harvard.edu/abs/2022A&A...667A..38A}
}

@ARTICLE{Romeo2020,
       author = {{Romeo}, Alessandro B.},
        title = "{From massive spirals to dwarf irregulars: a new set of tight scaling relations for cold gas and stars driven by disc gravitational instability}",
      journal = {\mnras},
         year = 2020,
        month = feb,
       volume = {491},
       number = {4},
        pages = {4843-4851},
          doi = {10.1093/mnras/stz3367},
archivePrefix = {arXiv},
       eprint = {1905.05752},
 primaryClass = {astro-ph.GA},
       adsurl = {https://ui.adsabs.harvard.edu/abs/2020MNRAS.491.4843R}
}

@ARTICLE{Murugeshan2019,
       author = {{Murugeshan}, Chandrashekar and {Kilborn}, Virginia and {Obreschkow}, Danail and {Glazebrook}, Karl and {Lutz}, Katharina and {D{\v{z}}ud{\v{z}}ar}, Robert and {D{\'e}nes}, Helga},
        title = "{Angular momentum regulates H I gas content and H I central hole size in the discs of spirals}",
      journal = {\mnras},
         year = 2019,
        month = feb,
       volume = {483},
       number = {2},
        pages = {2398-2412},
          doi = {10.1093/mnras/sty3265},
archivePrefix = {arXiv},
       eprint = {1811.11386},
 primaryClass = {astro-ph.GA},
       adsurl = {https://ui.adsabs.harvard.edu/abs/2019MNRAS.483.2398M}
}

@ARTICLE{Serra2015,
       author = {{Serra}, Paolo and {Westmeier}, Tobias and {Giese}, Nadine and {Jurek}, Russell and {Fl{\"o}er}, Lars and {Popping}, Attila and {Winkel}, Benjamin and {van der Hulst}, Thijs and {Meyer}, Martin and {Koribalski}, B{\"a}rbel S. and {Staveley-Smith}, Lister and {Courtois}, H{\'e}l{\`e}ne},
        title = "{SOFIA: a flexible source finder for 3D spectral line data}",
      journal = {\mnras},
         year = 2015,
        month = apr,
       volume = {448},
       number = {2},
        pages = {1922-1929},
          doi = {10.1093/mnras/stv079},
archivePrefix = {arXiv},
       eprint = {1501.03906},
 primaryClass = {astro-ph.IM},
       adsurl = {https://ui.adsabs.harvard.edu/abs/2015MNRAS.448.1922S}
}

@ARTICLE{Westmeier2021MNRAS,
       author = {{Westmeier}, T. and {Kitaeff}, S. and {Pallot}, D. and {Serra}, P. and {van der Hulst}, J.~M. and {Jurek}, R.~J. and {Elagali}, A. and {For}, B.-Q. and {Kleiner}, D. and {Koribalski}, B.~S. and {Lee-Waddell}, K. and {Mould}, J.~R. and {Reynolds}, T.~N. and {Rhee}, J. and {Staveley-Smith}, L.},
        title = "{SOFIA 2 - an automated, parallel H I source finding pipeline for the WALLABY survey}",
      journal = {\mnras},
         year = 2021,
        month = sep,
       volume = {506},
       number = {3},
        pages = {3962-3976},
          doi = {10.1093/mnras/stab1881},
archivePrefix = {arXiv},
       eprint = {2106.15789},
 primaryClass = {astro-ph.IM},
       adsurl = {https://ui.adsabs.harvard.edu/abs/2021MNRAS.506.3962W}
}

@ARTICLE{Cluver2014,
       author = {{Cluver}, M.~E. and {Jarrett}, T.~H. and {Hopkins}, A.~M. and {Driver}, S.~P. and {Liske}, J. and {Gunawardhana}, M.~L.~P. and {Taylor}, E.~N. and {Robotham}, A.~S.~G. and {Alpaslan}, M. and {Baldry}, I. and {Brown}, M.~J.~I. and {Peacock}, J.~A. and {Popescu}, C.~C. and {Tuffs}, R.~J. and {Bauer}, A.~E. and {Bland-Hawthorn}, J. and {Colless}, M. and {Holwerda}, B.~W. and {Lara-L{\'o}pez}, M.~A. and {Leschinski}, K. and {L{\'o}pez-S{\'a}nchez}, A.~R. and {Norberg}, P. and {Owers}, M.~S. and {Wang}, L. and {Wilkins}, S.~M.},
        title = "{Galaxy and Mass Assembly (GAMA): Mid-infrared Properties and Empirical Relations from WISE}",
      journal = {\apj},
         year = 2014,
        month = feb,
       volume = {782},
       number = {2},
          eid = {90},
        pages = {90},
          doi = {10.1088/0004-637X/782/2/90},
archivePrefix = {arXiv},
       eprint = {1401.0837},
 primaryClass = {astro-ph.GA},
       adsurl = {https://ui.adsabs.harvard.edu/abs/2014ApJ...782...90C}
}

@ARTICLE{Bekki2008,
       author = {{Bekki}, Kenji},
        title = "{Formation of blue compact dwarf galaxies from merging and interacting gas-rich dwarfs}",
      journal = {\mnras},
         year = 2008,
        month = jul,
       volume = {388},
       number = {1},
        pages = {L10-L14},
          doi = {10.1111/j.1745-3933.2008.00489.x},
archivePrefix = {arXiv},
       eprint = {0805.0662},
 primaryClass = {astro-ph},
       adsurl = {https://ui.adsabs.harvard.edu/abs/2008MNRAS.388L..10B}
}

@ARTICLE{Krumholz2009,
       author = {{Krumholz}, Mark R. and {McKee}, Christopher F. and {Tumlinson}, Jason},
        title = "{The Atomic-to-Molecular Transition in Galaxies. II: H I and H$_{2}$ Column Densities}",
      journal = {\apj},
         year = 2009,
        month = mar,
       volume = {693},
       number = {1},
        pages = {216-235},
          doi = {10.1088/0004-637X/693/1/216},
archivePrefix = {arXiv},
       eprint = {0811.0004},
 primaryClass = {astro-ph},
       adsurl = {https://ui.adsabs.harvard.edu/abs/2009ApJ...693..216K}
}

@ARTICLE{Ostriker2010,
       author = {{Ostriker}, Eve C. and {McKee}, Christopher F. and {Leroy}, Adam K.},
        title = "{Regulation of Star Formation Rates in Multiphase Galactic Disks: A Thermal/Dynamical Equilibrium Model}",
      journal = {\apj},
         year = 2010,
        month = oct,
       volume = {721},
       number = {2},
        pages = {975-994},
          doi = {10.1088/0004-637X/721/2/975},
archivePrefix = {arXiv},
       eprint = {1008.0410},
 primaryClass = {astro-ph.CO},
       adsurl = {https://ui.adsabs.harvard.edu/abs/2010ApJ...721..975O}
}

@ARTICLE{Stilp2013,
       author = {{Stilp}, Adrienne M. and {Dalcanton}, Julianne J. and {Warren}, Steven R. and {Skillman}, Evan and {Ott}, J{\"u}rgen and {Koribalski}, B{\"a}rbel},
        title = "{Global H I Kinematics in Dwarf Galaxies}",
      journal = {\apj},
         year = 2013,
        month = mar,
       volume = {765},
       number = {2},
          eid = {136},
        pages = {136},
          doi = {10.1088/0004-637X/765/2/136},
archivePrefix = {arXiv},
       eprint = {1301.1989},
 primaryClass = {astro-ph.GA},
       adsurl = {https://ui.adsabs.harvard.edu/abs/2013ApJ...765..136S}
}

@ARTICLE{Kleiner2017,
       author = {{Kleiner}, Dane and {Pimbblet}, Kevin A. and {Jones}, D. Heath and {Koribalski}, B{\"a}rbel S. and {Serra}, Paolo},
        title = "{Evidence for H I replenishment in massive galaxies through gas accretion from the cosmic web}",
      journal = {\mnras},
         year = 2017,
        month = apr,
       volume = {466},
       number = {4},
        pages = {4692-4710},
          doi = {10.1093/mnras/stw3328},
archivePrefix = {arXiv},
       eprint = {1701.03467},
 primaryClass = {astro-ph.GA},
       adsurl = {https://ui.adsabs.harvard.edu/abs/2017MNRAS.466.4692K}
}

@ARTICLE{Healy2024,
       author = {{Healy}, J. and {de Blok}, W.~J.~G. and {Maccagni}, F.~M. and {Amram}, P. and {Chemin}, L. and {Combes}, F. and {Holwerda}, B.~W. and {Kamphuis}, P. and {Pisano}, D.~J. and {Schinnerer}, E. and {Spekkens}, K. and {Verdes-Montenegro}, L. and {Walter}, F. and {Adams}, E.~A.~K. and {Gibson}, B.~K. and {Kleiner}, D. and {Veronese}, S. and {Zabel}, N. and {English}, J. and {Carignan}, C.},
        title = "{Possible origins of anomalous H I gas around MHONGOOSE galaxy, NGC 5068}",
      journal = {\aap},
         year = 2024,
        month = jul,
       volume = {687},
          eid = {A254},
        pages = {A254},
          doi = {10.1051/0004-6361/202347475},
archivePrefix = {arXiv},
       eprint = {2402.13749},
 primaryClass = {astro-ph.GA},
       adsurl = {https://ui.adsabs.harvard.edu/abs/2024A&A...687A.254H}
}

@ARTICLE{Lambert_2020,
       author = {{Lambert}, Trystan S. and {Kraan-Korteweg}, R.~C. and {Jarrett}, T.~H. and {Macri}, L.~M.},
        title = "{The 2MASS redshift survey galaxy group catalogue derived from a graph-theory based friends-of-friends algorithm}",
      journal = {\mnras},
         year = 2020,
        month = sep,
       volume = {497},
       number = {3},
        pages = {2954-2973},
          doi = {10.1093/mnras/staa1946},
archivePrefix = {arXiv},
       eprint = {2007.00581},
 primaryClass = {astro-ph.GA},
       adsurl = {https://ui.adsabs.harvard.edu/abs/2020MNRAS.497.2954L}
}

@dataset{Lambert_2023,
       author = {{Lambert}, T.~S. and {Kraan-Korteweg}, R.~C. and {Jarrett}, T.~H. and {Macri}, L.~M.},
        title = "{VizieR Online Data Catalog: 2MASS redshift survey galaxy group catalogue (Lambert+, 2020)}",
 howpublished = {VizieR On-line Data Catalog: J/MNRAS/497/2954. Originally published in: 2020MNRAS.497.2954L},
         year = 2023,
        month = jul,
          eid = {J/MNRAS/497/2954},
       adsurl = {https://ui.adsabs.harvard.edu/abs/2023yCat..74972954L}
}

@article{Carvalho2024,
	author = {{Carvalho}, Matheus C. and {Naguleswaran}, Bavithra and {Barmby}, Pauline and {Gorski}, Mark and {Köenig}, Sabine and {Holwerda}, Benne and {Young}, Jason},
	title = {A multi-wavelength overview of the giant spiral UGC 2885},
	DOI= "10.1051/0004-6361/202450916",
	url= "https://doi.org/10.1051/0004-6361/202450916",
	journal = {A&A},
	year = 2024,
	volume = 692,
	pages = "A105",
}

@ARTICLE{Sandstrom2013,
       author = {{Sandstrom}, K.~M. and {Leroy}, A.~K. and {Walter}, F. and {Bolatto}, A.~D. and {Croxall}, K.~V. and {Draine}, B.~T. and {Wilson}, C.~D. and {Wolfire}, M. and {Calzetti}, D. and {Kennicutt}, R.~C. and {Aniano}, G. and {Donovan Meyer}, J. and {Usero}, A. and {Bigiel}, F. and {Brinks}, E. and {de Blok}, W.~J.~G. and {Crocker}, A. and {Dale}, D. and {Engelbracht}, C.~W. and {Galametz}, M. and {Groves}, B. and {Hunt}, L.~K. and {Koda}, J. and {Kreckel}, K. and {Linz}, H. and {Meidt}, S. and {Pellegrini}, E. and {Rix}, H.-W. and {Roussel}, H. and {Schinnerer}, E. and {Schruba}, A. and {Schuster}, K.-F. and {Skibba}, R. and {van der Laan}, T. and {Appleton}, P. and {Armus}, L. and {Brandl}, B. and {Gordon}, K. and {Hinz}, J. and {Krause}, O. and {Montiel}, E. and {Sauvage}, M. and {Schmiedeke}, A. and {Smith}, J.~D.~T. and {Vigroux}, L.},
        title = "{The CO-to-H$_{2}$ Conversion Factor and Dust-to-gas Ratio on Kiloparsec Scales in Nearby Galaxies}",
      journal = {\apj},
         year = 2013,
        month = nov,
       volume = {777},
       number = {1},
          eid = {5},
        pages = {5},
          doi = {10.1088/0004-637X/777/1/5},
archivePrefix = {arXiv},
       eprint = {1212.1208},
 primaryClass = {astro-ph.CO},
       adsurl = {https://ui.adsabs.harvard.edu/abs/2013ApJ...777....5S}
}

@ARTICLE{Holwerda2011,
       author = {{Holwerda}, B.~W. and {Pirzkal}, N. and {de Blok}, W.~J.~G. and {Bouchard}, A. and {Blyth}, S.-L. and {van der Heyden}, K.~J. and {Elson}, E.~C.},
        title = "{Quantified H I morphology - I. Multi-wavelength analysis of the THINGS galaxies}",
      journal = {\mnras},
         year = 2011,
        month = oct,
       volume = {416},
       number = {4},
        pages = {2401-2414},
          doi = {10.1111/j.1365-2966.2011.18938.x},
archivePrefix = {arXiv},
       eprint = {1104.3291},
 primaryClass = {astro-ph.CO},
       adsurl = {https://ui.adsabs.harvard.edu/abs/2011MNRAS.416.2401H}
}

@ARTICLE{Holwerda_2012,
       author = {{Holwerda}, B.~W. and {Pirzkal}, N. and {Heiner}, J.~S.},
        title = "{Quantified H I morphology - VI. The morphology of extended discs in UV and H I}",
      journal = {\mnras},
         year = 2012,
        month = dec,
       volume = {427},
       number = {4},
        pages = {3159-3175},
          doi = {10.1111/j.1365-2966.2012.21975.x},
archivePrefix = {arXiv},
       eprint = {1207.4916},
 primaryClass = {astro-ph.CO},
       adsurl = {https://ui.adsabs.harvard.edu/abs/2012MNRAS.427.3159H}
}

@ARTICLE{Das_2013,
       author = {{Das}, M.},
        title = "{Giant Low Surface Brightness Galaxies: Evolution in Isolation}",
      journal = {Journal of Astrophysics and Astronomy},
         year = 2013,
        month = mar,
       volume = {34},
       number = {1},
        pages = {19-31},
          doi = {10.1007/s12036-013-9166-8},
archivePrefix = {arXiv},
       eprint = {1310.6495},
 primaryClass = {astro-ph.CO},
       adsurl = {https://ui.adsabs.harvard.edu/abs/2013JApA...34...19D}
}

@ARTICLE{Bigiel_2010,
       author = {{Bigiel}, F. and {Leroy}, A. and {Seibert}, M. and {Walter}, F. and {Blitz}, L. and {Thilker}, D. and {Madore}, B.},
        title = "{Tightly Correlated H I and FUV Emission in the Outskirts of M83}",
      journal = {\apjl},
         year = 2010,
        month = sep,
       volume = {720},
       number = {1},
        pages = {L31-L35},
          doi = {10.1088/2041-8205/720/1/L31},
archivePrefix = {arXiv},
       eprint = {1007.4550},
 primaryClass = {astro-ph.CO},
       adsurl = {https://ui.adsabs.harvard.edu/abs/2010ApJ...720L..31B}
}

@ARTICLE{Das_2021,
       author = {{Das}, M. and {Yadav}, J. and {Patra}, N. and {Dwarakanath}, K.~S. and {McGaugh}, S.~S. and {Schombert}, J. and {Rahna}, P.~T. and {Murthy}, J.},
        title = "{A comparison of the UV and HI properties of the extended UV (XUV) disk galaxies NGC 2541, NGC 5832 and ESO406-042}",
      journal = {Journal of Astrophysics and Astronomy},
         year = 2021,
        month = oct,
       volume = {42},
       number = {2},
          eid = {85},
        pages = {85},
          doi = {10.1007/s12036-021-09749-9},
archivePrefix = {arXiv},
       eprint = {2106.05520},
 primaryClass = {astro-ph.GA},
       adsurl = {https://ui.adsabs.harvard.edu/abs/2021JApA...42...85D}
}

@ARTICLE{Radburn-Smith_2014,
       author = {{Radburn-Smith}, David J. and {de Jong}, Roelof S. and {Streich}, David and {Bell}, Eric F. and {Dalcanton}, Julianne J. and {Dolphin}, Andrew E. and {Stilp}, Adrienne M. and {Monachesi}, Antonela and {Holwerda}, Benne W. and {Bailin}, Jeremy},
        title = "{Constraining the Age of the NGC 4565 H I Disk Warp: Determining the Origin of Gas Warps}",
      journal = {\apj},
         year = 2014,
        month = jan,
       volume = {780},
       number = {1},
          eid = {105},
        pages = {105},
          doi = {10.1088/0004-637X/780/1/105},
archivePrefix = {arXiv},
       eprint = {1312.2008},
 primaryClass = {astro-ph.GA},
       adsurl = {https://ui.adsabs.harvard.edu/abs/2014ApJ...780..105R}
}

@ARTICLE{Matthews2001,
       author = {{Matthews}, L.~D. and {van Driel}, W. and {Monnier-Ragaigne}, D.},
        title = "{H I observations of giant low surface brightness galaxies}",
      journal = {\aap},
         year = 2001,
        month = jan,
       volume = {365},
        pages = {1-10},
          doi = {10.1051/0004-6361:20000002},
archivePrefix = {arXiv},
       eprint = {astro-ph/0010075},
 primaryClass = {astro-ph},
       adsurl = {https://ui.adsabs.harvard.edu/abs/2001A&A...365....1M}
}




\appendix

\section{Notes on the sample galaxies}

All 5 sample galaxies show some asymmetry in their optical and \hi\ morphology which can indicate recent interactions. Here we present notes on the \hi\ and optical morphology of our sample galaxies and their environment. 

\subsection{ESO055-G013}

\begin{compactitem}
\item Two neighbouring galaxies within $\sim$ 0.5\,deg, PGC 14802 ($v_\mathrm{sys} = 7374$\,\kms) and PGC 14869 ($v_\mathrm{sys} = 10\,274$\,\kms).
\item The projected distance between ESO055-G013 and PGC 14802 is $\sim$ 946\,kpc, assuming that they are at the same distance. 
\end{compactitem}

\subsection{ESO208-G026}

\begin{compactitem}
\item The morphology of the \co\ disc (Fig.~\ref{fig:overlays}) indicates a warp in the \co\ disc. This is also supported by a $\sim 15^{\circ}$ change in the position angle of the outer rings in the \textsc{3DBarolo} model of the \co\ disc.  
\item No neighbouring galaxies. 
\end{compactitem}

\subsection{ESO378-G003}

\begin{compactitem}
\item A member of the NGC 3557 galaxy group \citep{Mathewson1992}; however, the galaxy is 3.76 degrees from NGC 3557 and most galaxies in this group are within 0.5 degrees from the central galaxy.  
\item This galaxy is 0.36 deg from the early-type galaxy NGC 3706, which corresponds to 268\,kpc projected distance between the two galaxies, assuming that they are at the same distance. The tidal index calculations confirm that the two galaxies are likely interacting. 
\item An interaction between these two galaxies may be able to explain the extended arms in the \hi\ distribution. 
\item There are clumpy star formation regions on the west and the south side of this galaxy (Fig.~\ref{fig:378_HI_map}), which overlap with slightly elevated \hi\ intensities.
\item One of the clumps to the south-east of the galaxy overlaps with the southern \hi\ tail. This galaxy could be a tidal-dwarf galaxy considering that it is relatively compact and the residuals of the \hi\ disc model indicate a velocity gradient and slightly higher velocity dispersion at the position of this clump.  

\begin{figure}
    \centering
    \includegraphics[width=0.5\textwidth]{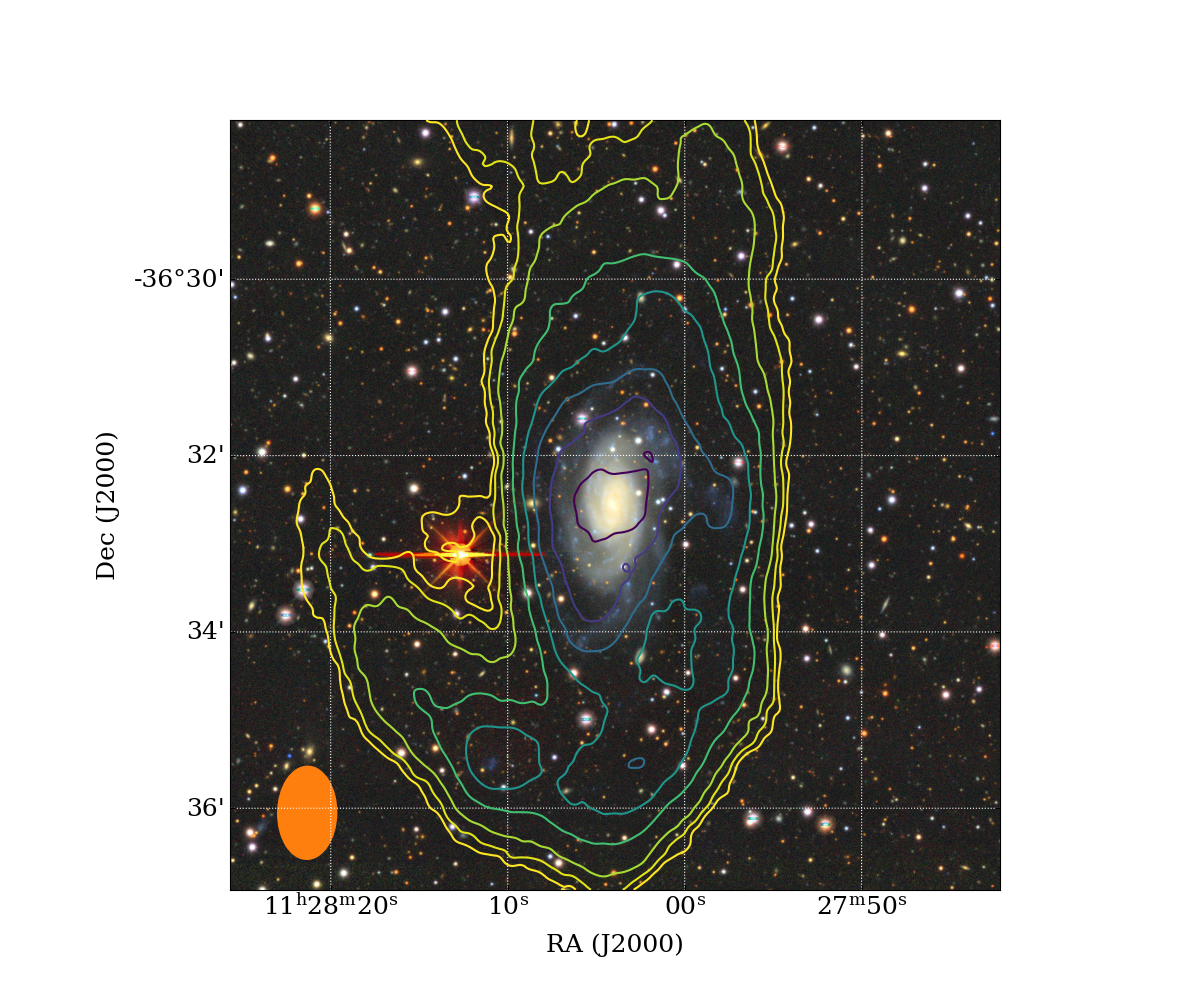}
    \caption{Legacy Sky Survey (DR10) colour image overlaid with the \hi\ column density map of ESO378-G003. Contours are 0.05, 0.1, 0.2, 0.4, 0.6, 0.8, 1.0, 1.2 $\times 10^{22}\,\mathrm{cm^{-2}}$. The orange ellipse shows the synthesised beam.}
    \label{fig:378_HI_map}
\end{figure}

\end{compactitem}

\subsection{ESO381-G005}

\begin{compactitem}
\item This galaxy has 2 small companion galaxies one to the north and one to the west \citep{Lauberts1982}. Both of these galaxies are within the \hi\ envelope of ESO381-G005 \citep{Weilbacher2003}.
\item A clump in the other part of the northern spiral arm (clump ``g'' in \citealt{Weilbacher2003}) is a tidal dwarf galaxy candidate.
\item This galaxy is classified as an M51-type galaxy \citep{Arp1987}, due to one of the companion galaxies appearing at the end of one of the spiral arms.
\item Based on the \hi\ morphology and the tidal index, this galaxy is clearly interacting with the galaxy to the south, PGC 629239. 
\item ESO381-G005 has an extended \hi\ tail which overlaps with both small companion galaxies to the east and extends directly towards PGC 629239 to the south (Fig.~\ref{fig:381_HI_map}). In addition, PGC 629239 has detected \hi\ with faint extensions to the north and the south. 
\end{compactitem}

\begin{figure}
    \centering
    \includegraphics[width=0.5\textwidth]{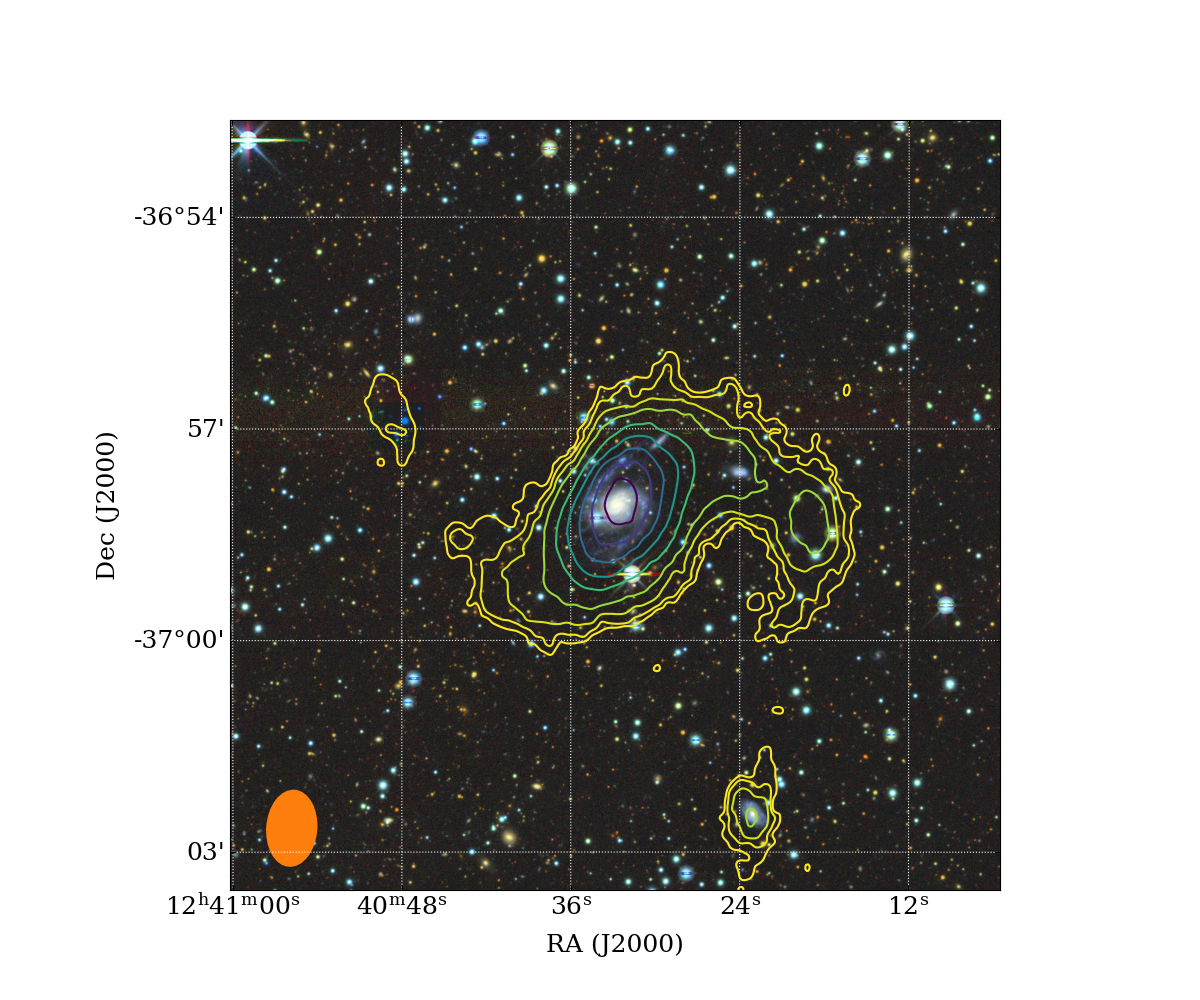}
    \caption{Legacy Sky Survey (DR10) colour image overlaid with the \hi\ column density map of ESO381-G005 and PGC 629239. Contours are 0.025, 0.05, 0.1, 0.2, 0.4, 0.6, 0.8, 1.0, 1.2 $\times 10^{22}\,\mathrm{cm^{-2}}$. The orange ellipse shows the synthesised beam.}
    \label{fig:381_HI_map}
\end{figure}

\subsection{ESO417-G018}

\begin{compactitem}
\item There is a neighbouring galaxy, ESO417-G019, at 0.64\,deg angular separation with $v_\mathrm{sys}= 4712 \pm 7$\,\kms, which corresponds to a projected separation of 757\,kpc.  
\end{compactitem}


\bsp	
\label{lastpage}
\end{document}